\documentclass[11pt,a4paper,aps,prd,superscriptaddress,nofootinbib,preprintnumbers]{revtex4-1}
\usepackage[utf8]{inputenc}
\usepackage{graphicx}
\usepackage{amssymb}
\usepackage{textcomp}
\usepackage{amsmath}
\usepackage{tabularx}
\usepackage{bm}
\usepackage{times}
\usepackage[dvipsnames,table]{xcolor}
\usepackage{hhline}
\usepackage{slashed}
\usepackage{multirow}
\usepackage{verbatim}
\usepackage{cancel}
\usepackage[normalem]{ulem}
\usepackage[sort&compress]{natbib}
\def\linkcolor{cyan!70!black}

\usepackage[colorlinks=true, pdfstartview=FitV, linkcolor=\linkcolor, citecolor=\linkcolor, urlcolor=\linkcolor]{hyperref}
\usepackage[capitalise]{cleveref}
\makeatletter
\AddToHook{cmd/appendix/before}{\crefalias{section}{appendix}}
\makeatother
\allowdisplaybreaks[4]

\usepackage{makecell}
\usepackage{orcidlink}

\usepackage{tikz} 
\usepackage{tkz-euclide}
\usetikzlibrary{backgrounds} 
\usetikzlibrary{decorations.pathmorphing}
\usetikzlibrary{arrows.meta}
\tikzset{
mystyle/.style={line width=1, baseline, scale=0.6, every node/.style={scale=1}},
v/.style={decorate, draw, decoration={snake, segment length=2.mm, amplitude=0.5mm}},
f/.style={draw, decoration={markings,mark=at position #1 with {\arrow[]{Latex[length=1.5mm,width=1.5mm]}}},
    postaction={decorate},node contents=#1},
f/.default=.6,
fb/.style={draw,decoration={markings,mark=at position #1 with {\arrowreversed[]{Latex[length=1.5mm,width=1.5mm]}}},
    postaction={decorate},node contents=#1},
fb/.default=.6,
s/.style={dashed,draw, decoration={markings,mark=at position #1 with {\arrow[]{Latex[length=1.5mm,width=1.5mm]}}},
    postaction={decorate},node contents=#1},
s/.default=.6,    
sb/.style={dashed,draw,decoration={markings,mark=at position #1 with {\arrowreversed[]{Latex[length=1.5mm,width=1.5mm]}}},
    postaction={decorate},node contents=#1},
sb/.default=.4,
snar/.style={dashed,draw,line width =1.25pt},
cross/.style={cross out, draw=black, minimum size=2*(#1-\pgflinewidth), inner sep=0pt, outer sep=0pt}, }

\newcommand{\calO}{{\mathcal{O}}}

\newcommand{\GeV}{{\rm GeV}}

\newcommand{\C}{ {\tt C} }
\newcommand{\tL}{ {\tt L} }
\newcommand{\tR}{ {\tt R} }

\begin{document}

\title{\boldmath Baryon-number-violating nucleon decays in SMEFT extended with a light scalar}
\author{Xiao-Dong Ma\,\orcidlink{0000-0001-7207-7793}}
\email{maxid@scnu.edu.cn}
\affiliation{State Key Laboratory of Nuclear Physics and Technology, Institute of Quantum Matter,\\
South China Normal University, Guangzhou 510006, China}
\affiliation{Guangdong Basic Research Center of Excellence for Structure and Fundamental Interactions of Matter,\\ Guangdong Provincial Key Laboratory of Nuclear Science, Guangzhou 510006, China}
\author{Michael A.~Schmidt\,\orcidlink{0000-0002-8792-5537}}
\email{m.schmidt@unsw.edu.au}
\affiliation{
Sydney Consortium for Particle Physics and Cosmology,\\
School of Physics, The University of New South Wales, Sydney, New South Wales 2052, Australia}
\author{Weihang Zhang\,\orcidlink{0009-0001-8826-2499}}
\email{charlie.zhang@unsw.edu.au}
\affiliation{
Sydney Consortium for Particle Physics and Cosmology,\\
School of Physics, The University of New South Wales, Sydney, New South Wales 2052, Australia}
\preprint{}
\begin{abstract}
New light particles have received considerable attention in recent years. Baryon-number-violating (BNV) nucleon decays involving such light particles are able to provide stringent constraints. They exhibit 
distinctive experimental signatures that merit thorough investigation.  
We systematically investigate BNV nucleon decay with a light scalar  in an effective field theory framework. Within this framework, we set stringent bounds on BNV operators using available experimental data and predict the occurrence of several BNV three-body nucleon decays.
We further study contributions to dinucleon to dilepton transitions in a nucleus mediated by the scalar, which complements single nucleon decay. Finally, we provide three ultraviolet-complete models that can generate different subsets of BNV operators in leading order.
Our theoretical framework will facilitate experimental searches for those exotic nucleon decays.

\end{abstract}
\preprint{CPPC-2025-09}
\maketitle

\tableofcontents

\newpage

%%%%%%%%%%%%%%%%%%%%%%%%%%%%%%
\section{Introduction}
%%%%%%%%%%%%%%%%%%%%%%%%%%%%%%

A few months ago, JUNO~\cite{JUNO:2015zny}, a 20 kton liquid scintillator detector, started collecting its first data. In the next years it will be followed by Hyper-Kamiokande~\cite{Hyper-Kamiokande:2018ofw}, a 188 kton water Cherenkov detector, and DUNE~\cite{DUNE:2020ypp}, a 68 kton liquid argon time-projection chamber, in addition to the two proposed detectors THEIA~\cite{Theia:2019non} (50 kton liquid argon) and ESSnuSB~\cite{ESSnuSB:2023ogw} (500 kton water Cherenkov detector). 
All of these detectors will search for baryon-number-violating (BNV) nucleon decays as one of their goals. In light of this, it is timely to revisit BNV signatures and to point out exotic BNV nucleon decays with novel signatures. 

BNV nucleon decays involving a new light particle form an intriguing class of exotic decay modes. They have recently been considered in both effective field theory (EFT) and ultraviolet (UV) models~\cite{Heeck:2019kgr,Heeck:2020nbq,Fridell:2023tpb,Domingo:2024qoj,Heeck:2025uwh,Helo:2025kgx}, including an axion-like particle~\cite{Li:2024liy,Fan:2025xhi}, a light sterile neutrino~\cite{Fajfer:2020tqf,Li:2025slp}, and a dark photon~\cite{Liao:2025vlj}. This new light particle could be connected to the neutron decay anomaly~\cite{Strumia:2021ybk,Fornal:2018eol}, dark matter~\cite{Fornal:2020poq}, and matter-antimatter asymmetry in the universe~\cite{Elor:2018twp}.\footnote{It may also be related to the neutrino mass mechanism, as suggested in~\cite{Heeck:2025uwh}. However, no concrete model  realizing this connection exists yet, and we plan to explore this possibility in future work.}
Given the vast interest of BNV interactions involving such a new light particle, this work aims to continue that effort by providing a systematic study of BNV nucleon decays with a light scalar particle, within both the EFT framework and UV-complete models.  

As nucleon decays are GeV-scale processes, they can be systematically considered via the EFT approach, where only the relevant degrees of freedom at each
characteristic energy scale are kept. 
For this purpose, we will start with the Standard Model effective field theory (SMEFT) extended with a light singlet scalar $\varphi$, named $\varphi$SMEFT. The $\varphi$SMEFT
is a natural framework to describe physics between the electroweak (EW) scale and the heavy new physics scale, and can be directly linked to the new physics models by integrating out new heavy particles.
At the EW scale, the full SM gauge symmetry $\rm SU(3)_c\otimes SU(2)_L\otimes U(1)_Y$ is broken into the subgroup $\rm SU(3)_c\otimes U(1)_{em}$. As the SM particles display a hierarchical mass spectrum, it motivates the construction of the low-energy EFT (LEFT), where the relevant degrees of freedom are the three charged leptons and neutrinos, and light $u,d,s,c,b$ quarks, while the heavy $W,Z$, Higgs, and top quark have been integrated out. In the LEFT framework extended by the scalar $\varphi$, termed $\varphi$LEFT, we construct the leading-order dimension-7 (dim-7) BNV operators and perform tree-level matching with the BNV $\varphi$SMEFT interactions at the EW scale. 
To deal with nucleon decay matrix elements induced by relevant $\varphi$LEFT interactions, a systematic and powerful approach is the chiral perturbation theory (ChPT)~\cite{Weinberg:1978kz,Gasser:1983yg,Gasser:1984gg} and its extension to the baryon sector~\cite{Jenkins:1990jv,Bijnens:1985kj}. The ChPT serves as an EFT of QCD at the nonperturbative energy scale below $\calO(\rm GeV)$, tailored to systematically organize the interactions among octet baryons and mesons. By employing the chiral framework for the BNV interactions developed in~\cite{Claudson:1981gh,Liao:2025vlj}, we 
formulate the general width expressions of two- and three-body nucleon decays involving $\varphi$ in terms of the $\varphi$LEFT and $\varphi$SMEFT Wilson coefficients.

Having established the general formalism, we proceed to investigate the experimental signatures of these
exotic nucleon decay modes and the constraints imposed by the existing data. Since $\varphi$ is assumed to be invisible in our analysis, the primary characteristic signature is the momentum distribution of visible SM particles, which encodes information about the scalar mass and the underlying interaction structure.    
With the aforementioned ingredients, we can employ the available data to study the relevant constraints.  
To do that, we first reinterpret the Super-Kamiokande (Super-K) experimental data~\cite{Super-Kamiokande:2015pys,Super-Kamiokande:2013rwg} to set limits on the two-body proton decay modes $p\to e^+(\mu^+)\varphi$ and the three-body neutron decay $n\to\bar\nu(\nu)\pi^0\varphi$ across the entire kinematically allowed mass range.
Together with the available bounds on the invisible neutron decay from the SNO experiment~\cite{SNO:2022trz} and recast bounds on three-body proton decays involving a $K^0$ or $K^+$ in~\cite{Fan:2025xhi},  
we can set stringent bounds on the Wilson coefficients of the $\varphi$LEFT and $\varphi$SMEFT operators.
Since these BNV interactions can induce dinucleon to dilepton transitions within a nucleus via a t-channel $\varphi$ mediator, we also explore the constraints from these $\Delta B=2$ processes, which complementarily constrain the relevant parameter space, especially when the mass of $\varphi$ exceeds that of the nucleon. 

The remainder of this paper is organized as follows. 
In \cref{sec:EFTs}, we construct the relevant leading-order $\varphi$SMEFT and $\varphi$LEFT BNV operators involving a light scalar, and perform the chiral matching to derive their hadronic counterparts. 
\cref{sec:nucleondecay} presents the general formalism for two- and three-body BNV nucleon decays with a light scalar, followed by a discussion of current experimental constraints on the associated Wilson coefficients and inverse decay widths. 
In \cref{sec:dinucleondecay}, we analyze the long-distance contribution to dinucleon decays mediated by the scalar and investigate the complementary constraints on EFT interactions when the scalar mass exceeds the neutron mass.  
Our conclusion is given in \cref{sec:conclusion}. 
Additionally, \cref{app:Gammaatm20} provides the general expressions of two- and three-body nucleon decays involving the scalar $\varphi$ in the final state in the limit of a massless scalar, while \cref{app:M2fordinucleon} lists the matrix elements for all possible dinucleon decay channels. 

%%%%%%%%%%%%%%%%%%%%%%%%%%%%%%
\section{Baryon number violation involving a light scalar in EFTs}
\label{sec:EFTs}
%%%%%%%%%%%%%%%%%%%%%%%%%%%%%%

%%%%%%%%%%%%%%%%%%%%%%%%%%%%%%
\subsection{SMEFT extended with a single light scalar ($\varphi$SMEFT)}
%%%%%%%%%%%%%%%%%%%%%%%%%%%%%%

\begin{table}[tb]
\center
\resizebox{0.9\linewidth}{!}{
\renewcommand{\arraystretch}{1.2}
\begin{tabular}{|c|l|c|l|}
\hline
\multicolumn{2}{|c|}{Dim 7: $\Delta(B-L)= 0$} &
\multicolumn{2}{c|}{Dim 8:  $\Delta(B+L)=0$}
\\
\hline
~~~$\calO_{LQdu\varphi} $~~~   
& $\epsilon_{\alpha\beta\gamma}\epsilon_{ij}(\overline{L^{i\C}} Q^{\alpha j} )
(\overline{d^{\beta\C}} u^{\gamma})\varphi $~~~ 
& ~~~$\calO_{LdudH\varphi}$~~~  
&$\epsilon_{\alpha\beta\gamma}
(\overline{L} d^{\alpha}\tilde{H})
(\overline{u^{\beta\C}} d^{\gamma}) 
\varphi$ ~~~  
\\
$\calO_{euQQ\varphi}$ 
& $\epsilon_{\alpha\beta\gamma}\epsilon_{ij}
(\overline{e^\C} u^{\alpha})(\overline{Q^{\beta i\C}} Q^{\gamma j})\varphi$
& $\calO_{LdddH\varphi}$
& $\epsilon_{\alpha\beta\gamma}
(\overline{L} d^{\alpha}H)
(\overline{d^{\beta\C}} d^{\gamma})\varphi $
\\
$\calO_{LQQQ\varphi}$
&$\epsilon_{\alpha\beta\gamma}
\epsilon_{ik}\epsilon_{jl}
(\overline{L^{i\C}}Q^{\alpha j})
(\overline{Q^{\beta k\C}}Q^{\gamma l})\varphi$
& $\calO_{eQddH\varphi}$
& $\epsilon_{\alpha\beta\gamma} 
(H^\dagger\overline{e} Q^{\alpha}) 
(\overline{d^{\beta\C}} d^{\gamma})\varphi$
\\
$\calO_{eudu\varphi}$
& $\epsilon_{\alpha\beta\gamma}
(\overline{e^\C} u^{\alpha})
(\overline{d^{\beta\C}} u^{\gamma})\varphi$
& $\calO_{LdQQH\varphi}$
& $\delta_{ij}\epsilon_{\alpha\beta\gamma} 
(\overline{L^i} d^{\alpha}) 
(H^\dagger \overline{Q^{\beta j \C}} Q^{\gamma}) \varphi$
\\
\hline
\end{tabular}
}
\caption{The $\varphi$SMEFT dim-7 and dim-8 BNV operators involving a light scalar. $\alpha,\beta,\gamma$ are color indices while the flavor indices are omitted for simplicity. $\tilde H=\epsilon H^*$ with $\epsilon_{ij}$ being the rank-2 totally antisymmetric tensor with $\epsilon_{12}=1$. }
\label{tab:SMEFTlike_ope}
\end{table}

We extend the SM with a single light singlet scalar $\varphi$, which can be a complex or real scalar field. The relevant effective field theory is $\varphi$SMEFT, which respects the full SM gauge symmetry $\rm SU(3)_c\otimes SU(2)_L \otimes U(1)_Y$. The leading BNV effective operators with the scalar field $\varphi$ arise at dimension 7 with $\Delta L=1$ [and thus $\Delta (B-L)=0$] and at dimension 8 with $\Delta L=-1$ or $\Delta (B+L)=0$. 
These operators are obtained by adding the scalar field $\varphi$ to the corresponding SMEFT dim-6 and dim-7 BNV operators~\cite{Grzadkowski:2010es,Lehman:2014jma,Liao:2016hru,Song:2023jqm}. 
The relevant\footnote{We do not include the higher dimensional operators with a covariant derivative like $\epsilon_{\alpha\beta\gamma}(\overline{L} \gamma^\mu Q^\alpha) (\overline{d^{\beta \C}} i\overleftrightarrow{D_\mu} d^\gamma)\varphi$,
because their contribution to nucleon decay is suppressed by $m_{\texttt N}/v$ due to the derivative interaction.} BNV operators are listed in \cref{tab:SMEFTlike_ope}, where the $Q$ and $L$ denote the left-handed SM quark and lepton doublets, $u$, $d$ and $e$ denote the right-handed up- and down-type quarks and charged leptons, and $H$ denotes the SM Higgs doublet.
In \cref{sec:UVmodels}, we will provide several UV-complete models that naturally generate these effective operators at tree level upon integrating out new heavy particles.

%%%%%%%%%%%%%%%%%%%%%%%%%%%%%%
\subsection{LEFT extended with a single light scalar ($\varphi$LEFT) }
%%%%%%%%%%%%%%%%%%%%%%%%%%%%%%

\begin{table}[t]
\center
\resizebox{\linewidth}{!}{
\renewcommand{\arraystretch}{1.3}
\begin{tabular}{|c|c|c|c|c|c|}
\hline
\multicolumn{3}{|c|}{$\Delta(B-L)= 0$} &
\multicolumn{3}{c|}{$\Delta(B+L)=0$}
\\
\hline
\multicolumn{2}{|c|}{Operators} 
& Matching (up vs down basis)
&\multicolumn{2}{c|}{Operators}
& Matching (up vs down basis)
\\
\hline
\multirow{3}*{ $\calO_{\varphi\nu\rm dud}^{\tL\tL}$  } 
& \multirow{3}*{ $\varphi (\overline{\nu_{\tL}^{\C}} d_{\tL}^\alpha) 
(\overline{u_{\tL}^{\beta \C}} d_{\tL}^\gamma)
\epsilon_{\alpha \beta \gamma}$ }
& $\makecell{C_{\varphi\nu\rm dud}^{\tL\tL,xyzw} = 
V_{ay}V_{bw}
( C^{xabz}_{LQQQ\varphi}  \\
+C^{xzab}_{LQQQ\varphi}
-C^{xzba}_{LQQQ\varphi})}$
& \multirow{3}*{ $\calO_{\varphi \bar{\ell} \rm ddd}^{\tR\tL}$}  
&\multirow{3}*{$\varphi (\overline{\ell_{\tL}} d_{\tR}^\alpha) 
(\overline{d_{\tL}^{\beta \C}} d_{\tL}^\gamma)
\epsilon_{\alpha \beta \gamma}$ }
& $\makecell{ C_{\varphi \bar{\ell}\rm ddd}^{\tR\tL,xyzw}=
\frac{v}{2\sqrt{2}}(V_{az}V_{bw}
\\
-V_{aw}V_{bz}) C_{LdQQH\varphi}^{xyab} }$ 
\\\cline{3-3}\cline{6-6}%
& 
& $\makecell{C_{\varphi\nu\rm dud}^{\tL\tL,xyzw} = V_{za}^*
( C^{xywa}_{LQQQ\varphi}  \\
+C^{xayw}_{LQQQ\varphi}
-C^{xawy}_{LQQQ\varphi})}$
& 
& 
& $\makecell{ C_{\varphi \bar{\ell}\rm ddd}^{\tR\tL,xyzw}=
\frac{v}{2\sqrt{2}} (C_{LdQQH\varphi}^{xyzw} -C_{LdQQH\varphi}^{xywz}) }$ 
\\%
\hline
\multirow{3}*{$\calO_{\varphi\ell\rm udu}^{\tL\tL}$} 
& \multirow{3}*{$\varphi (\overline{\ell_{\tL}^{\C}} u_{\tL}^\alpha) 
(\overline{d_{\tL}^{\beta \C}} u_{\tL}^\gamma)
\epsilon_{\alpha \beta \gamma}$ } 
& $\makecell{C_{\varphi\ell\rm udu}^{\tL\tL,xyzw} = V_{az}
( C^{xywa}_{LQQQ\varphi}  \\
+C^{xayw}_{LQQQ\varphi}
-C^{xawy}_{LQQQ\varphi})}$
& \multirow{3}*{$\calO_{\varphi\bar{\nu}\rm dud}^{\tR\tL}$}
&\multirow{3}*{ $\varphi (\overline{\nu_{\tL}} d_{\tR}^\alpha) 
(\overline{u_{\tL}^{\beta \C}} d_{\tL}^\gamma)
\epsilon_{\alpha \beta \gamma}$ }
& $C_{\varphi\bar{\nu}\rm dud}^{\tR\tL,xyzw} =
\frac{v}{\sqrt{2}}V_{aw}C_{LdQQH\varphi}^{xyza}$
\\\cline{3-3}\cline{6-6}%
& 
& $\makecell{C_{\varphi\ell\rm udu}^{\tL\tL,xyzw} = 
V_{ya}^* V_{wb}^*
( C^{xabz}_{LQQQ\varphi}  \\
+C^{xzab}_{LQQQ\varphi}
-C^{xzba}_{LQQQ\varphi})}$
& 
& 
& $C_{\varphi\bar{\nu}\rm dud}^{\tR\tL,xyzw} =
\frac{v}{\sqrt{2}}V_{za}^* C_{LdQQH\varphi}^{xyaw}$
\\%
\hline
$\calO_{\varphi\ell\rm duu}^{\tR\tL}$   
&$\varphi(\overline{\ell_{\tR}^{\C}} d_{\tR}^\alpha) 
(\overline{u_{\tL}^{\beta \C}} u_{\tL}^\gamma)
\epsilon_{\alpha \beta \gamma}$  
& ---
&$\calO_{\varphi\bar{\nu}\rm udd}^{\tR\tL}$  
&$\varphi(\overline{\nu_{\tL}} u_{\tR}^\alpha) 
(\overline{d_{\tL}^{\beta \C}} d_{\tL}^\gamma)
\epsilon_{\alpha \beta \gamma}$ 
& ---
\\%
\hline
\multirow{2}*{$\calO_{\varphi\ell\rm udu}^{\tR\tL}$ } 
& \multirow{2}*{$\varphi (\overline{\ell_{\tR}^{\C}} u_{\tR}^\alpha) 
(\overline{d_{\tL}^{\beta \C}} u_{\tL}^\gamma)
\epsilon_{\alpha \beta \gamma}$}
& $C_{\varphi\ell\rm udu}^{\tR\tL,xyzw} =
- 2 V_{az}C_{euQQ\varphi}^{xyaw}$ 
& \multirow{2}*{$\calO_{\varphi\bar{\ell}\rm ddd}^{\tL\tR}$ }
& \multirow{2}*{$\varphi (\overline{\ell_{\tR}} d_{\tL}^\alpha) 
(\overline{d_{\tR}^{\beta \C}} d_{\tR}^\gamma)
\epsilon_{\alpha \beta \gamma}$ }
& $C_{\varphi\bar{\ell}\rm ddd}^{\tL\tR,xyzw} = 
\frac{v}{\sqrt{2}}V_{ay}C_{eQddH\varphi}^{xazw}$
\\\cline{3-3}\cline{6-6}%
& 
& $C_{\varphi\ell\rm udu}^{\tR\tL,xyzw} =
- 2 V_{wa}^* C_{euQQ\varphi}^{xyza}$ 
& 
& 
& $C_{\varphi\bar{\ell}\rm ddd}^{\tL\tR,xyzw} = 
\frac{v}{\sqrt{2}} C_{eQddH\varphi}^{xyzw}$
\\%
\hline
$\calO_{\varphi\ell\rm duu}^{\tL\tR}$  
&  $\varphi (\overline{\ell_{\tL}^{\C}} d_{\tL}^\alpha) 
(\overline{u_{\tR}^{\beta \C}} u_{\tR}^\gamma)
\epsilon_{\alpha \beta \gamma}$ 
& ---
& $\calO_{\varphi \bar{\ell}\rm ddd}^{\tL\tL}$  
& $\varphi (\overline{\ell_{\tR}} d_{\tL}^\alpha) (\overline{d_{\tL}^{\beta \C}} d_{\tL}^\gamma)\epsilon_{\alpha \beta \gamma}$ 
& ---
\\%
\hline
\multirow{2}*{$\calO_{\varphi\ell\rm udu}^{\tL\tR}$}    
& \multirow{2}*{$\varphi (\overline{\ell_{\tL}^{\C}} u_{\tL}^\alpha) 
(\overline{d_{\tR}^{\beta \C}} u_{\tR}^\gamma)
\epsilon_{\alpha \beta \gamma}$ } 
& $C_{\varphi\ell\rm udu}^{\tL\tR,xyzw} = -  C_{LQdu\varphi}^{xyzw}$ 
&$\calO_{\varphi\bar{\nu}\rm dud}^{\tR\tR}$  
& $\varphi (\overline{\nu_{\tL}} d_{\tR}^\alpha) 
(\overline{u_{\tR}^{\beta \C}} d_{\tR}^\gamma)
\epsilon_{\alpha \beta \gamma}$ 
& $C_{\varphi\bar{\nu}\rm dud}^{\tR\tR,xyzw} = 
\frac{v}{\sqrt{2}}C_{LdudH\varphi}^{xyzw} $
\\\cline{3-6}%
& 
& $C_{\varphi\ell\rm udu}^{\tL\tR,xyzw} =
- V_{y a}^* C_{LQdu\varphi}^{xazw}$ 
& $\calO_{\varphi \bar{\ell}\rm ddd}^{\tR\tR}$  
& $\varphi (\overline{\ell_{\tL}} d_{\tR}^\alpha) 
(\overline{d_{\tR}^{\beta \C}} d_{\tR}^\gamma)
\epsilon_{\alpha \beta \gamma}$ 
& $C_{\varphi \bar{\ell}\rm ddd}^{\tR\tR,xyzw} =
\frac{v}{\sqrt{2}}C_{LdddH\varphi}^{xyzw}$
\\% 
\hline
\multirow{2}*{$\calO_{\varphi\nu\rm ddu}^{\tL\tR}$}  
& \multirow{2}*{$\varphi (\overline{\nu_{\tL}^{\C}} d_{\tL}^\alpha) 
(\overline{d_{\tR}^{\beta \C}} u_{\tR}^\gamma)
\epsilon_{\alpha \beta \gamma}$} 
& $C_{\varphi\nu\rm ddu}^{\tL\tR,xyzw} =
V_{ay}C_{LQdu\varphi}^{xazw}$
&
&
&
\\\cline{3-3}% 
& 
& $C_{\varphi\nu\rm ddu}^{\tL\tR,xyzw} = C_{LQdu\varphi}^{xyzw}$
& 
& 
& 
\\%
\cline{1-3}
$\calO_{\varphi\nu\rm udd}^{\tL\tR}$
& $\varphi (\overline{\nu_{\tL}^{\C}} u_{\tL}^\alpha) 
(\overline{d_{\tR}^{\beta \C}} d_{\tR}^\gamma)
\epsilon_{\alpha \beta \gamma}$
& ---
&  &  & 
\\%
\cline{1-3}
$\calO_{\varphi\ell\rm udu}^{\tR\tR}$  
&  $\varphi (\overline{\ell_{\tR}^{\C}} u_{\tR}^\alpha) 
(\overline{d_{\tR}^{\beta \C}} u_{\tR}^\gamma)
\epsilon_{\alpha \beta \gamma}$ 
& $\calO_{\varphi\ell\rm udu}^{\tR\tR,xyzw} = C_{eudu\varphi}^{xyzw}$
&  & & 
\\
\hline
\end{tabular}
}
\caption{The $\varphi$LEFT dim-7 BNV operators involving a light scalar and their matching onto the $\varphi$SMEFT dim-7 and dim-8 BNV interactions presented in Tab.~\ref{tab:SMEFTlike_ope}. 
$\alpha,\beta,\gamma$ are color indices while the $x,y,z,w$ denote the flavor indices. 
The matching is given in both up (upper cell) and down (lower cell) flavor bases. }
\label{tab:dim7ope}
\end{table}

Below the electroweak scale, physics is described by $\varphi$LEFT, low-energy effective theory extended with the light scalar $\varphi$. 
Effective operators in $\varphi$LEFT are invariant under QCD and QED, $\rm SU(3)_c\otimes U(1)_{em}$. The lowest dimensional BNV operators with a scalar field $\varphi$ arise at dimension 7. They are generated by the $\varphi$SMEFT operators in \cref{tab:SMEFTlike_ope}. \Cref{tab:dim7ope} presents $\Delta (B-L)=0$ operators on the left and $\Delta (B+L)=0$ operators on the right, where $u_{\tL,\tR}$ and $d_{\tL,\tR}$ denote light up- and down-type quarks and $\ell_{\tL,\tR}$ and $\nu_\tL$ denote charged leptons and neutrinos. Left-chiral fields are labelled with a subscript $\tL$ and right-chiral fields with a subscript $\tR$. 

Next, we consider tree-level matching of $\varphi\rm SMEFT$ interactions onto the $\varphi\rm LEFT$ interactions at the electroweak scale. This matching depends on the choice of quark flavor basis. We consider two possibilities: up-quark or down-quark flavor basis. We neglect neutrino masses in this analysis, and choose charged lepton flavor eigenstates to be mass eigenstates. 
In the up-quark flavor basis, both the left- and right-handed up-type quark fields, as well as the right-handed down-type quark fields, are already mass eigenstates. In this case, the weak eigenstates $d'_{\tL}$ and mass eigenstates $d_{\tL}$ of the left-handed down-type quarks are related by the Cabibbo-Kobayashi-Maskawa (CKM) matrix $V$~\cite{Cabibbo:1963yz, Kobayashi:1973fv} through $d'_{\tL}=V d_{\tL}$.
Conversely, in the down-quark flavor basis, the weak and mass eigenstates of left-handed up-type quarks are related by $u_\tL'=V^\dagger u_\tL$.
The results are listed in the third and sixth columns of \cref{tab:dim7ope}. The SM Higgs field in the dim-8 $\Delta (B+L)=0$ $\varphi$SMEFT operators is replaced by its vacuum expectation value (VEV) $v=(\sqrt{2} G_F)^{-1/2}\simeq 246$ GeV. The top row of each cell shows the result in the up-type quark basis and the bottom row the result in the down-type quark basis. If the matching does not depend on the basis, the result is presented in a single row. Note that the operators $\calO_{\varphi \ell\rm duu}^{\tR \tL (\tL \tR)}$, $\calO_{\varphi \nu\rm udd}^{\tL\tR}$, $\calO_{\varphi\bar \nu\rm udd}^{\tR \tL}$, and $\calO_{\varphi\bar\ell\rm ddd}^{\tL\tL}$ are not generated in the matching to dim-7 or dim-8 $\varphi$SMEFT operators.

%%%%%%%%%%%%%%%%%%%%%%%%%%%%%%
\subsection{Baryon chiral perturbation theory}
%%%%%%%%%%%%%%%%%%%%%%%%%%%%%%

To calculate the nucleon decay matrix elements based on the above quark-level operators, a systematic and reliable way is to work within the framework of baryon ChPT. 
We focus on the $\varphi$LEFT operators involving only light $u,d,s$ quarks, and employ the spurion field method to identify their chiral realizations. 
We denote the octet pseudoscalar field by
$\Sigma(x) = \xi^2(x) = \exp[i\sqrt{2}\Pi(x)/F_0]$
and baryon field by $B(x)$, with
\begin{align}
\Pi(x)  =   
\begin{pmatrix}
\frac{\pi^0}{\sqrt{2}}+\frac{\eta}{\sqrt{6}} & \pi^+ & K^+
\\
\pi^- & -\frac{\pi^0}{\sqrt{2}}+\frac{\eta}{\sqrt{6}} & K^0
\\
K^- & \bar{K}^0 & -\sqrt{\frac{2}{3}}\eta
\end{pmatrix},~
B(x) =
\begin{pmatrix}
{\Sigma^{0}\over \sqrt{2}}+{\Lambda^0 \over \sqrt{6}}  & \Sigma^+ & p 
\\
\Sigma^- & -{\Sigma^{0} \over \sqrt{2}}+{\Lambda^0 \over \sqrt{6}} &  n 
\\ 
\Xi^- & \Xi^0 & - \sqrt{2\over 3}\Lambda^0
\end{pmatrix},   
\end{align}
where $F_0 = (86.2\pm 0.5)\rm MeV$ is the pion decay constant in the chiral limit.
Then the relevant leading order BNV chiral Lagrangian terms are~\cite{Claudson:1981gh,Fan:2024gzc,Liao:2025vlj},
\begin{align}
{\cal L}_B^{\Delta B=1} &=
c_1 {\rm Tr}\big[ 
{\cal P}_{  \bar{\pmb{3}}_\tL \otimes \pmb{3}_\tR} \xi B_\tL \xi -
{\cal P}_{\pmb{3}_\tL \otimes \bar{\pmb{3}}_\tR} \xi^\dagger B_\tR \xi^\dagger 
 \big]
 +c_2  {\rm Tr}\big[ 
{\cal P}_{\pmb{8}_\tL \otimes \pmb{1}_\tR}\xi B_\tL \xi^\dagger
- {\cal P}_{ \pmb{1}_\tL \otimes  \pmb{8}_\tR} \xi^\dagger B_\tR \xi
\big],
\label{eq:LBlM}
\end{align}
where $c_1$ and $c_2$ are the low energy constants in the chiral matching and the recent LQCD results lead to $c_1=-0.01257(111)\,{\rm GeV}^3$ and $c_2=0.01269(107)\,{\rm GeV}^3$~\cite{Yoo:2021gql}. The low-energy constants $c_1$ and $c_2$ are often denoted $\alpha$ and $\beta$, respectively.
Here ${\cal P}_i$ are spurion matrices consisting of the lepton components and the scalar $\varphi$ multiplied by their Wilson coefficients~\cite{Fan:2024gzc}:
\begin{subequations}
\label{eq:spu}
\begin{align}
{\cal P}_{\bar{\pmb{3}}_\tL \otimes \pmb{3}_\tR} & = \varphi 
\begin{pmatrix}
C^{\tL\tR,x}_{\varphi\nu uds}\overline{\nu_{\tL x}^{\C}} 
& {\cellcolor{gray!15}C^{\tL\tR,x}_{\varphi\bar{\ell} dds} \overline{\ell_{\tR x}} }
&{\cellcolor{gray!15}{C^{\tL\tR,x}_{\varphi\bar{\ell}sds} \overline{\ell_{\tR x}}} } 
\\[1pt]
C^{\tL\tR,x}_{\varphi\ell usu} \overline{\ell_{\tL x}^{\C}}
& C^{\tL\tR,x}_{\varphi\nu dsu}  \overline{\nu_{\tL x}^{\C}} 
&C^{\tL\tR,x}_{\varphi\nu ssu}  \overline{\nu_{\tL x}^{\C}} 
\\[1pt]
C^{\tL\tR,x}_{\varphi\ell uud} \overline{\ell_{\tL x}^{\C}} 
& C^{\tL\tR,x}_{\varphi\nu dud}  \overline{\nu_{\tL x}^{\C}} 
&C^{\tL\tR,x}_{\varphi\nu sud} \overline{\nu_{\tL x}^{\C}} 
\end{pmatrix}, 
\\
{\cal P}_{\pmb{3}_\tL \otimes \bar{\pmb{3}}_\tR} & = \varphi
\begin{pmatrix}	
{\cellcolor{gray!15}C^{\tR\tL,x}_{\varphi\bar{\nu}uds}\overline{\nu_{\tL x}}  } 
& {\cellcolor{gray!15}C^{\tR\tL,x}_{\varphi\bar{\ell} dds} \overline{\ell_{\tL x}} }
& {\cellcolor{gray!15}C^{\tR\tL,x}_{\varphi\bar{\ell} sds} \overline{\ell_{\tL x}}} 
\\[1pt]
C^{\tR\tL,x}_{\varphi\ell usu} \overline{\ell_{\tR x}^{\C}}
& {\cellcolor{gray!15}C^{\tR\tL,x}_{\varphi\bar{\nu} dsu}  \overline{\nu_{\tL x}} }
&  {\cellcolor{gray!15}C^{\tR\tL,x}_{\varphi\bar{\nu} ssu}  \overline{\nu_{\tL x}}} 
\\[1pt]
C^{\tR\tL,x}_{\varphi\ell uud} \overline{\ell_{\tR x}^{\C}} 
&{\cellcolor{gray!15}C^{\tR\tL,x}_{\varphi\bar{\nu} dud}  \overline{\nu_{\tL x}} }
&{\cellcolor{gray!15}C^{\tR\tL,x}_{\varphi\bar{\nu} sud} \overline{\nu_{\tL x}} }
\end{pmatrix},
\\
{\cal P}_{\pmb{8}_\tL \otimes \pmb{1}_\tR} & = \varphi
\begin{pmatrix}
0  & {\cellcolor{gray!15}{C^{\tL\tL,x}_{\varphi\bar{\ell} dds} \overline{\ell_{\tR x}}}}
&  {\cellcolor{gray!15}C^{\tL\tL,x}_{\varphi\bar{\ell} sds} \overline{\ell_{\tR x}} }
\\[1pt]
C^{\tL\tL,x}_{\varphi\ell usu} \overline{\ell_{\tL x}^{\C}} 
& C^{\tL\tL,x}_{\varphi\nu dsu} \overline{\nu_{\tL x}^{\C}} 
& C^{\tL\tL,x}_{\varphi\nu ssu} \overline{\nu_{\tL x}^{\C}} 
\\[1pt]
C^{\tL\tL,x}_{\varphi\ell uud} \overline{\ell_{\tL x}^{\C}} 
& C^{\tL\tL,x}_{\varphi\nu dud} \overline{\nu_{\tL x}^{\C}} 
& C^{\tL\tL,x}_{\varphi\nu sud} \overline{\nu_{\tL x}^{\C}} 
\end{pmatrix},
\\
{\cal P}_{\pmb{1}_\tL \otimes \pmb{8}_\tR}& =\varphi 
\begin{pmatrix}
0  & {\cellcolor{gray!15}C^{\tR\tR,x}_{\varphi\bar{\ell} dds} \overline{\ell_{\tL x}}}
&  {\cellcolor{gray!15}C^{\tR\tR,x}_{\varphi\bar{\ell} sds} \overline{\ell_{\tL x}}} 
\\[1pt]
C^{\tR\tR,x}_{\varphi\ell usu} \overline{\ell_{\tR x}^{\C}} 
&  {\cellcolor{gray!15}C^{\tR\tR,x}_{\varphi\bar{\nu} dsu}  \overline{\nu_{\tL x}}} 
&  {\cellcolor{gray!15}C^{\tR\tR,x}_{\varphi\bar{\nu} ssu}  \overline{\nu_{\tL x}}}  
\\[1pt]
C^{\tR\tR,x}_{\varphi\ell uud} \overline{\ell_{\tR x}^{\C}} 
&  {\cellcolor{gray!15}C^{\tR\tR,x}_{\varphi\bar{\nu} dud}  \overline{\nu_{\tL x}} }
&  {\cellcolor{gray!15}C^{\tR\tR,x}_{\varphi\bar{\nu} sud} \overline{\nu_{\tL x}} }
\end{pmatrix}.
\end{align}
\end{subequations}
Note that the subscript quark flavor indices in the $\varphi$LEFT Wilson coefficients are organized to follow the pattern of chiral irreducible representation. 
Gray-shaded terms describe interactions with $\Delta (B+L)=0$, while the other terms describe interactions with $\Delta (B-L)=0$.
Expanding \cref{eq:LBlM} after taking into account spurions in \cref{eq:spu} leads to the following baryon octet-lepton-$\varphi$ contact interactions,  
\begin{align}
{\cal L}_{l B \varphi} \supset 
& \Big[ ( c_1 C^{\tL \tR,x}_{\varphi\ell uud} + 
 c_2 C^{\tL \tL,x}_{\varphi\ell uud } )
\overline{\ell_{\tL x}^{\C}} p_{\tL}
+( c_1 C^{\tL \tR,x}_{\varphi\ell usu} + 
c_2 C^{\tL \tL,x}_{\varphi\ell usu} )
\overline{\ell_{\tL x}^{\C}} \Sigma^+_{\tL}
\notag\\
& +\frac{1}{\sqrt{2}} \big[ c_1 
(C^{\tL \tR,x}_{\varphi\nu uds} 
- C^{\tL \tR,x}_{\varphi\nu dsu})
 - c_2 C^{\tL \tL,x}_{\varphi\nu dsu} \big]
\overline{\nu_{\tL x}^{\C}} \Sigma^0_{\tL}
\notag\\
& +\frac{1}{\sqrt{6}} \big[ c_1 
(C^{\tL \tR,x}_{\varphi\nu uds} + C^{\tL \tR,x}_{\varphi\nu dsu} 
- 2 C^{\tL \tR,x}_{\varphi\nu sud} )
+ c_2 ( C^{\tL \tL,x}_{\varphi\nu dsu} 
- 2 C^{\tL \tL,x}_{\varphi \nu sud}) \big] 
 \overline{\nu_{\tL x}^{\C}} \Lambda^0_{\tL}
\notag\\
& +( c_1 C^{\tL \tR,x}_{\varphi\nu dud} + 
c_2 C^{\tL \tL,x}_{\varphi\nu dud} )
\overline{\nu_{\tL x}^{\C}} n_{\tL} 
+ ( c_1 C^{\tL \tR,x}_{\varphi \bar\ell dds} 
+ c_2 C^{\tL \tL,x}_{\varphi \bar\ell dds} ) 
\overline{\ell_{\tR x}} \Sigma_\tL^-
\notag\\
& +( c_1 C^{\tL \tR,x}_{\varphi\nu ssu} + 
c_2 C^{\tL \tL,x}_{\varphi\nu ssu} )
\overline{\nu^{\C}_{\tL x}} \Xi^0_{\tL} 
+ ( c_1 C^{\tL \tR,x}_{\varphi \bar\ell sds} +  c_2 C^{\tL \tL,x}_{\varphi \bar\ell sds}) 
\overline{\ell_{\tR x}} \Xi_\tL^- \Big] \varphi
 - (\tL,\nu_\tL^{\C})\leftrightarrow (\tR,\nu_\tL).
\label{eq:VertexBl}
\end{align}
The last line containing $\Xi^0$ and $\Xi^-$ can not contribute to nucleon decays at leading order and the associated $\varphi$LEFT Wilson coefficients $ C^{\tL \tR(\tL\tL),x}_{\varphi\nu ssu}$ and $C^{\tL \tR(\tL\tL),x}_{\varphi \bar\ell sds}$  (and their chirality partners) will be omitted in the following analysis.

Expanding to first order in the meson fields, we obtain all operators contributing to nucleon decays that contain one meson and one charged lepton
\begin{align}
\mathcal{L}_{\ell\texttt{N}M\varphi} 
\supset
& \frac{i}{\sqrt{2}F_0} 
\Big[
\frac{1}{\sqrt{2}}(c_1  C^{\tL \tR}_{\varphi\ell uud} +c_2 C^{\tL \tL}_{\varphi\ell uud}) \overline{\ell^\C_\tL}p_\tL \pi^0
- \frac{1}{\sqrt{6}}(c_1  C^{\tL \tR}_{\varphi\ell uud} - 3 c_2 C^{\tL \tL}_{\varphi\ell uud}) \overline{\ell^\C_\tL}p_\tL \eta 
\notag\\
&+ (c_1  C^{\tL \tR}_{\varphi\ell usu} -c_2 C^{\tL \tL}_{\varphi\ell usu}) \overline{\ell^\C_\tL}p_\tL \overline{K^0} 
+ (c_1  C^{\tL \tR}_{\varphi\ell uud} +c_2 C^{\tL \tL}_{\varphi\ell uud}) \overline{\ell^\C_\tL}n_\tL \pi^+ 
\notag\\
& + (c_1  C^{\tL \tR}_{\varphi\overline{\ell}dds} -c_2 C^{\tL \tL}_{\varphi\overline{\ell}dds}) \overline{\ell_\tR}n_\tL K^-
\Big]\varphi + \tL \leftrightarrow \tR,
\label{eq:VertexBellM}
\end{align}
and one meson and one neutrino
\begin{align}
\mathcal{L}_{\nu \texttt{N}M\varphi} 
\supset & \frac{i}{\sqrt{2}F_0}
\Big[ (c_1  C^{\tL \tR}_{\varphi\nu dud} +c_2 C^{\tL \tL}_{\varphi\nu dud}) \overline{\nu^\C_\tL}p_\tL \pi^- 
+(c_1  (C^{\tL \tR}_{\varphi\nu sud} + C^{\tL \tR}_{\varphi\nu uds}) +c_2 C^{\tL \tL}_{\varphi\nu sud}) \overline{\nu^\C_\tL}p_\tL K^- 
\notag\\
& - \frac{1}{\sqrt{2}}(c_1  C^{\tL \tR}_{\varphi\nu dud} +c_2 C^{\tL \tL}_{\varphi\nu dud}) \overline{\nu^\C_\tL}n_\tL \pi^0 
- \frac{1}{\sqrt{6}}(c_1  C^{\tL \tR}_{\varphi\nu dud} - 3 c_2 C^{\tL \tL}_{\varphi\nu dud}) \overline{\nu^\C_\tL}n_\tL \eta 
\notag\\
&+ \big[c_1 (C^{\tL \tR}_{\varphi\nu sud} + C^{\tL \tR}_{\varphi\nu dsu}) + c_2 (C^{\tL \tL}_{\varphi\nu sud} - C^{\tL \tL}_{\varphi\nu dsu})\big] \overline{\nu^\C_\tL}n_\tL \overline{K^0}
\Big]\varphi
+ (\tL,\nu_\tL^{\C})\leftrightarrow (\tR,\nu_\tL).
\label{eq:VertexBvM}
\end{align}
Additionally, the conventional baryon-meson interactions in the SM are also relevant to the nucleon decays~\cite{Bijnens:1985kj}, 
\begin{align}
\label{LBBM}
{\cal L}_B
= & {\rm Tr}[\bar B (i \slashed{D} -M) B] 
+  \frac{D}{2} {\rm Tr}(\bar B \gamma^\mu \gamma_5\{u_\mu,B\}) 
+ \frac{F}{2} {\rm Tr}(\bar B \gamma^\mu \gamma_5 [u_\mu,B])
\nonumber
\\
\supset & 
\frac{D-F}{2F_0} 
 \big[
  \overline{\Sigma^0} \gamma^\mu \gamma_5 p \, \partial_\mu K^-
- \overline{\Sigma^0} \gamma^\mu \gamma_5 n\, \partial_\mu \overline{K^0} 
+ \sqrt{2}\big(\overline{\Sigma^+}\gamma^\mu \gamma_5 p \, \partial_\mu \overline{K^0} 
+ \overline{\Sigma^-} \gamma^\mu \gamma_5 n \, \partial_\mu K^- \big) 
\big]
\notag\\ 
& 
+ \frac{3F-D}{2\sqrt{3}F_0} 
\big(
  \overline{p} \gamma^\mu \gamma_5 p \, \partial_\mu \eta 
+ \overline{n} \gamma^\mu \gamma_5 n \, \partial_\mu \eta \big)
-\frac{D+3F}{2\sqrt{3}F_0} 
\big(
  \overline{\Lambda^0} \gamma^\mu \gamma_5 p \, \partial_\mu K^- 
+ \overline{\Lambda^0} \gamma^\mu \gamma_5 n \, \partial_\mu \overline{K^0}
\big)
\notag\\ 
&
+
\frac{D+F}{2F_0} 
\left[
  \overline{p} \gamma^\mu \gamma_5 p\, \partial_\mu \pi^0
- \overline{n} \gamma^\mu \gamma_5 n\, \partial_\mu \pi^0
+ \sqrt{2} \big(\overline{n} \gamma^\mu \gamma_5 p \, \partial_\mu \pi^- 
+ \overline{p} \gamma^\mu \gamma_5 n \, \partial_\mu \pi^+ \big)
\right],
\end{align}
where we have excluded operators that do not contain any nucleons.

%%%%%%%%%%%%%%%%%%%%%%%%%%%%%%
\subsection{Renormalization group corrections}
%%%%%%%%%%%%%%%%%%%%%%%%%%%%%%

As the BNV nucleon decay searches place very stringent limits on the scale of new physics operators, it is important to include renormalization group (RG) corrections in the analysis. In our analysis in \cref{sec:analysis}, we choose to consider operators with first and second generation flavor indices which directly contribute to BNV nucleon decays. In this scenario, the leading RG corrections originate from gluon and top loops and are universal. They can be translated from the relevant dim-6 and -7 BNV operators in SMEFT and dim-6 BNV operators in LEFT~\cite{Abbott:1980zj,Alonso:2014zka,Liao:2016hru,Jenkins:2017jig}\footnote{See \cite{Banik:2025wpi} for a recent calculation of the two-loop RG equations for dim-6 BNV operators in SMEFT.} and are described by
\begin{align}
\mu\frac{d C}{d\mu} = -\gamma_3\frac{\alpha_3}{2\pi} C-\gamma_t\frac{\alpha_t}{2\pi} C,
\end{align}  
where $C$ denotes the Wilson coefficient,  $\alpha_3=\frac{g_3^2}{4\pi}$, $\alpha_t=\frac{y_t^2}{4\pi}$, and $\gamma_3=2$, $\gamma_t=0$ for the BNV $\varphi$LEFT and dim-7 $\varphi$SMEFT operators listed in \cref{tab:SMEFTlike_ope,tab:dim7ope}, and $\gamma_t=-\frac12$ for dim-8 $\varphi$SMEFT operators. The RG evolution of $\alpha_{3,t}$ is described by
\begin{align}
\mu\frac{d\alpha_3}{d\mu} &= -\frac{\beta_{33}}{2\pi} \alpha_3^2,
&
\mu\frac{d\alpha_t}{d\mu} &=  -\frac{\beta_{tt}}{2\pi} \alpha_t^2 -\frac{\beta_{t3}}{2\pi} \alpha_3 \alpha_t,
\end{align}
with $\beta_{33}=11-\frac23 n_f$, $\beta_{tt}=-\frac92$, and $\beta_{t3}=8$.
At one-loop order, the equations can be solved analytically below the top quark mass 
\begin{align}
C(\mu) = C(\mu_0) \left(\frac{\alpha_s(\mu)}{\alpha_s(\mu_0)}\right)^{\gamma_3/\beta_{33}} .
\end{align}
Using $\alpha_s=0.1074$~\cite{Huang:2020hdv} at the top quark pole mass $M_t=173.21$ GeV as input, we find using RunDec~\cite{Chetyrkin:2000yt,Herren:2017osy}
\begin{align}
C(\Lambda_\chi)= 1.32\,C(M_t),
\end{align}
where $\Lambda_\chi\approx1.2\,\rm GeV$ is the chiral symmetry breaking scale. 
A numerical solution above the top quark mass threshold shows
\begin{subequations} 
\begin{align}
C^{(7)}(M_t) &= 1.35\, C^{(7)}(10^9 \mathrm{GeV}), &
C^{(7)}(M_t) &= 1.12\, C^{(7)}(10^4 \mathrm{GeV}), \\
C^{(8)}(M_t) &= 1.22\, C^{(8)}(10^7 \mathrm{GeV}),  & 
C^{(8)}(M_t) &= 1.10\, C^{(8)}(10^4 \mathrm{GeV}). 
\end{align}
\end{subequations}

%%%%%%%%%%%%%%%%%%%%%%%%%%%%%%
\section{Nucleon decays involving a new light scalar}
\label{sec:nucleondecay}
%%%%%%%%%%%%%%%%%%%%%%%%%%%%%%

\begin{table}[t]
\center
\resizebox{\linewidth}{!}{
\renewcommand{\arraystretch}{1.3}
\begin{tabular}{|l|c c|c c|c c|c c|c c c c c| }
\hline
\multirow{2}*{\rotatebox[origin=c]{90}{Ope.}}
&$~\calO_{\varphi\ell uud}^{\tL\tR,x}~$ 
&$~\calO_{\varphi\ell uud}^{\tL\tL,x}~$ 
&$~\calO_{\varphi\ell usu}^{\tL\tR,x}~$ 
&$~\calO_{\varphi\ell usu}^{\tL\tL,x}~$ 
&\cellcolor{gray!15}$~\calO_{\varphi\bar\ell dds}^{\tL\tR,x}~$
&\cellcolor{gray!15}$~\calO_{\varphi\bar\ell dds}^{\tL\tL,x}~$
&$~\calO_{\varphi\nu dud}^{\tL\tR,x}~$ 
&$~\calO_{\varphi\nu dud}^{\tL\tL,x}~$
&$~\calO_{\varphi\nu uds}^{\tL\tR,x}~$ 
&$~\calO_{\varphi\nu dsu}^{\tL\tR,x}~$ 
&$~\calO_{\varphi\nu sud}^{\tL\tR,x}~$
&$~\calO_{\varphi\nu dsu}^{\tL\tL,x}~$ 
&$~\calO_{\varphi\nu sud}^{\tL\tL,x}~$
\\
&$~\calO_{\varphi\ell uud}^{\tR\tL,x}~$ 
&$~\calO_{\varphi\ell uud}^{\tR\tR,x}~$ 
&$~\calO_{\varphi\ell usu}^{\tR\tL,x}~$ 
&$~\calO_{\varphi\ell usu}^{\tR\tR,x}~$ 
&\cellcolor{gray!15}$~\calO_{\varphi\bar\ell dds}^{\tR\tL,x}~$
&\cellcolor{gray!15}$~\calO_{\varphi\bar\ell dds}^{\tR\tR,x}~$
&\cellcolor{gray!15}$\calO_{\varphi\bar\nu dud}^{\tR\tL,x}~$ 
&\cellcolor{gray!15}$\calO_{\varphi\bar\nu dud}^{\tR\tR,x}~$
&\cellcolor{gray!15}$~\calO_{\varphi\bar\nu uds}^{\tR\tL,x}~$ 
&\cellcolor{gray!15}$~\calO_{\varphi\bar\nu dsu}^{\tR\tL,x}~$ 
&\cellcolor{gray!15}$~\calO_{\varphi\bar\nu sud}^{\tR\tL,x}~$
&\cellcolor{gray!15}$~\calO_{\varphi\bar\nu dsu}^{\tR\tR,x}~$ 
&\cellcolor{gray!15}$~\calO_{\varphi\bar\nu sud}^{\tR\tR,x}~$
\\\hline
\multirow{8}*{\rotatebox[origin=c]{90}{Nucleon decay modes}}
&\multicolumn{2}{c|}{$p\to e^+\varphi$ (\checkmark)} 
&\multicolumn{2}{c|}{$p\to e^+ K^{0}\varphi$  (\checkmark)}
&\multicolumn{2}{c|}{\cellcolor{gray!15}$n\to e^- K^+\varphi$}
&\multicolumn{2}{c|}{$n\to \bar\nu_x (\colorbox{gray!15}{$\nu_x$}) \varphi$  (\checkmark)}
&\multicolumn{5}{c|}{$p\to \bar\nu_x (\colorbox{gray!15}{$\nu_x$}) K^{+}\varphi$  (\checkmark)}
\\%
&\multicolumn{2}{c|}{$p\to \mu^+\varphi$  (\checkmark)} 
&\multicolumn{2}{c|}{$p\to \mu^+ K^{0}\varphi$ (\checkmark)}
&\multicolumn{2}{c|}{\cellcolor{gray!15}$n\to \mu^- K^+\varphi$}
&\multicolumn{2}{c|}{$p\to \bar\nu_x (\colorbox{gray!15}{$\nu_x$}) \pi^{+}\varphi$}
&\multicolumn{5}{c|}{$n\to \bar\nu_x (\colorbox{gray!15}{$\nu_x$}) K^0\varphi$} 
\\%
& \multicolumn{2}{c|}{$p\to e^+\pi^0\varphi $}
& & & &
& \multicolumn{2}{c|}{$n\to \bar\nu_x (\colorbox{gray!15}{$\nu_x$}) \pi^0\varphi$ (\checkmark)}
& & & & &
\\%
& \multicolumn{2}{c|}{$p\to \mu^+ \pi^0\varphi$}
& & & &
& \multicolumn{2}{c|}{$n\to \bar\nu_x (\colorbox{gray!15}{$\nu_x$}) \eta \varphi$} 
& & & & &
\\%
& \multicolumn{2}{c|}{$p\to e^+ \eta\varphi$}
& & & & & & & & & & &
\\%
& \multicolumn{2}{c|}{$p\to \mu^+ \eta\varphi$} & & & & & & & & & & &
\\%
& \multicolumn{2}{c|}{$n\to e^+ \pi^-\varphi$}
& & & & & & & & & & &
\\%
& \multicolumn{2}{c|}{$n\to \mu^+ \pi^-\varphi$} & & & & & & & & & & &
\\
\hline
\end{tabular} }
\caption{Summary of BNV nucleon decay modes involving a light scalar $\varphi$ and their corresponding $\varphi$LEFT operators that can induce them. The gray shading indicates nucleon decay modes with $\Delta (B+L)=0$ and checkmarks indicate nucleon decays with experimental constraints. } 
\label{tab:opeandprocess}
\end{table}

We are now ready to calculate the nucleon decay rates based on the effective interactions developed in the previous section. In \cref{tab:opeandprocess}, we list all possible two- and three-body nucleon decays involving the $\varphi$ that can be generated at leading order, along with the corresponding $\varphi$LEFT operators responsible for inducing them. 
For the processes marked with a checkmark, either direct experimental constraints or indirect bounds obtained from the reanalysis of existing experimental data are available.  
In the subsequent subsections, we will use these processes to constrain the associated Wilson coefficients in each sector, and then apply the resulting bounds to derive limits on the inverse decay widths of other processes that have not yet been experimentally constrained. 

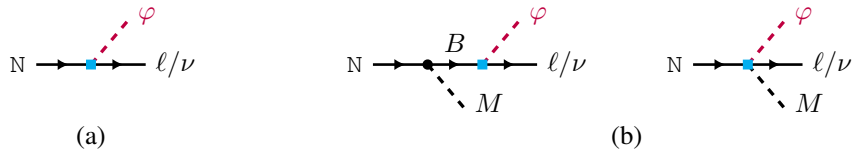
\begin{figure}[b]
\centering
\begin{tikzpicture}[mystyle,scale=0.8]
\begin{scope}[shift={(1,1)}]
\draw[f] (0, 0)node[left]{$\texttt{N}$} -- (1.5,0);
\draw[f] (1.5, 0) -- (3,0) node[right]{$\ell/\nu$};
\draw[snar, purple] (1.5,0) -- (2.5,1.2) node[right,yshift = 2 pt]{$\varphi$};
\filldraw [cyan] (1.4,-0.1) rectangle (1.6,0.1);
\node at (1.5,-2) {(a)};
\end{scope}
\end{tikzpicture}\quad\quad\quad\quad\quad 
\begin{tikzpicture}[mystyle,scale=0.8]
\begin{scope}[shift={(1,1)}]
\draw[f] (0, 0)node[left]{$\texttt{N}$} -- (1.5,0);
\draw[f] (1.5, 0) -- (3,0) node[midway,yshift = 8 pt]{$B$};
\draw[snar, black] (1.5,0) -- (2.5,-1.2) node[right,yshift = 2 pt]{$M$};
\draw[f] (3.0, 0) -- (4.5,0) node[right]{$\ell/\nu$};
\draw[snar, purple] (3,0) -- (4,1.2) node[right,yshift = 2 pt]{$\varphi$};
\filldraw [black] (1.5,0) circle (3pt);
\filldraw [cyan] (2.9,-0.1) rectangle (3.1,0.1);
\end{scope}
\end{tikzpicture}
\begin{tikzpicture}[mystyle,scale=0.8]
\begin{scope}[shift={(1,1)}]
\node at (1,-2) {(b)};
\end{scope}
\end{tikzpicture}
\begin{tikzpicture}[mystyle,scale=0.8]
\begin{scope}[shift={(1,1)}]
\draw[f] (0, 0)node[left]{$\texttt{N}$} -- (1.5,0);
\draw[f] (1.5, 0) -- (3,0) node[right]{$\ell/\nu$};
\draw[snar, purple] (1.5,0) -- (2.5,1.2) node[right,yshift = 2 pt]{$\varphi$};
\draw[snar, black] (1.5,0) -- (2.5,-1.2) node[right,yshift = 2 pt]{$M$};
\filldraw [cyan] (1.4,-0.1) rectangle (1.6,0.1);
\end{scope}
\end{tikzpicture}
\caption{Diagrams for BNV two-body (a) and three-body (b) nucleon decays involving a scalar. The cyan square (black blob) represents the insertion of a BNV (usual) chiral vertex.}
\label{fig:N2lMphi}
\end{figure}

%%%%%%%%%%%%%%%%%%%%%%%%%%%%%%
\subsection{General expression of nucleon decays involving the $\varphi$} 
%%%%%%%%%%%%%%%%%%%%%%%%%%%%%%

In terms of the hadronic effective interactions given in the previous section, the leading order Feynman diagrams responsible for all relevant two- and three-body nucleon decays are shown in \cref{fig:N2lMphi}.
For the three-body channels, the baryon-exchange diagrams mediated by octet baryons contribute at the same order as the contact diagram and therefore must be considered together. 
For a general two-body nucleon decay (${\texttt N}=p,n$), ${\texttt N}\to l \varphi$ with $l=\ell^+,\bar\nu,\nu$, the Lagrangian terms for it take the following general form
\begin{align}
 {\cal L}_{{\texttt N}\to l \varphi} = \overline{l}[ C_{{\texttt N}l}^\tL P_\tL + C_{{\texttt N}l}^\tR P_\tR ]{\texttt N} \varphi,  
 \label{eq:N2lphi}
\end{align}
where the coefficients $C_{\texttt N l}^{\tL(\tR)}$ can be read off from \cref{eq:VertexBl} and are summarized in the third column of \cref{tab:N2lMphi_vertices}. From the above vertices, the spin-summed and -averaged squared matrix element is 
\begin{align}
 \overline{|{\cal M}_{{\texttt N} \to l\varphi}|^2} = 
 \frac{1}{2}(m_{\texttt N}^2 + m_l^2- m_\varphi^2)
 \left( |C_{{\texttt N}l}^{\tL}|^2 + |C_{{\texttt N}l}^{\tR}|^2\right)
 +
 2 m_{\texttt N} m_l \, \Re\left(C_{{\texttt N} l}^{\tL}C_{{\texttt N} l}^{\tR\,*} \right). 
\end{align}
Then the two-body decay width becomes, 
\begin{align}
 \Gamma({\texttt N}\to l \varphi) = 
 \frac{\overline{|{\cal M}_{{\texttt N}\to l \varphi }|^2}}{16\pi m_{\texttt N}}
 \lambda^{1/2}(1,x_l,x_\varphi), 
\end{align}
where $\lambda(x,y,z)\equiv x^2+y^2+z^2-2xy-2yz-2zx$ 
is the triangle function, and
$x_{l,\varphi} \equiv m_{l,\varphi}^2/m_{\texttt N}^2$.

\begin{table}[t]
\center
\resizebox{\linewidth}{!}{
\renewcommand{\arraystretch}{1.4}
\begin{tabular}{|c| c| c | c |}
\hline
$\texttt{N}\to lM\varphi$
& $F_0\,C_{\texttt{N} \to BM}$
& \multicolumn{1}{c|}{$C^{\tL}_{Bl}$ (upper cell) and $C^{\tR}_{Bl}$ (lower cell)}
& \multicolumn{1}{c|}{$F_0\,C^{\tL}_{\texttt{N}lM}$ (upper cell) and $F_0\,C^{\tR}_{\texttt{N}lM}$ (lower cell)} 
\\\hline%p->ell+ pi0
\multirow{2}*{$p\to \ell_x^+ \pi^0\varphi$ } 
& $C_{p\to p\pi^0}$
& $c_1  C^{\tL \tR,x}_{\varphi\ell uud} + c_2 C^{\tL \tL,x}_{\varphi\ell uud }$
& $\frac{1}{2} (c_1  C^{\tL\tR,x}_{\varphi\ell uud} +c_2 C^{\tL\tL,x}_{\varphi\ell uud})$
\\\hhline{~---}
& $\frac{D+F}{2}$
& $-( c_1  C^{\tR \tL,x}_{\varphi\ell uud} +c_2 C^{\tR \tR,x}_{\varphi\ell uud} )$
& $\frac{1}{2} (c_1  C^{\tR\tL,x}_{\varphi\ell uud} +c_2 C^{\tR\tR,x}_{\varphi\ell uud})$
\\\hline %p->ell+ eta
\multirow{2}*{$p\to \ell_x^+ \eta\varphi$}
& $C_{p\to p\eta}$
& $c_1  C^{\tL \tR,x}_{\varphi\ell uud} +c_2 C^{\tL \tL,x}_{\varphi\ell uud }$
& $-\frac{1}{2\sqrt{3}} (c_1  C^{\tL\tR,x}_{\varphi\ell uud} - 3c_2 C^{\tL\tL,x}_{\varphi\ell uud})$
\\\hhline{~---}
& $-\frac{D-3F}{2\sqrt{3}}$
& $-( c_1  C^{\tR \tL,x}_{\varphi\ell uud} +c_2 C^{\tR \tR,x}_{\varphi\ell uud} )$
& $-\frac{1}{2\sqrt{3}} (c_1  C^{\tR\tL,x}_{\varphi\ell uud} - 3c_2 C^{\tR\tR,x}_{\varphi\ell uud})$
\\\hline %p->ell+ K0
\multirow{2}*{$p\to \ell_x^+ K^0\varphi$}
& $C_{p\to \Sigma^+ K^0}$
& $c_1  C^{\tL \tR,x}_{\varphi\ell usu} +c_2 C^{\tL \tL,x}_{\varphi\ell usu}$
& $\frac{1}{\sqrt{2}} (c_1  C^{\tL\tR,x}_{\varphi\ell usu} - c_2 C^{\tL\tL,x}_{\varphi\ell usu})$
\\\hhline{~---}
& $\frac{D-F}{\sqrt{2}}$
& $-(c_1  C^{\tR \tL,x}_{\varphi\ell usu} +c_2  C^{\tR \tR,x}_{\varphi\ell usu} )$
& $\frac{1}{\sqrt{2}} (c_1  C^{\tR\tL,x}_{\varphi\ell usu} -c_2 C^{\tR\tR,x}_{\varphi\ell usu})$
\\\hline %p->nu(bar) pi+ 
\multirow{2}*{$\makecell{ p \to \bar\nu_x \pi^+\varphi \\
\mbox{ (\colorbox{gray!15}{$p \to \nu_x\pi^+\varphi$})} }$}
& $C_{p\to n\pi^+}$
& $c_1  C^{\tL \tR,x}_{\varphi\nu dud} +c_2 C^{\tL \tL,x}_{\varphi \nu dud} $
& $\frac{1}{\sqrt{2}} (c_1  C^{\tL\tR,x}_{\varphi\nu dud}+c_2 C^{\tL\tL,x}_{\varphi\nu dud})$
\\\hhline{~---}
& $\frac{D+F}{\sqrt{2}}$
& \cellcolor{gray!15}$-(c_1  C^{\tR \tL,x}_{\varphi\bar{\nu}dud} +c_2 C^{\tR \tR,x}_{\varphi\bar{\nu}dud})$
& \cellcolor{gray!15}$\frac{1}{\sqrt{2}} (c_1  C_{\varphi\bar\nu dud}^{\tR\tL,x} +c_2  C_{\varphi\bar\nu dud}^{\tR\tR,x})$
\\\hline %p->nu(bar) K+
\multirow{4}*{$\makecell{ p \to \bar\nu_x K^+\varphi \\
\mbox{~(\colorbox{gray!15}{$p \to \nu_x K^+\varphi$})~} }$}
& $C_{p\to \Lambda^0 K^+}$
& $\frac{1}{\sqrt{6}} \left[c_1  (C^{\tL \tR,x}_{\varphi\nu uds} + C^{\tL \tR,x}_{\varphi\nu dsu} 
- 2 C^{\tL \tR,x}_{\varphi\nu sud} )
+c_2 ( C^{\tL \tL,x}_{\varphi\nu dsu} - 2 C^{\tL \tL,x}_{\varphi\nu sud}) \right]$
&  \multirow{2}*{$\frac{1}{\sqrt{2}}
\left[ c_1  (C^{\tL\tR,x}_{\varphi\nu uds} + C^{\tL\tR,x}_{\varphi\nu sud}) 
+c_2  C^{\tL\tL,x}_{\varphi\nu sud} \right]$}
\\\hhline{~--}
& $-\frac{D+3F}{2\sqrt{3}}$
& \cellcolor{gray!15}$-\frac{1}{\sqrt{6}} \left[c_1  (C^{\tR \tL,x}_{\varphi\bar{\nu}uds} 
+ C^{\tR \tL,x}_{\varphi\bar{\nu}dsu} - 2 C^{\tR \tL,x}_{\varphi\bar{\nu} sud} )
+c_2 (C^{\tR \tR,x}_{\varphi\bar{\nu}dsu} - 2 C^{\tR \tR,x}_{\varphi\bar{\nu} sud} ) \right]$
& 
\\\hhline{~---}
& $C_{p\to \Sigma^0 K^+}$
& $\frac{1}{\sqrt{2}} \left[c_1  (C^{\tL \tR,x}_{\varphi\nu uds} 
- C^{\tL \tR,x}_{\varphi\nu dsu})-c_2 C^{\tL \tL,x}_{\varphi\nu dsu} \right]$
& \cellcolor{gray!15}
\\\hhline{~--}
& $\frac{D-F}{2}$
& \cellcolor{gray!15}$-\frac{1}{\sqrt{2}} \left[c_1  (C^{\tR \tL,x}_{\varphi\bar{\nu}uds} 
- C^{\tR \tL,x}_{\varphi\bar{\nu}dsu}) -c_2 C^{\tR \tR,x}_{\varphi\bar{\nu}dsu} \right]$
& \multirow{-2}*{\cellcolor{gray!15}{$\frac{1}{\sqrt{2}} \left[ c_1  (C^{\tR\tL,x}_{\varphi\bar\nu uds} 
+ C^{\tR\tL,x}_{\varphi\bar\nu sud} ) +c_2 C^{\tR\tR,x}_{\varphi\bar\nu sud}\right]$}}
\\\hline %n->ell+ pi-
\multirow{2}*{$n\to \ell_x^+ \pi^-\varphi$}
& $C_{n\to p \pi^-}$
& $c_1  C^{\tL \tR,x}_{\varphi\ell uud} +c_2 C^{\tL \tL,x}_{\varphi\ell uud }$
&  $\frac{1}{\sqrt{2}} (c_1  C^{\tL\tR,x}_{\varphi\ell uud} +c_2 C^{\tL\tL,x}_{\varphi\ell uud})$
\\\hhline{~---}
& $\frac{D+F}{\sqrt{2}}$
& $-( c_1  C^{\tR \tL,x}_{\varphi\ell uud} +c_2 C^{\tR \tR,x}_{\varphi\ell uud} )$
& $\frac{1}{\sqrt{2}} (c_1  C^{\tR\tL,x}_{\varphi\ell uud} +c_2 C^{\tR\tR,x}_{\varphi\ell uud})$
\\\hline %n->nu(bar) pi0
\multirow{2}*{$\makecell{ n \to\bar\nu_x \pi^0\varphi  \\
\mbox{ (\colorbox{gray!15}{$n \to\nu_x\pi^0\varphi $})} }$}
& $C_{n\to n \pi^0}$
& $c_1  C^{\tL \tR,x}_{\varphi\nu dud} +c_2 C^{\tL \tL,x}_{\varphi\nu dud} $
& $ -\frac{1}{2} (c_1  C^{\tL\tR,x}_{\varphi\nu dud} +c_2 C^{\tL\tL,x}_{\varphi\nu dud})$
\\\hhline{~---}
& $-\frac{D+F}{2}$
& \cellcolor{gray!15}$-(c_1  C^{\tR \tL,x}_{\varphi\bar{\nu}dud} +c_2 C^{\tR \tR,x}_{\varphi\bar{\nu}dud})$
& \cellcolor{gray!15}{$-\frac{1}{2} (c_1  C^{\tR\tL,x}_{\varphi\bar\nu dud} +c_2 C^{\tR\tR,x}_{\varphi\bar\nu dud})$}
\\\hline %n->nu(bar) eta
\multirow{2}*{$\makecell{ n\to\bar\nu_x\eta\varphi \\
\mbox{ (\colorbox{gray!15}{$n\to\nu_x\eta\varphi$})} }$}
&  $C_{n\to n \eta}$
&  $c_1  C^{\tL \tR,x}_{\varphi\nu dud} + 
 c_2 C^{\tL \tL,x}_{\varphi\nu dud} $
& $- \frac{1}{2\sqrt{3}} (c_1  C^{\tL\tR,x}_{\varphi\nu dud} - 3c_2  C^{\tL\tL,x}_{\varphi\nu dud} )$
\\\hhline{~---}
& $-\frac{D-3F}{2\sqrt{3}}$
& \cellcolor{gray!15}$-(c_1  C^{\tR \tL,x}_{\varphi\bar{\nu}dud} +c_2  C^{\tR \tR,x}_{\varphi\bar{\nu}dud})$
& \cellcolor{gray!15}{$- \frac{1}{2\sqrt{3}} (c_1  C^{\tR\tL,x}_{\varphi\bar\nu dud} 
- 3c_2  C^{\tR\tR,x}_{\varphi\bar\nu dud} )$}
\\\hline %n->nu(bar) K0
\multirow{4}*{$\makecell{ n \to \bar\nu_x K^0\varphi \\
\mbox{ (\colorbox{gray!15}{$n \to \nu_x K^0\varphi$})} }$}
& $C_{n\to \Lambda^0 K^0}$
& $\frac{1}{\sqrt{6}} \left[c_1  (C^{\tL \tR,x}_{\varphi\nu uds} + C^{\tL \tR,x}_{\varphi\nu dsu} 
- 2 C^{\tL \tR,x}_{\varphi\nu sud} )
+c_2 (C^{\tL \tL,x}_{\varphi\nu dsu} - 2 C^{\tL \tL,x}_{\varphi\nu sud}) \right]$
& \multirow{2}*{$\frac{1}{\sqrt{2}} \left[c_1 ( C^{\tL\tR,x}_{\varphi\nu sud} + C^{\tL\tR,x}_{\varphi\nu dsu})
+c_2 (C^{\tL\tL,x}_{\varphi\nu sud} - C^{\tL\tL,x}_{\nu dsu}) \right]$}
\\\hhline{~--}
& $-\frac{D+3F}{2\sqrt{3}}$
& \cellcolor{gray!15}$~-\frac{1}{\sqrt{6}} 
\left[c_1  (C^{\tR \tL,x}_{\varphi\bar{\nu}uds} + C^{\tR \tL,x}_{\varphi\bar{\nu}dsu} 
- 2 C^{\tR \tL,x}_{\varphi\bar{\nu} sud} )+c_2 (C^{\tR \tR,x}_{\varphi\bar{\nu}dsu}  
- 2 C^{\tR \tR,x}_{\varphi\bar{\nu} sud} ) \right]~$
& 
\\\hhline{~---}
& $C_{n\to \Sigma^0 K^0}$
& $\frac{1}{\sqrt{2}} \left[c_1  (C^{\tL \tR,x}_{\varphi\nu uds} - C^{\tL \tR,x}_{\varphi\nu dsu}) -c_2  C^{\tL \tL,x}_{\varphi\nu dsu} \right]$
& \cellcolor{gray!15}
\\\hhline{~--}
& $-\frac{D-F}{2}$
& \cellcolor{gray!15}$-\frac{1}{\sqrt{2}} 
\left[c_1  (C^{\tR \tL,x}_{\varphi\bar{\nu}uds} - C^{\tR \tL,x}_{\varphi\bar{\nu}dsu}) 
-c_2 C^{\tR \tR,x}_{\varphi\bar{\nu}dsu} \right]$
& \multirow{-2}*{\cellcolor{gray!15}{$\frac{1}{\sqrt{2}} 
\left[ c_1 ( C^{\tR\tL,x}_{\varphi\bar\nu sud} + C^{\tR\tL,x}_{\varphi\bar\nu dsu}) 
+c_2 (C^{\tR\tR,x}_{\varphi\bar\nu sud} - C^{\tR\tR,x}_{\varphi\bar\nu dsu}) \right]$}}
\\\hline %n->ell- K+
\cellcolor{gray!15}
& $C_{n\to \Sigma^- K^+}$
&\cellcolor{gray!15} $c_1  C^{\tL \tR,x}_{\varphi\bar{\ell} dds} +c_2  C^{\tL \tL,x}_{\varphi\bar{\ell} dds}$
& \cellcolor{gray!15}{$\frac{1}{\sqrt{2}} 
(c_1  C^{\tL\tR,x}_{\varphi\bar\ell dds} -c_2 C^{\tL\tL,x}_{\varphi\bar\ell dds})$}
\\\hhline{~---}
\multirow{-2}*{\cellcolor{gray!15}{$n\to \ell_x^- K^+\varphi$}}
& $\frac{D-F}{\sqrt{2}}$
&\cellcolor{gray!15} $-(c_1  C^{\tR \tL,x}_{\varphi\bar{\ell} dds}
+ c_2 C^{\tR \tR,x}_{\varphi\bar{\ell} dds} )$
&\cellcolor{gray!15}{$\frac{1}{\sqrt{2}} 
(c_1  C^{\tR\tL,x}_{\varphi\bar\ell dds} -c_2 C^{\tR\tR,x}_{\varphi\bar\ell dds})$}
\\\hline
\end{tabular} }
\caption{The explicit expression for the relevant vertices in the three-body nucleon decays involving an invisible light scalar particle.
The white and gray backgrounds mark the processes with $\Delta(B-L)=0$ and $\Delta(B+L)=0$, respectively.}
\label{tab:N2lMphi_vertices}
\end{table}

For three-body nucleon decays, each channel involves three vertices. Generally, the relevant Lagrangian terms for the process $\texttt N\to l M \varphi$ can be written as 
\begin{align}
 {\cal L}_{{\texttt N}\to l M \varphi} =
C_{\texttt{N}\to BM} \overline{B}\gamma_\mu \gamma_5 \texttt{N} \, \partial^\mu \bar M
 + \overline{l}[ C_{Bl}^\tL P_\tL + C_{Bl}^\tR P_\tR ] B \varphi
+  i\,\overline{l}[ C_{{\texttt N}lM}^\tL P_\tL + C_{{\texttt N}lM}^\tR P_\tR ]{\texttt N}\, \bar M  \varphi,  
\end{align}
where $C_{\texttt N\to BM}$ parametrizes the coupling from the standard baryon ChPT Lagrangian given in \cref{LBBM}.
$C_{B l}^{\tL,\tR}$ and $C_{\texttt{N}l M}^{\tL,\tR}$ can be identified from the BNV interactions given in \cref{eq:VertexBl,eq:VertexBellM,eq:VertexBvM}. 
For each specific process, we provide the relevant ingredients in the third and fourth columns in \cref{tab:N2lMphi_vertices}.
The decay amplitude for $\texttt N(p)\to l(p_l) + M(p_M) + \varphi(p_\varphi)$ from the two diagrams given in \cref{fig:N2lMphi}\,(b) is calculated to take the following form 
\begin{align}
\label{eq:MN2lMphi}
{\cal M}_{\texttt N\to l M\varphi} 
= i \bar u_l\left[
D_{\texttt{N} lM}^{{\tt S},\tL} P_\tL 
+ D_{\texttt{N} lM}^{{\tt S},\tR} P_\tR 
+ D_{\texttt{N} lM}^{{\tt V},\tL} m_\texttt{N}^{-1} \slashed{p}_B P_\tL 
+ D_{\texttt{N} lM}^{{\tt V},\tR} m_\texttt{N}^{-1} \slashed{p}_B P_\tR 
\right]u_\texttt{N}. 
\end{align}
By denoting $s\equiv (p_l + p_\varphi)^2$ and $t \equiv (p-p_l)^2 = (p_M +p_\varphi)^2$, the
$D$-coefficients above are related to the $C$-parameters in \cref{tab:N2lMphi_vertices} in the manner, 
\begin{subequations}
\begin{align}
D_{\texttt{N} lM}^{{\tt S},\tL(\tR)} & = C_{{\texttt N} lM}^{\tL(\tR)}
\pm \sum_B \frac{m_B m_{\texttt N} + s}{m_B^2 - s} 
C_{\texttt N\to B M} C_{B l}^{\tL(\tR)},
\\
D_{\texttt{N} lM}^{{\tt V},\tL(\tR)} & = 
\pm \sum_B \frac{m_B m_{\texttt N} + m_{\texttt N}^2 }{m_B^2 - s} 
C_{\texttt N\to B M} C_{B l}^{\tR(\tL)},  
\end{align}
\end{subequations}
where the sum is over all possible intermediate baryon states $B$ as indicated by the coefficients $C_{\texttt{N}\to BM}$ in the second column of \cref{tab:N2lMphi_vertices}.

From \cref{eq:MN2lMphi}, the spin-summed and -averaged matrix element squared takes a compact form: 
\begin{align}
 \overline{|{\cal M}_{\texttt{N}\to lM\varphi}|^2} & = 
 \frac{1}{2}(m_\texttt{N}^2 + m_l^2 - t)
\left(|D_{\texttt{N} lM}^{{\tt S},\tL}|^2 + |D_{\texttt{N} lM}^{{\tt S},\tR}|^2\right)
\nonumber
\\
& + \frac{1}{2 m_\texttt{N}^2} \left[(m_\texttt{N}^2 - m_M^2)(m_l^2 - m_\varphi^2) 
+ s (s+t - m_M^2 - m_\varphi^2) \right] 
\left(|D_{\texttt{N} lM}^{{\tt V},\tL}|^2 
+ |D_{\texttt{N} lM}^{{\tt V},\tR}|^2\right)
\nonumber
\\
& + 2 m_l m_\texttt{N} \, \Re\left(
D_{\texttt{N} lM}^{{\tt S},\tL}D_{\texttt{N} lM}^{{\tt S},\tR\,*} 
+ \frac{s}{m_\texttt{N}^2}\, D_{\texttt{N} lM}^{{\tt V},\tL}D_{\texttt{N} lM}^{{\tt V},\tR\,*} \right)
\nonumber
\\
& + \frac{ m_l}{m_\texttt{N} } (m_\texttt{N}^2 + s - m_M^2)\,\Re\left(
D_{\texttt{N} lM}^{{\tt S},\tL}D_{\texttt{N} lM}^{{\tt V},\tL\,*} 
+ D_{\texttt{N} lM}^{{\tt S},\tR}D_{\texttt{N} lM}^{{\tt V},\tR\,*}
\right) 
\nonumber
\\
& + (s+ m_l^2 - m_\varphi^2) \,\Re\left(
D_{\texttt{N} lM}^{{\tt S},\tL}D_{\texttt{N} lM}^{{\tt V},\tR\,*} 
+ D_{\texttt{N} lM}^{{\tt S},\tR} D_{\texttt{N} lM}^{{\tt V},\tL\,*} \right). 
\end{align}
The decay width then reads 
\begin{align}
\Gamma_{\texttt{N}\to lM\varphi} =
\frac{1}{256\pi^3 m_{\texttt N}^3} 
\int_{s_-}^{s_+} d s 
\int_{t_-}^{t_+} d t 
\overline{|{\cal M}_{\texttt{N}\to lM\varphi}|^2},
\end{align}
where the integration limits are 
\begin{align}
& s_- = (m_l +m_\varphi)^2,\,
s_+ = (m_{\texttt N} - m_M)^2,
\notag\\
& t_\pm = (E_2^* + E_3^*)^2 - \Big(\sqrt{E_2^{*2} - m_\varphi^2} \mp \sqrt{E_3^{*2} - m_M^2} \Big)^2, 
\notag\\
& E_2^* \equiv \frac{s - m_l^2 + m_\varphi^2}{2\sqrt{s} },\,
E_3^* \equiv \frac{ m_{\texttt N}^2 - s - m_M^2}{2\sqrt{s}}.
\end{align}
For the nucleon decay modes listed in \cref{tab:opeandprocess}, the decay widths can be expressed in terms of $\varphi$LEFT Wilson coefficients after performing the above phase space integration. 
In \cref{app:Gammaatm20}, we summarize the decay width expressions in the limit of vanishing $m_\varphi$. 
For a given model and a negligible $m_\varphi$, 
one can easily obtain the relevant nucleon decay width in terms of the model parameters by identifying the relevant Wilson coefficients.

%%%%%%%%%%%%%%%%%%%%%%%%%%%%%%
\subsection{Momentum distribution}\label{subsec:momentum distribution}
%%%%%%%%%%%%%%%%%%%%%%%%%%%%%%

For three-body decay modes, the momentum distribution of the final-state charged leptons or mesons encodes rich information that is directly related to the experimental measurement. 
The momentum distribution can be used to differentiate between interaction structures and determine the scalar mass value once a definite signal is observed.  
Hence, we derive the differential decay width expression as a function of the  energies of the final-state lepton ($E_l$) and meson ($E_M$). 
In the centre of mass frame of the mother nucleon, 
\begin{align}
\frac{d^2\Gamma}{d E_l d E_M}
= \frac{ 1}{64\pi^3 m_{\texttt N}} 
\overline{|{\cal M}_{\texttt{N}\to lM\varphi}|^2},
\end{align}
and the invariant-mass variables $s= m_{\texttt N}^2 +m_M^2 - 2 m_{\texttt N} E_M$
and $t = m_{\texttt N}^2 +m_l^2 - 2 m_{\texttt N} E_l$.
The integration limits are 
\begin{subequations}
\begin{align}
E_{M,\pm} & =  \frac{1}{2 t} 
\left[ (t+m_M^2 - m_\varphi^2)
(m_{\texttt N} - E_l) \pm |\pmb{p}_l| \lambda^{1/2}(t,m_M^2, m_\varphi^2) \right], 
\\
m_l \leq  E_l & \leq 
\frac{ m_{\texttt N}^2 + m_l^2 - (m_M+m_\varphi)^2} {2 m_{\texttt N} }. 
\end{align}
\end{subequations}
Changing the integration order of $E_M$ and $E_l$ implies the simple exchange of the subscripts $M\leftrightarrow l$ in the above integration limits, i.e., 
\begin{subequations}
\begin{align}
E_{l,\pm} & =  \frac{1}{2 s} 
\left[ (s + m_l^2 - m_\varphi^2) (m_{\texttt N}-E_M) \pm |\pmb{p}_M|\lambda^{1/2}(s,m_l^2, m_\varphi^2) \right],
\\
m_M \leq  E_M & \leq 
\frac{ m_{\texttt N}^2 + m_M^2 - (m_l+m_\varphi)^2} {2 m_{\texttt N} }. 
\end{align}
\end{subequations}
To study the distribution against the three-momenta of either the lepton or the meson, we can define the normalized width
\begin{align}
\frac{d\tilde \Gamma} {d |\pmb{p}_l|}
= \frac{1}{\Gamma }\frac{|\pmb{p}_l|}{E_l} \int d E_M \frac{d^2\Gamma}{d E_l d E_M}, \quad 
\frac{d\tilde \Gamma} {d |\pmb{p}_M|}
= \frac{1}{\Gamma }\frac{|\pmb{p}_M|}{E_M} \int d E_l \frac{d^2\Gamma}{d E_l d E_M}.
\end{align}

In addition to the single momentum distribution, the double differential distribution $d^2\tilde\Gamma/d|\mathbf{p}_\ell| d|\mathbf{p}_M|$ also provides valuable information, as it is unique to those three-body modes involving both a charged lepton and a meson. 
This observable can be used complementarily in future experimental analysis to help identify these exotic channels and their underlying interaction structures. Here we focus on the discussion of the first case. 

\begin{figure}[t]
\centering
\includegraphics[width=0.32\linewidth]
{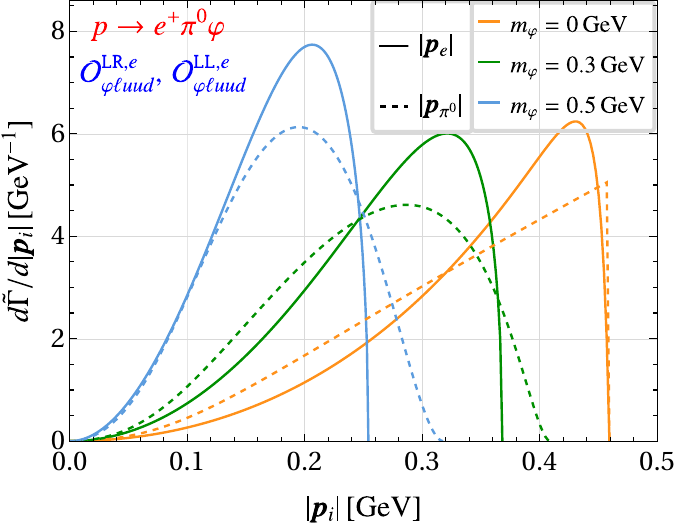}
\includegraphics[width=0.32\linewidth]{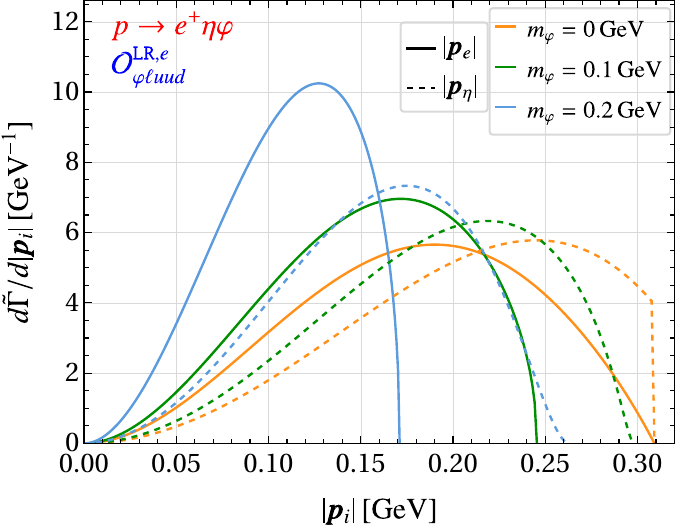}
\includegraphics[width=0.32\linewidth]{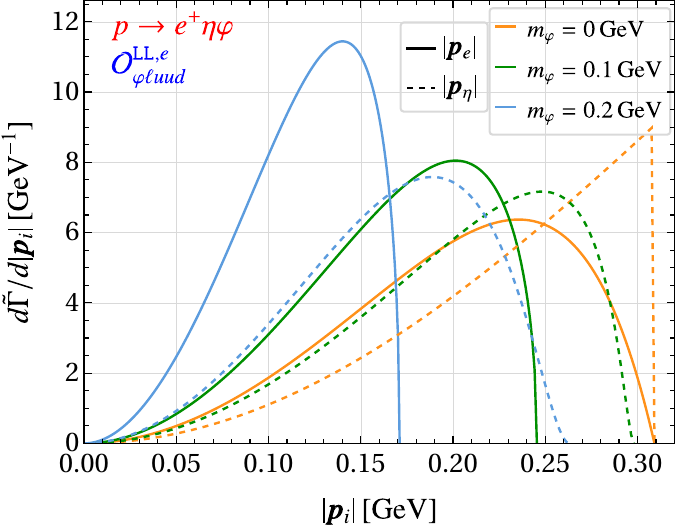}
\\
\vspace{0.2em}
\includegraphics[width=0.32\linewidth]{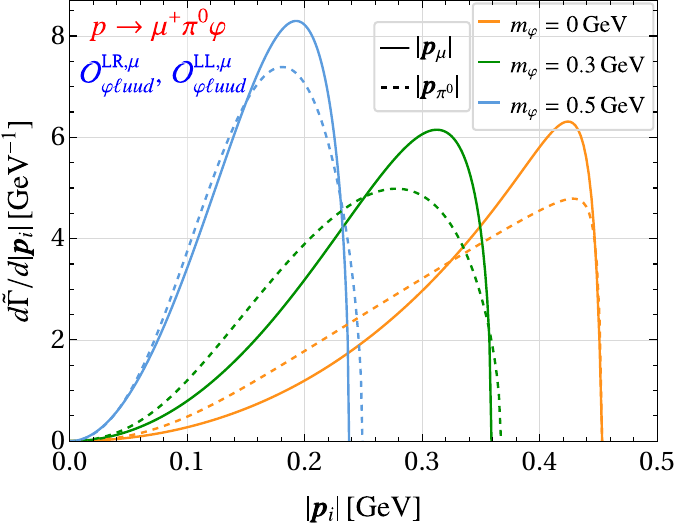}
\includegraphics[width=0.32\linewidth]{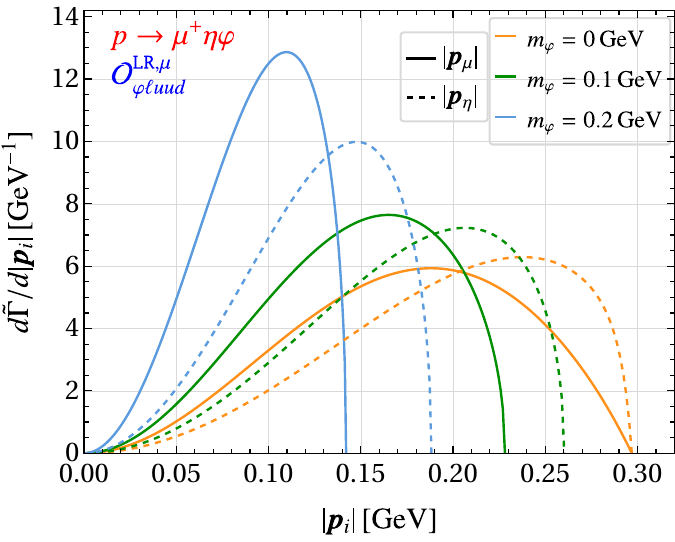}
\includegraphics[width=0.32\linewidth]{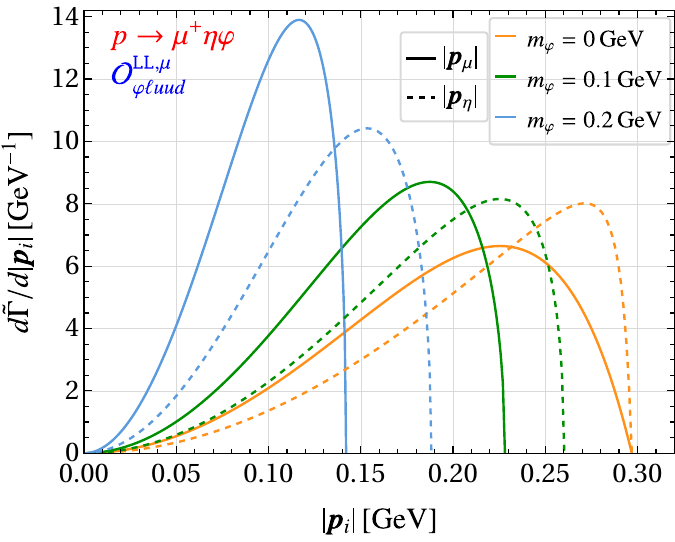}
\\
\vspace{0.2em}
\includegraphics[width=0.32\linewidth]{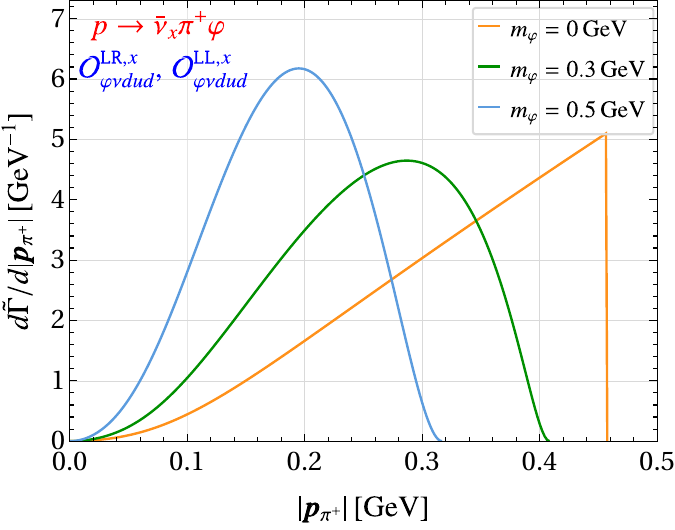}
\includegraphics[width=0.32\linewidth]{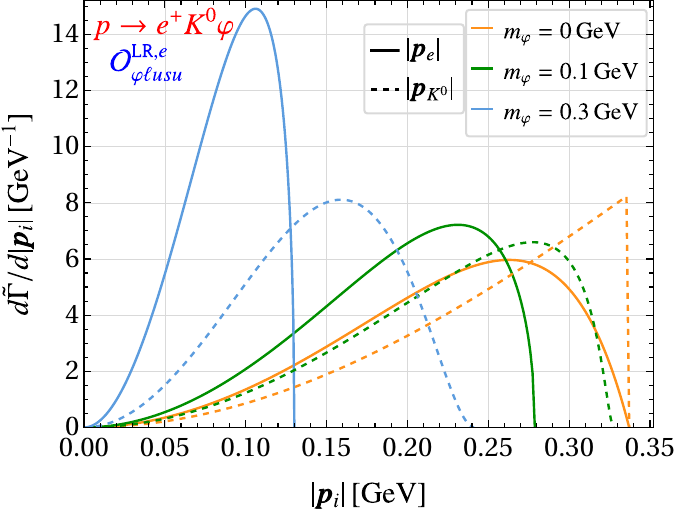}
\includegraphics[width=0.32\linewidth]{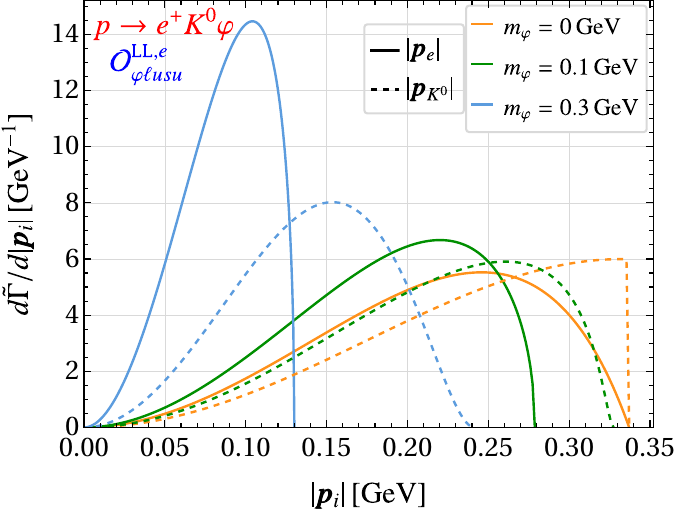}
\\
\vspace{0.2em}
\includegraphics[width=0.32\linewidth]{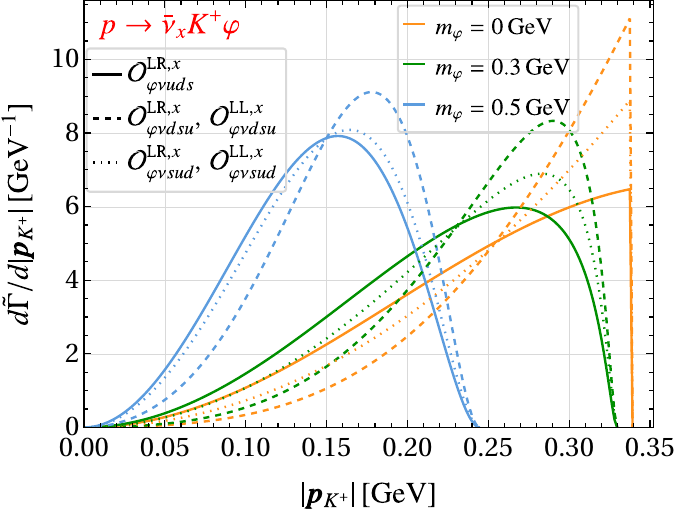}
\includegraphics[width=0.32\linewidth]{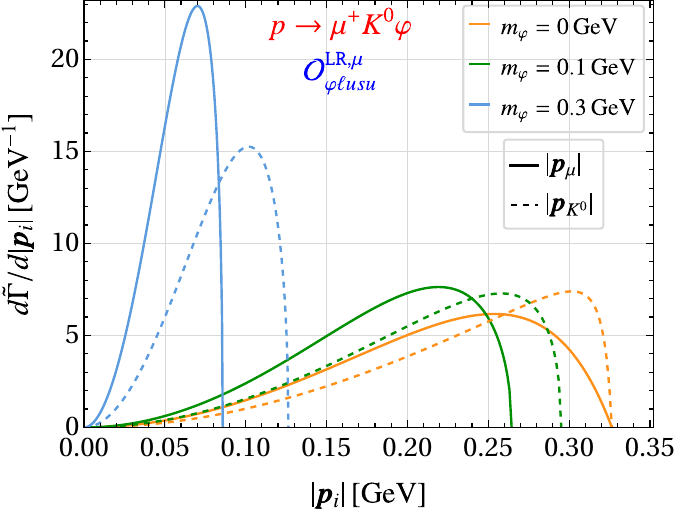}
\includegraphics[width=0.32\linewidth]{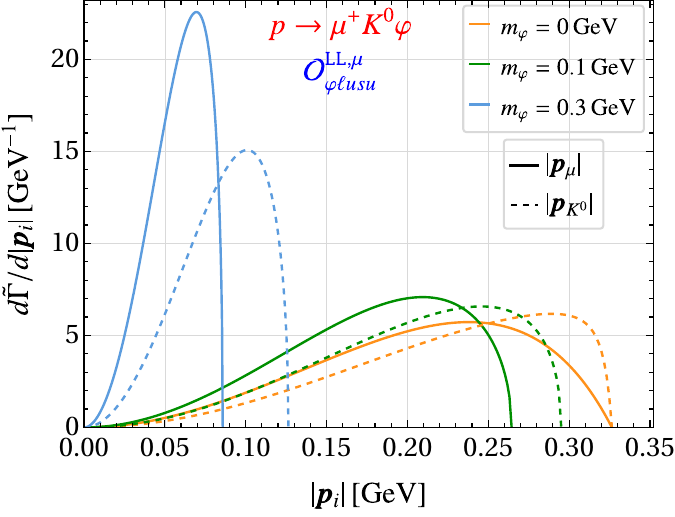}
\caption{The normalized $|\pmb{p}_i|$-distribution for various three-body proton decay modes with the insertion of relevant $\varphi$LEFT interactions. In each process, we consider three benchmark values of the scalar mass $m_\varphi$: 0 (orange), 0.1 GeV (green), and 0.3 GeV (blue).}
\label{fig:pdistribution}
\end{figure}

\cref{fig:pdistribution} shows the normalized momentum distributions of the charged leptons ($e^+,\mu^+$) and pseudoscalar mesons ($\pi, \eta, K$) for the eight three-body proton decay modes with the insertion of each relevant operator. 
These distributions are important to consider when considering experimental searches, because the number of signal events crucially depends on the momenta of the final state particles, which is discussed in more detail below, when recasting the existing Super-K searches for proton decay to obtain bounds on $p\to \ell^+ \varphi$. 

Due to isospin symmetry, the neutron modes $n\to \ell_x^+ \pi^-\varphi$ 
and $n\to \bar\nu_x(\nu_x)\pi^0\varphi$ have the same momentum distributions as those of  $p\to \ell_x^+ \pi^0\varphi$ and $p\to \bar\nu_x(\nu_x)\pi^+\varphi$, respectively, which is also evident from the results in \cref{tab:N2lMphi_vertices}.
The neutron decays $n\to \bar\nu_x(\nu_x)\eta\varphi$ have momentum distributions similar to those of  $p\to \ell_x^+ \eta\varphi$, when we neglect charged-lepton masses.
For $n\to \ell_x^- K^+ \varphi$, the distributions can be directly obtained from those of $p\to \ell_x^+ K^0\varphi$ with the exchange of operators $\calO_{\varphi\ell usu}^{\tL\tR(\tL\tL),x}\leftrightarrow
\calO_{\varphi\bar\ell dds}^{\tL\tR(\tL\tL),x}$.
Thus, we do not consider the neutron case here for simplicity. 
Finally, all processes share the same distributions for the chirality-flipped operators with $\tL\leftrightarrow \tR$ and thus we do not show them either. 

As can be seen from the plots in \cref{fig:pdistribution}, the processes $p\to\ell^+_x \pi^0\varphi$, $p \to \bar\nu_x \pi^+\varphi$, and $p\to\bar\nu_x K^+\varphi$ have the same distributions for the pair of operators with $\tL\tR$ and $\tL\tL$ chiralities, as they always appear in the combination of $c_1  C^{\tL\tR}_i +c_2 C^{\tL\tL}_i$, while for the processes containing an $\eta$ or a $K^0$, the distributions from the two operators with $\tL\tR$ and $\tL\tL$ are generally different since the contact and noncontact contributions involve different combinations of $C^{\tL\tR}_i$ and $C^{\tL\tL}_i$.
Therefore, these distributions can be used to partially distinguish operator structures once a positive signal is observed. 
Moreover, these distributions are sensitive to the scalar mass due to kinematic constraints, and the endpoint of the distribution can be used to determine the scalar mass. 

%%%%%%%%%%%%%%%%%%%%%%%%%%%%%%
\subsection{Existing constraints on BNV nucleon decay}
%%%%%%%%%%%%%%%%%%%%%%%%%%%%%%

We explore constraints on the relevant $\varphi$LEFT and $\varphi$SMEFT Wilson coefficients by employing available experimental data.
First, for the two-body proton decays $p \to e^+(\mu^+) + \varphi$, the Super-K experiment has conducted searches for a massless boson $\varphi$ and set the bounds $\Gamma^{-1}({p \to e^+ + \varphi}) > 7.9 \times 10^{32}$ yr and
$\Gamma^{-1}(p \to \mu^+ + \varphi)> 4.1 \times 10^{32}$ yr~\cite{Super-Kamiokande:2015pys}, respectively.
We will reinterpret their data and set a bound for a general massive scalar $\varphi$. 
A similar reinterpretation will be carried out for the neutron decay $n\to \bar\nu (\nu)\pi^0\varphi$ based on the data in~\cite{Super-Kamiokande:2013rwg}.
Secondly, the invisible two-body neutron decay $n\to \bar\nu (\nu)+\varphi$ is based on the detection of gamma rays in the $5-10$ MeV window from an excited nucleus and thus applies to a massive scalar $\varphi$, as long as the nucleon is kinematically possible. We adopt the most recent limit reported by the SNO$+$ experiment, $\Gamma^{-1}(n\to inv.)> 0.9\times 10^{30}\,{\rm yr}$~\cite{SNO:2022trz}.
This bound is expected to improve to $\Gamma^{-1}_{\tt JUNO}(n\to inv.)> 5\times 10^{31}\,{\rm yr}$ in the JUNO experiment~\cite{JUNO:2024pur} after 2-years of data collection.

There is no dedicated experimental search for the three-body modes with an invisible scalar. In these cases, we resort to inclusive searches and a recast of a related two-body decay search.
Nucleon decays involving a positively charged lepton, $\texttt{N}\to e^+(\mu^+) +M + \varphi$, 
and a pseudoscalar meson $M$ can be constrained using the inclusive limits 
$\Gamma^{-1}(\texttt{N}\to e^+ + \text{anything})>0.6\times 10^{30}$ yr~\cite{Learned:1979gp} and 
$\Gamma^{-1}(\texttt{N}\to \mu^+ + \text{anything})>12\times 10^{30}$ yr~\cite{Cherry:1981uq}. 
Lastly, in the decay $p\to\bar\nu(\nu) K^+\varphi$, the neutrino and scalar $\varphi$ escape the detector undetected. 
The existing search for $p\to\bar\nu K^+$ can be directly reinterpreted as constraint on
$p \to \bar\nu(\nu) K^+ \varphi$, because the charged kaon is below the Cherenkov threshold, and thus the Super-K search only detects the decaying charged kaon. 
Consequently, we adopt the current Super-K limit~\cite{Super-Kamiokande:2014otb} to set $\Gamma^{-1}(p \to \bar\nu(\nu) K^+ \varphi) >5.9 \times 10^{33}\,\rm yr$.

In addition to the direct experimental bounds mentioned above, Ref.~\cite{Fan:2025xhi} interprets other available Super-K experimental data as bounds on several three-body proton modes involving an invisible ALP ($a$). We adopt an averaged bound on $p\to \mu^+ K^0 a$ derived from two similar chiral irreducible representation operators in that work, along with the inclusive bound, as complementary constraints on $p\to \mu^+ K^0 \varphi$ in our analysis. For each $\varphi$LEFT Wilson coefficient $C_i$, we define an effective scale through $\Lambda_{\tt eff} \equiv C_i^{-1/3}$ and set the bound on $\Lambda_{\tt eff}$ as a function of $m_\varphi$ by requiring $\Gamma_{\tt theory}^{-1}\gtrsim \Gamma_{\tt exp}^{-1}$. 

Using the processes listed above, it is possible to constrain all $\varphi$LEFT operators, listed in \cref{tab:opeandprocess}, except for the operators in the third column $\calO_{\varphi \bar\ell d d s}$, which contribute to $n \to e^-(\mu^-) K^+ \varphi$.
In the remainder of this subsection we recast data from existing Super-K searches $p \to e^+(\mu^+) \varphi$~\cite{Super-Kamiokande:2015pys}, where $\varphi$ is a massless boson, to obtain a limit for a massive scalar $\varphi$. Similarly, we consider $n \to \bar\nu \pi^0$~\cite{Super-Kamiokande:2013rwg} to set lifetime constraints for the three-body decay channel $n \to \bar\nu (\nu) \pi^0 \varphi$.

%%%%%%%%%%%%%%%%%%%%%%%%%%%%%%
\subsubsection{Two-body proton decay $p \to \ell^+ \varphi$}
%%%%%%%%%%%%%%%%%%%%%%%%%%%%%%

\begin{figure}[tb!]
\centering
\includegraphics[width=0.45\linewidth]{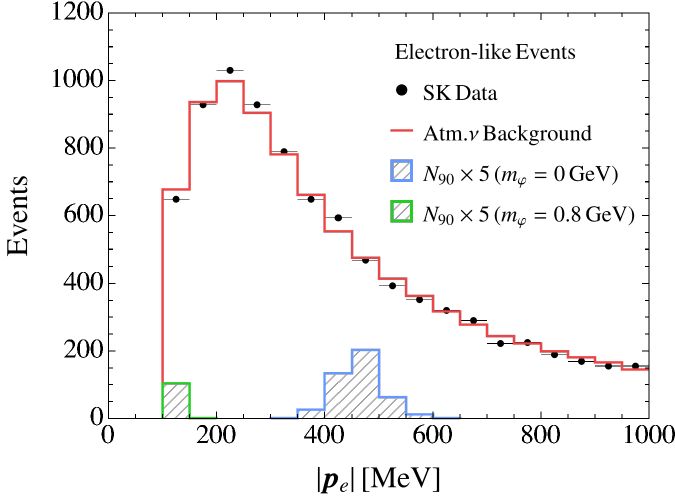}
\includegraphics[width=0.45\linewidth]{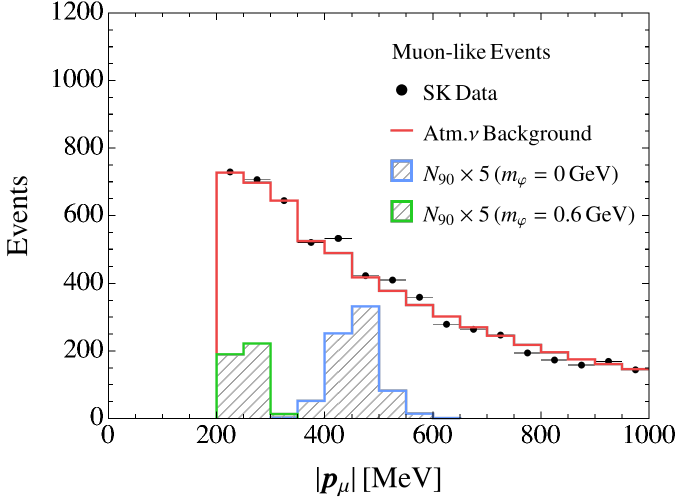}
\caption{Binned Super-K data (black) as a function of the lepton momentum together with the atmospheric neutrino background (red), and two allowed signal distributions at 90\% confidence level (hatched histograms) for a massless and a heavy scalar respectively, scaled by a factor of five. The left plot shows electron-like events and the right plot muon-like events.}
\label{fig:SuperKData}
\end{figure}

The search for proton decay with a single charged lepton and an invisible massless particle by the Super-K experiment provides a stringent lifetime constraint for the massless $\varphi$ case. We derive the constraints for a general massive $\varphi$ by performing the same spectral fit analysis on their data. The binned data as a function of the lepton momentum is shown in \cref{fig:SuperKData} together with the atmospheric neutrino background (red), and two allowed proton decay signals. Events in each bin are assumed to follow a Poisson distribution, and systematic errors are modeled by Gaussian distributions.
Hence, the $\chi^2$ function used is~\cite{Super-Kamiokande:2015pys}
\begin{subequations}
\begin{align}
\chi^2 & = 2\sum_{i=1}^{n} \left( N_i^{\mathrm{exp}} + N_i^{\mathrm{obs}} \left[ \ln \dfrac{N_i^{\mathrm{obs}}}{N_i^{\mathrm{exp}}} - 1 \right] \right) + \sum_{j=1}^{\mathrm{Syserr}} \left(\dfrac{\epsilon_j}{\sigma_j}\right)^2, \\
N_i^{\mathrm{exp}} & = \left[ \alpha N_i^{\mathrm{background}} + \beta N_i^{\mathrm{signal}} \right] \bigg( 1 + \sum_{j=1}^{\mathrm{Syserr}} f_i^j \frac{\epsilon_j}{\sigma_j}
\bigg),
\end{align}
\end{subequations}
where $i$ denotes individual bins, the index $j$ labels the systematic errors, $\epsilon_j$ the fit error parameters, $\sigma_j$ their $1\sigma$ uncertainties, and $f_i^j$ the corresponding fractional change in the $i^{\rm th}$ bin.
The parameters $\alpha$ and $\beta$ denote the normalizations of the background and the signal, respectively.
The terms $N_i^{\mathrm{obs}}$ and $N_i^{\mathrm{background}}$ refer to the number of observed events and background events, respectively, and are extracted by digitizing the graphs from the Super-K publication~\cite{Super-Kamiokande:2015pys}. 

The number of expected signal events $N_i^{\mathrm{signal}}$ is obtained from simulations of decay lepton momentum distributions. As most nucleon decays originate from the bound nucleus $^{16}\rm O$, we consider various nuclear effects. Nuclear binding energies are taken into account by assuming the quasi-invariant mass of the parent nucleon to be a Gaussian distribution with mean $\mu = 899.3 \, (922.8) \, \rm MeV$ and standard deviation $\sigma = 10.2 \, (3.82) \, \rm MeV$ for the $s(p)$-state, taken from~\cite{Takhistov:2016iun,Shiozawa:1999gi}. For large $\varphi$ masses, a fraction of the bound nuclei becomes physically unable to decay due to the nuclear binding energy, which leads to a scaling of the exposure. Fermi momentum distributions of the nucleus are obtained from Ref.~\cite{Takhistov:2016iun,Nakamura:1976mb} and treated as a Lorentz boost to the decay lepton. Fermi surface momentum, correlated decays, and nuclear de-excitation effects are not included since they do not produce large contributions. The lepton momentum distribution is simulated by generating $8\times 10^6$ isotropic bound proton decay events ($2\times 10^6$ $s$-state decays and $6\times 10^6$ $p$-state decays), each with a randomly selected quasi-invariant mass, nucleus Fermi momentum, and decay angle, along with $2\times 10^6$ free Hydrogen decay events. 

Detector resolution effects, estimated to be $\pm(2.5/\sqrt{|\pmb{p}_e|(\GeV)}\,+\,0.5)\%$ for electrons and $\pm 3\%$ for muons~\cite{Takhistov:2016iun}, are approximated by a Gaussian distribution around the lepton momentum. The resulting distribution is then normalized to the total number of background events and used as $N_i^{\mathrm{signal}}$. 
We include all systematic errors listed in~\cite{Super-Kamiokande:2015pys}, but we assume global errors over all momentum bins and all Super-K phases, i.e.~$f_i^j=\sigma_j$.
Nuclear correlated decay was considered to contribute only 10\% systematic uncertainty, since only 10\% of nucleon decays are affected. 

The $\chi^2$ minimization is carried out over the parameters $\alpha$ and $\beta$, and the global minimum is defined as the best fit. 
The lower bound on the lifetime derives from the number of allowed signal events at 90\% confidence level $N_{90}$, which is calculated from the value of $\beta$ at the 90\% confidence level. For some $\varphi$ masses, parts of the lepton momentum distribution will fall outside of the selection criteria of the Super-K search, and therefore be undetectable. To account for this, $N_{90}$ is scaled according to the proportion of events that can be detected. The lifetime constraint is then computed according to
\begin{align}
\Gamma^{-1} = \dfrac{N_{\rm nucleon}}{N_{90}}\sum_{\rm SK=1}^4 \lambda_{\rm SK} \cdot \epsilon_{\rm SK}\;, 
\end{align}
where $\lambda_{\rm SK}$ and $\epsilon_{\rm SK}$ denotes the exposure and detection efficiency of each Super-K phase and $N_{\rm nucleon}$ is the number of nucleons per kiloton of water, corresponding to $3.3 \times 10^{32}$ and $2.7 \times 10^{32}$ for the search of proton and neutron decay, respectively.

We validate our analysis by comparing to the Super-K lifetime constraints in the massless $\varphi$ limit. For the positron channel, our analysis produces a lower lifetime bound of $180\times10^{31}\rm yr$ assuming free decays, which reduces to $99\times10^{31}\rm yr$ upon including nuclear effects, in agreement with Super-K's $79\times10^{31}\rm yr$ within 25\%. Similarly for the anti-muon channel, our analysis produces a lifetime limit of $125\times10^{31}\rm yr$ which improves to $52\times10^{31}\rm yr$ with the inclusion of nuclear effects, within 27\% of Super-K's $41\times10^{31}\rm yr$.

\begin{figure}[tb!]
\centering
\includegraphics[width=0.45\linewidth]{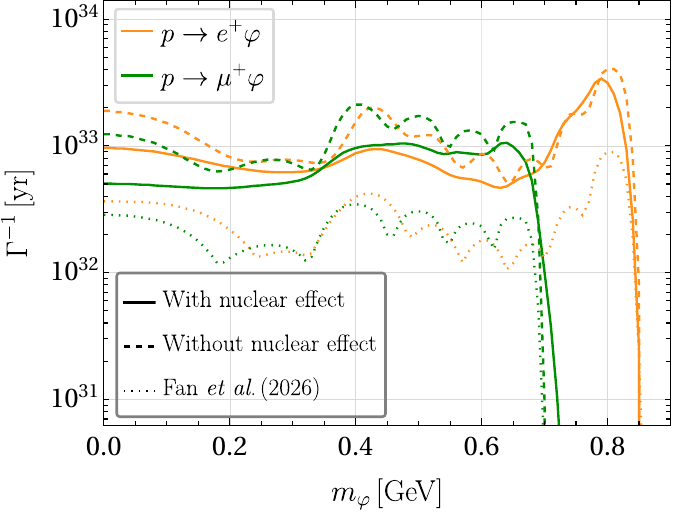}
\caption{
Recast Super-K constraints on the lifetime of two-body proton decays $p\to e^+(\mu^+)\varphi$ for a massive scalar $\varphi$. Results 
including nuclear effects are shown as solid curves, while those without considering nuclear effects are shown as dashed curves. The dotted curves represent previous results from~\cite{Fan:2025xhi}.
}
\label{fig:SuperKbounds}
\end{figure}

Our results are shown in \cref{fig:SuperKbounds}.
We note a sharp increase in the lifetime limit around $m_\varphi = 0.8\, \GeV$ for the positron channel. This is because the majority of decay events for this mass region reside in the first momentum bin, which has a large excess of atmospheric neutrino background events compared to the number of detected events.

%%%%%%%%%%%%%%%%%%%%%%%%%%%%%%
\subsubsection{Three-body neutron decay $n \to \bar\nu (\nu) \pi^0 \varphi$}
%%%%%%%%%%%%%%%%%%%%%%%%%%%%%%

\begin{figure}[tb!]
\centering
\includegraphics[width=0.465\linewidth]{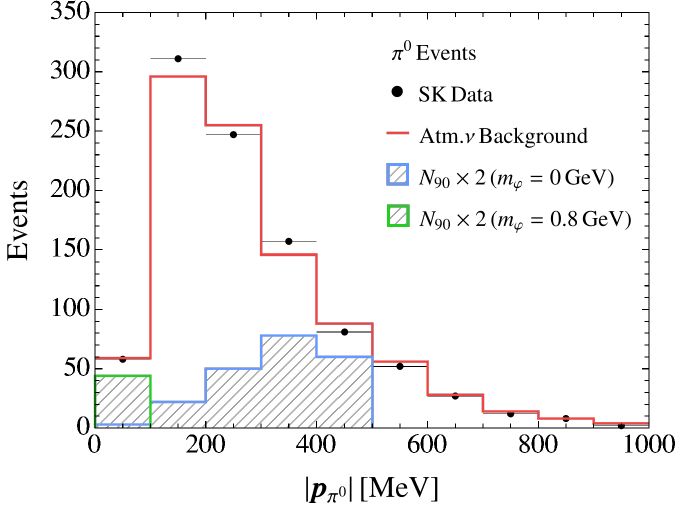}
\includegraphics[width=0.45\linewidth]{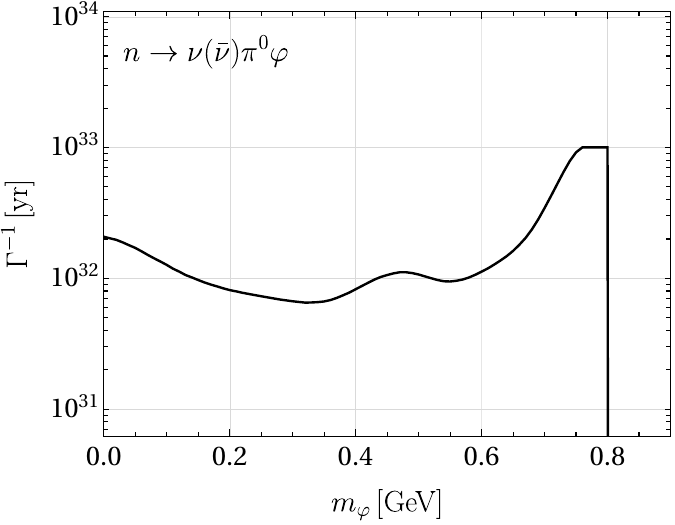}
\caption{Left: Binned Super-K data (black) as a function of the pion momentum together with the atmospheric neutrino background (red), and two allowed signal distributions at 90\% confidence level (hatched histograms) for a massless and a heavy scalar respectively, scaled by a factor of three.
Right: Recast lifetime constraint on the three-body neutron decay $n\to \bar\nu(\nu) \pi^0 \varphi$ as a function of the scalar mass $m_\varphi$.}
\label{fig:SuperKn2vpi0}
\end{figure}

We recast Super-K's search for BNV neutron decay with a neutrino and a pion~\cite{Super-Kamiokande:2013rwg}\footnote{The $p \to \bar{\nu} \pi^+$ channel is not considered because 
charged pions interact hadronically when traveling through the detector, an effect that we cannot account for, and more importantly provides a weaker constraint on the lifetime.} to construct lifetime constraints for three-body decay channels. The binned data are shown in the left panel of \cref{fig:SuperKn2vpi0}. The analysis performed is largely similar to the two-body decay case, with the three-body momentum distributions calculated in \cref{subsec:momentum distribution}. Nuclear effects are ignored because they result in broadening of the momentum distribution which is less pronounced in the three-body momentum distribution. The analysis is validated by performing a two-body $n \to \bar{\nu} \pi^0$ decay fit on the published data. Without nuclear effects, the lifetime bound obtained is $20 \times 10^{32}\rm yr$, which differs from Super-K's $11 \times 10^{32}\rm yr$ by roughly a factor of two, consistent with the previous analyses. We therefore do not simulate nuclear effects because they are not needed for the three-body analysis. 
Lifetime constraints obtained for the $n \to \bar{\nu} (\nu) \pi^0 \varphi$ channel are shown in the right panel of \cref{fig:SuperKn2vpi0}. We observe a flat peak in the lifetime constraint below $m_\varphi = 0.8\, \GeV$, due to all decay events occupying the first momentum bin.

Hyper-K will have 1.87 Megaton $\cdot$ years of exposure over ten years, a factor of 6.8 larger than the data used by the Super-K search for $p\to \ell^+\varphi$ and a factor of 10.8 larger than the $n\to \bar\nu (\nu) \pi^0 \varphi$ data, and hence we expect improvements in the sensitivity to the partial lifetimes of proton decay of 2.6 and 3.3, respectively.

%%%%%%%%%%%%%%%%%%%%%%%%%%%%%%
\subsection{Analysis}
\label{sec:analysis}
%%%%%%%%%%%%%%%%%%%%%%%%%%%%%%

%%%%%%%%%%%%%%%%%%%%%%%%%%%%%%
\subsubsection{Single operator limit in $\varphi$LEFT}
%%%%%%%%%%%%%%%%%%%%%%%%%%%%%%

\begin{figure}[t]
\centering
\includegraphics[width=0.45\linewidth]{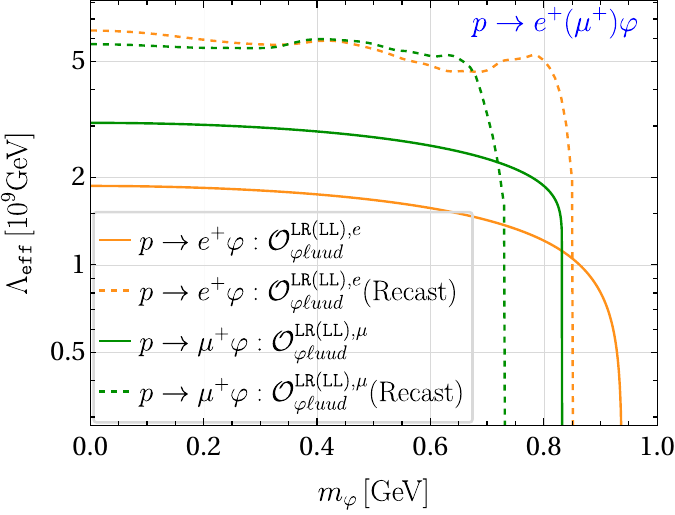}
\\
\vspace{0.2em}
\includegraphics[width=0.45\linewidth]{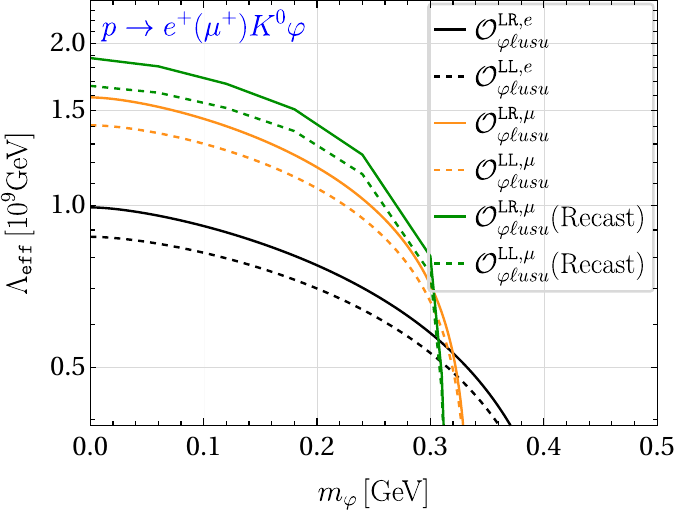}
\quad
\includegraphics[width=0.45\linewidth]{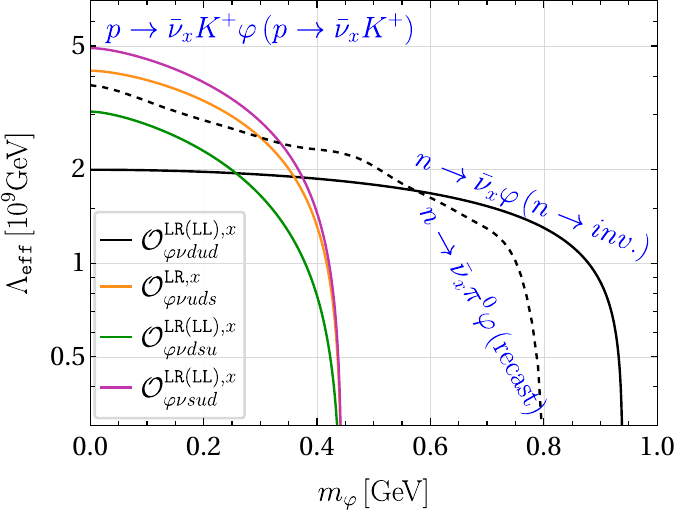}
\caption{  
Lower bounds on the effective scales associated with dim-7 $\varphi$LEFT operators based on available experimental data. }
\label{fig:1ope_bound}
\end{figure}

To illustrate the current exclusion limits, we present the constraints on scenarios with a single BNV operator. Taking advantage of the partial lifetime bounds of the two- and three-body nucleon decay modes detailed in the above subsection,  the constraints on the relevant Wilson coefficients are shown in \cref{fig:1ope_bound}.
We only show the bounds for operators with ${\tt LR}$ and ${\tt LL}$ chiralities, as their chirality partners with ${\tt RL}$ and ${\tt RR}$ receive the same constraints and can be directly inferred.  
The top panel shows the constraints on the effective scale related to operators $\calO_{\varphi\ell uud}^{\tL\tR(\tL\tL)}$ based on the bounds on $p\to e^+(\mu^+)\varphi$ from both the inclusive searches (solid curves)
and the recast analysis (dashed curves). As can be seen, the recast bounds are stronger for $m_\varphi \lesssim 0.86 (0.7)\,\rm GeV$ in the case with a charged $e^+$ ($\mu^+$), while the inclusive bounds dominate when $m_\varphi \gtrsim 0.86 (0.7)\,\rm GeV$, making them complementary with each other. 
The bottom-left panel demonstrates the constraints for operators $\calO_{\varphi\ell usu}^{\tL\tR(\tL\tL)}$ with a strange quark based on the bounds on $p\to e^+(\mu^+)K^0\varphi$. 
For the $e^+$ case, only the inclusive bound is used to set the constraint on $\calO_{\varphi\ell usu}^{\tL\tR,e}$ ($\calO_{\varphi\ell usu}^{\tL\tL,e}$) shown as black solid (dashed) curve. 
However, for the $\mu^+$ case, both the inclusive (orange) and recast (green) limits from~\cite{Fan:2025xhi} are taken into account. 
Last, the bottom-right panel collects the constraints on operators $\calO_{\varphi\nu dud}^{\tL\tR(\tL\tL)}$ and $\calO_{\varphi\nu uds}^{\tL\tR},\calO_{\varphi\nu  dsu}^{\tL\tR(\tL\tL)},\calO_{\varphi\nu sud}^{\tL\tR(\tL\tL)}$ from the decay modes $n\to\bar\nu\varphi(+\pi^0)$ and $p\to \bar\nu K^+\varphi$,
respectively. It is clear that the nucleon decays constrain the effective scale at the level of $\calO(10^9\rm GeV)$ for a wide range of the scalar mass. 
 
\begin{figure}[t]
\centering
\includegraphics[width=0.45\linewidth]{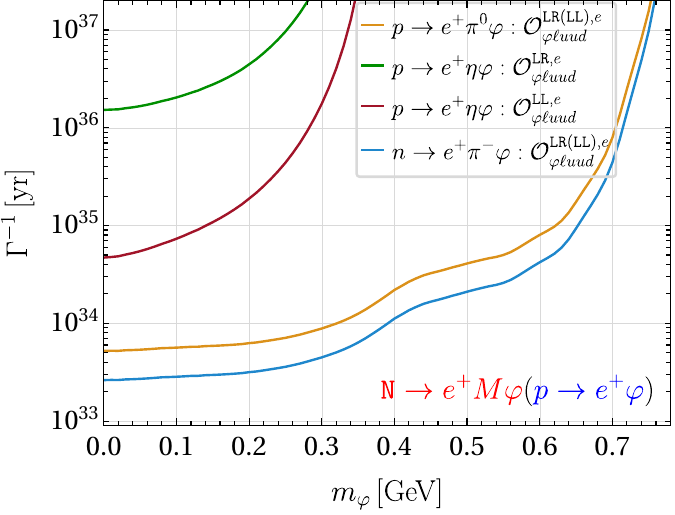}\quad
\includegraphics[width=0.45\linewidth]{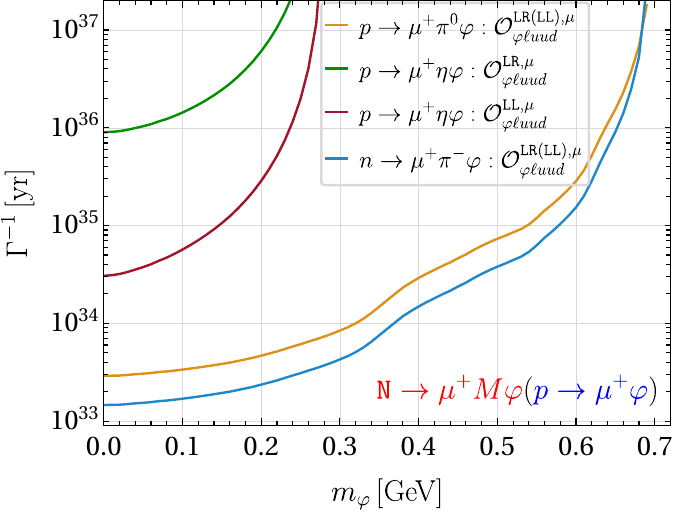}
\\
\vspace{0.2em}
\includegraphics[width=0.45\linewidth]{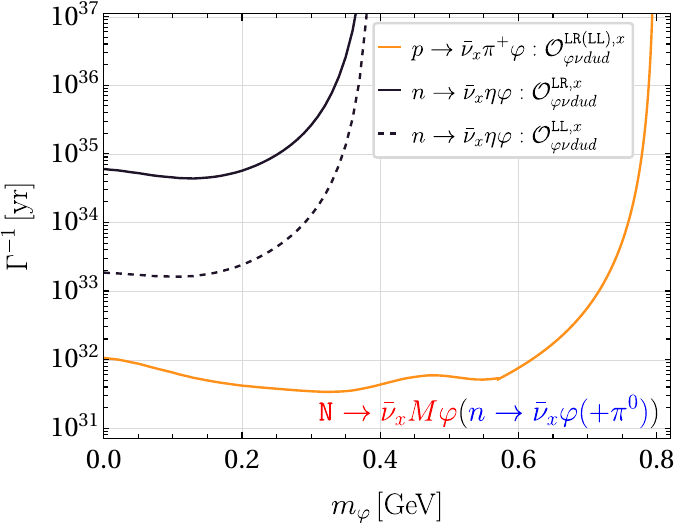}\quad 
\includegraphics[width=0.45\linewidth]{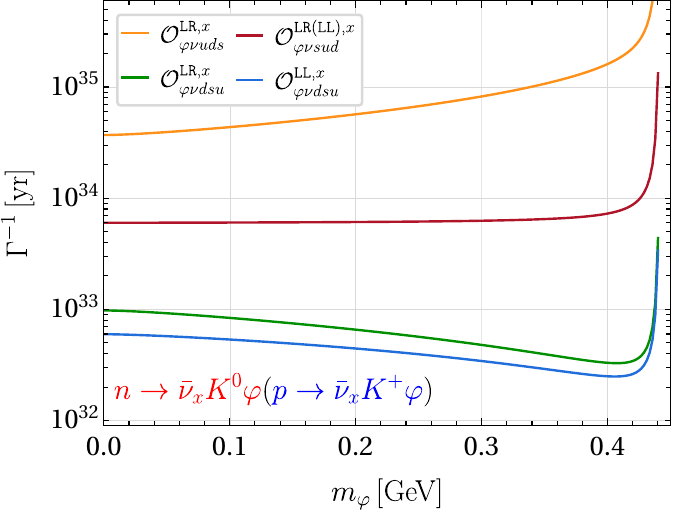}
\caption{  
Derived lower limits on the inverse lifetimes of three-body nucleon decays involving a light scalar $\varphi$ based on the results in \cref{fig:1ope_bound}. }
\label{fig:newlimits}
\end{figure}

Next, we employ the constraints in \cref{fig:1ope_bound} to derive stringent bounds on the partial lifetimes of the relevant three-body nucleon decay modes, the results are displayed in \cref{fig:newlimits}.
For the $p\to e^+(\mu^+)\varphi$ and $p\to e^+(\mu^+)K^0\varphi$ processes, we take the combined stronger constraints on each operator as our input. 
The top-left panel shows the bounds on $\Gamma^{-1}$ of $p\to e^+\pi^0(\eta)\varphi$ (orange, green, and purple curves) and $n\to e^+\pi^-\varphi$ (blue curve) based on the constraints on the Wilson coefficients from $p\to e^+\varphi$. 
Notice that the bound on $p\to e^+\pi^0\varphi$ is two times that of  $n\to e^+\pi^-\varphi$ due to an isospin relation for the matrix elements: $\sqrt{2}\langle \pi^0|\calO_{\varphi\ell uud}^{\tL\tR(\tL\tL)}|p\rangle = \langle \pi^-|\calO_{\varphi\ell uud}^{\tL\tR(\tL\tL)}|n\rangle$. The top-right panel shows the similar results for the $\mu^+$ case. The bottom two panels show the results for the three-body modes with a neutrino in the final state.
The bounds on the processes $p\to\bar\nu\pi^+\varphi$ and $n\to \bar\nu\eta\varphi$ in the bottom-left panel are based on the combined limit on $\Lambda_{\tt eff}$ from $n\to\bar\nu\pi^0\varphi$ and $n\to \bar\nu\varphi$, while the bounds on $n\to\bar\nu K^0\varphi$ shown in the bottom-right panel is derived from that of $p\to\bar\nu K^+\varphi$.    
As can be seen, the derived bounds on $\Gamma^{-1}$ are larger than $\calO(10^{32}\,\rm yrs)$ to $\calO(10^{36}\,\rm yrs)$ across difference processes and involved operators in the limit of $m_\varphi \to 0$.
Some of them may be surpassed at the next-generation experiments.

%%%%%%%%%%%%%%%%%%%%%%%%%%%%%%
\subsubsection{Two-dimensional constraints on operators with first-generation quarks in $\varphi$LEFT}
%%%%%%%%%%%%%%%%%%%%%%%%%%%%%%

\begin{figure}[t]
\centering
\includegraphics[width=0.45\linewidth]{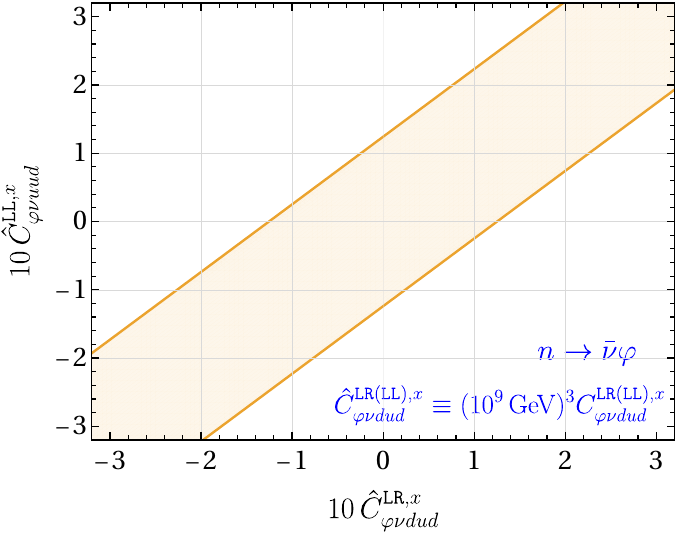}
\\
\vspace{0.2em}
\includegraphics[width=0.45\linewidth]{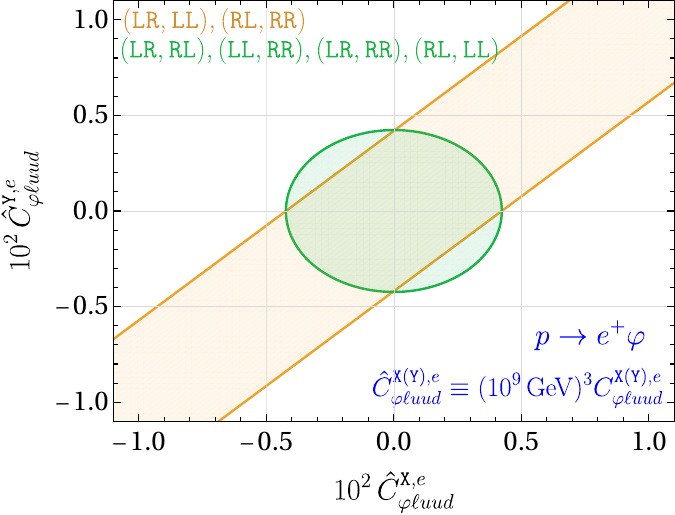}\quad
\includegraphics[width=0.45\linewidth]{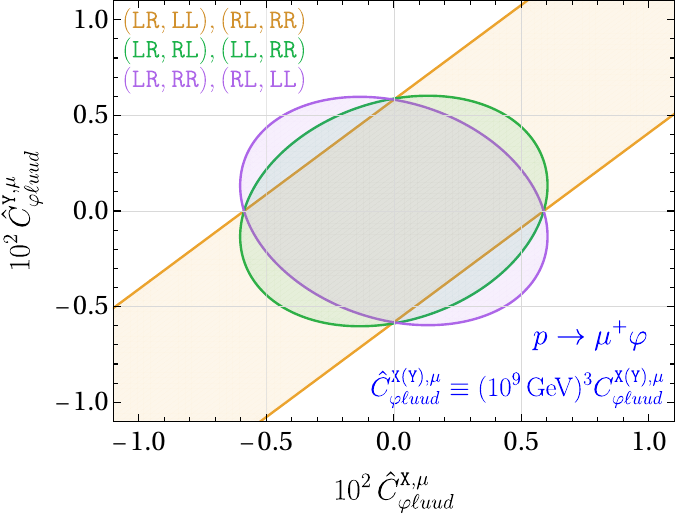}
\caption{  
Constraints in the plane of two Wilson coefficients  that involve only first-generation quarks for a massless scalar $\varphi$.}
\label{fig:2ope_bound}
\end{figure}

In this part, we consider the constraints in 
the two-dimensional plane with two nonvanishing Wilson coefficients at a time.
We restrict ourselves to the operators involving only first-generation quarks and focus on the two-body nucleon decays, $p\to e^+(\mu^+)\varphi$ and $n\to \bar\nu\varphi$, as they provide the most stringent constraints.
For illustration, we consider a massless scalar $\varphi$ and study the constraints on the dimensionless Wilson coefficients defined by $\hat C_i^j \equiv (10^9{\rm GeV})^3 C_i^j$, which are also assumed to be real numbers.
The bottom-left (right) panel of \cref{fig:2ope_bound} shows the results for all possible independent pairs of Wilson coefficients related to the process $p\to e^+\varphi$ ($p\to \mu^+\varphi$).
For $p\to e^+\varphi$, the degeneracy between the pairs $(C_{\varphi\ell uud}^{\tL\tR,e},C_{\varphi\ell uud}^{\tR\tL,e})$ and $(C_{\varphi\ell uud}^{\tL\tL,e},C_{\varphi\ell uud}^{\tR\tR,e})$ is due to the involved LECs $|c_1|\approx |c_2|$, while 
the further degeneracy with the pair $(C_{\varphi\ell uud}^{\tL\tR,e},C_{\varphi\ell uud}^{\tR\tR,e})$ or its chirality partner is due to the negligible electron mass. Similar behaviors appear in the $\mu^+$ case with the exception that the degeneracy of the pair $(C_{\varphi\ell uud}^{\tL\tR,\mu},C_{\varphi\ell uud}^{\tR\tR,\mu})$ and its chirality partner is broken due to the non-negligible  muon mass. Lastly, for $n\to \bar\nu \varphi$, only two Wilson coefficients are relevant and are constrained to take a band-like shape, as shown in the top panel. 

%%%%%%%%%%%%%%%%%%%%%%%%%%%%%%
\subsubsection{Constraints on $\varphi$SMEFT interactions}
%%%%%%%%%%%%%%%%%%%%%%%%%%%%%%

\begin{figure}[t]
\centering
\includegraphics[width=0.45\linewidth]{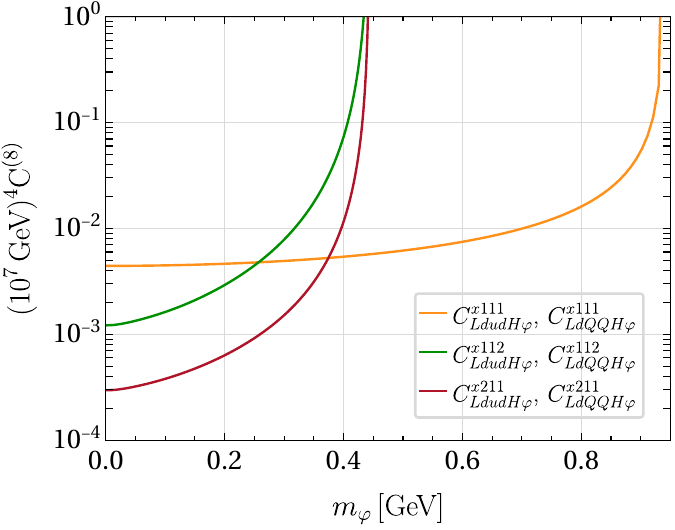}
\\
\vspace{0.2em}
\includegraphics[width=0.45\linewidth]{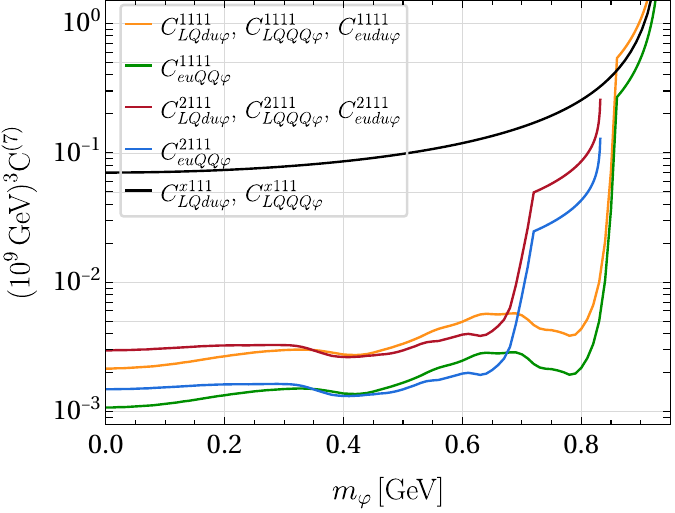}\quad
\includegraphics[width=0.45\linewidth]{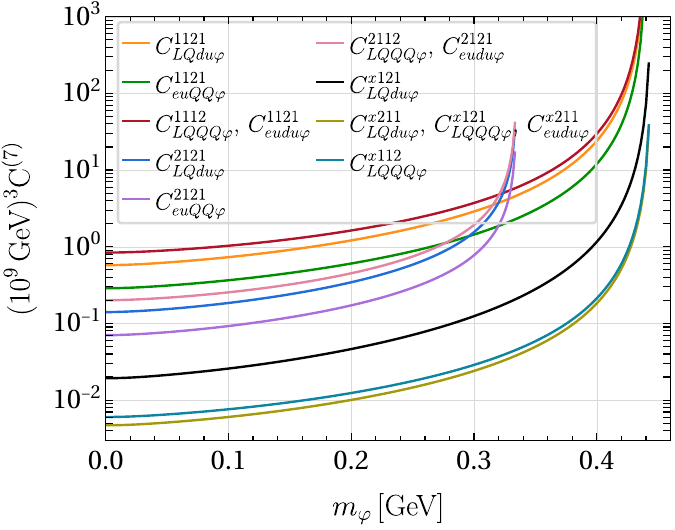}
\caption{  
Upper bounds on the dimensionless Wilson coefficients of dim-7 (bottom panels) and dim-8 (top panel) $\varphi$SMEFT operators.}
\label{fig:1ope_bound_SMEFT}
\end{figure}

In this subsection, we consider constraints on $\varphi$SMEFT interactions. 
For this purpose, we approximate the CKM matrix as a unit matrix and set bounds on dimensionless Wilson coefficients $(10^9\,{\rm GeV})^3C^{(7)}$ [$(10^7\,{\rm GeV})^4 C^{(8)}$] at a new physics scale $\Lambda_{\tt NP}=10^9\,[10^7]\,\rm GeV$ for dim-7 [dim-8] $\varphi$SMEFT operators. 
Similarly to \cref{fig:1ope_bound} for the $\varphi$LEFT case, we consider one operator at a time and employ the same set of processes and bounds on them to constrain the $\varphi$SMEFT Wilson coefficients. For $p \to e^+(\mu^+)\varphi$, we take the most stringent  combined limit throughout the $m_\varphi$ range as input; whereas for $p\to \mu^+ K^0\varphi$, only the inclusive limit is applied. 
Since the new physics scale is expected to be much higher, we also incorporate the RG running effect from $\Lambda_{\tt NP}$ to $\Lambda_\chi$ in our analysis. 

\cref{fig:1ope_bound_SMEFT} shows the final upper limits on the Wilson coefficients associated with the first and second generation of quarks. 
In the bottom-left panel, the constraints on the Wilson coefficients with lepton generation indices 1, 2, and $x$ are obtained from $p\to e^+\varphi$, $p\to \mu^+\varphi$, and $n\to \bar\nu_x \varphi$ processes, respectively; while in the bottom-right panel, they correspond, respectively, to $p\to e^+K^0\varphi$, $p\to \mu^+K^0\varphi$, and $p\to \bar\nu_x K^+ \varphi$ processes.
For the top panel, the constraints on the dim-8 $\varphi$SMEFT Wilson coefficients without (with) a second-generation quark are derived from the $n\to \nu_x \varphi$ ($p\to \nu_x K^+ \varphi$) process.
As observed, nucleon decays place stringent bounds on these couplings over a wide range of scalar masses $m_\varphi$. 
It is straightforward to infer (approximate) constraints on Wilson coefficients involving other flavor combinations, such as third-generation quarks, by multiplying the decay rate with the relevant off-diagonal CKM matrix elements. This results in slightly weaker constraints. As there are many operator combinations, we do not show two-dimensional figures, but note that constraints on a pair of $\varphi$SMEFT operators can be obtained by rescaling the bounds presented in \cref{fig:2ope_bound} with the appropriate RG and matching factors from \cref{tab:dim7ope}.

%%%%%%%%%%%%%%%%%%%%%%%%%%%%%%
\section{Long distance contribution to dinucleon decay }
\label{sec:dinucleondecay}
%%%%%%%%%%%%%%%%%%%%%%%%%%%%%%

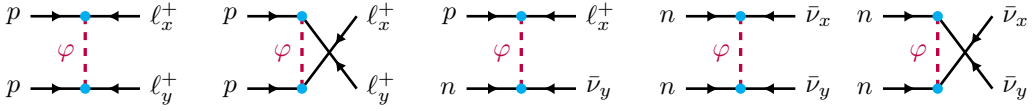
\begin{figure}[t]
\centering
\begin{tikzpicture}[mystyle,scale=0.8]
\begin{scope}[shift={(1,1)}]
\draw[f] (0, 0)node[left]{$p$} -- (1.5,0);
\draw[f] (3, 0)node[right]{$\ell_x^+$} -- (1.5,0) ;
\draw[snar, purple] (1.5,0) -- (1.5,-2) node[left,midway]{$\varphi$};
\draw[f] (0, -2)node[left]{$p$} -- (1.5,-2);
\draw[f] (3, -2)node[right]{$\ell_y^+$} -- (1.5,-2);
\filldraw [cyan] (1.5,0) circle (3pt);
\filldraw [cyan] (1.5,-2) circle (3pt);
\end{scope}
\end{tikzpicture}\quad
\begin{tikzpicture}[mystyle,scale=0.8]
\begin{scope}[shift={(1,1)}]
\draw[f] (0, 0)node[left]{$p$} -- (1.5,0);
\draw[fb] (1.5, 0) -- (3,-2) node[right]{$\ell_y^+$};
\draw[snar, purple] (1.5,0) -- (1.5,-2) node[left,midway]{$\varphi$};
\draw[f] (0, -2)node[left]{$p$} -- (1.5,-2);
\draw[fb] (1.5, -2) -- (3,0) node[right]{$\ell_x^+$};
\filldraw [cyan] (1.5,0) circle (3pt);
\filldraw [cyan] (1.5,-2) circle (3pt);
\end{scope}
\end{tikzpicture}\quad
\begin{tikzpicture}[mystyle,scale=0.8]
\begin{scope}[shift={(1,1)}]
\draw[f] (0, 0)node[left]{$p$} -- (1.5,0);
\draw[f] (3, 0)node[right]{$\ell_x^+$} -- (1.5,0);
\draw[snar, purple] (1.5,0) -- (1.5,-2) node[left,midway]{$\varphi$};
\draw[f] (0, -2)node[left]{$n$} -- (1.5,-2);
\draw[f] (3, -2)node[right]{$\bar\nu_y$} -- (1.5,-2);
\filldraw [cyan] (1.5,0) circle (3pt);
\filldraw [cyan] (1.5,-2) circle (3pt);
\end{scope}
\end{tikzpicture}\quad 
\begin{tikzpicture}[mystyle,scale=0.8]
\begin{scope}[shift={(1,1)}]
\draw[f] (0, 0)node[left]{$n$} -- (1.5,0);
\draw[f] (3, 0)node[right]{$\bar\nu_x$} -- (1.5,0);
\draw[snar, purple] (1.5,0) -- (1.5,-2) node[left,midway]{$\varphi$};
\draw[f] (0, -2)node[left]{$n$} -- (1.5,-2);
\draw[f] (3, -2)node[right]{$\bar\nu_y$} -- (1.5,-2);
\filldraw [cyan] (1.5,0) circle (3pt);
\filldraw [cyan] (1.5,-2) circle (3pt);
\end{scope}
\end{tikzpicture}
\begin{tikzpicture}[mystyle,scale=0.8]
\begin{scope}[shift={(1,1)}]
\draw[f] (0, 0)node[left]{$n$} -- (1.5,0);
\draw[fb] (1.5, 0) -- (3,-2) node[right]{$\bar\nu_y$};
\draw[snar, purple] (1.5,0) -- (1.5,-2) node[left,midway]{$\varphi$};
\draw[f] (0, -2)node[left]{$n$} -- (1.5,-2);
\draw[fb] (1.5, -2) -- (3,0) node[right]{$\bar\nu_x$};
\filldraw [cyan] (1.5,0) circle (3pt);
\filldraw [cyan] (1.5,-2) circle (3pt);
\end{scope}
\end{tikzpicture}
\caption{The t- and u-channel diagrams for dinucleon decays.}
\label{fig:NN2ll}
\end{figure}

The $\Delta B=2$ dinucleon decay processes for stable nuclei represent another golden channel to probe those exotic interactions when $\varphi$ is a real scalar, especially when its mass is beyond the nucleon decay threshold. 
We focus on the two-body final states with two leptons and neglect the three-body processes that include an additional pseudoscalar meson.
Such processes are generated via t- and u-channel diagrams mediating through the scalar particle as shown in \cref{fig:NN2ll},
which involve the insertion of two $\Delta B=1$ interactions in \cref{eq:N2lphi} within the $\varphi$LEFT framework.
Denoting collectively ${\texttt N}_1{\texttt N}_2\in\{pp, pn, nn\}$ and $l_x l_y\in \{\ell^+_x\ell^{+}_y,\ell^+_x\bar\nu_y(\nu_y),\bar\nu_x\bar\nu_y(\nu_x\nu_y)\}$, the decay rate for dinucleon ${\texttt N}_1{\texttt N}_2$ to dilepton $l_x l_y$ transition, ${\texttt N}_1(p_1)+{\texttt N}_2(p_2) \to l_x(k_1) +l_y(k_2)$, in a nucleus can be estimated in the following way~\cite{Goity:1994dq,He:2021mrt},
\begin{align}
 \Gamma_{{\texttt N}_1{\texttt N}_2\to l_x l_y} =
 \frac{1}{(2\pi)^3
 \sqrt{\rho_{{\texttt N}_1}\rho_{{\texttt N}_2}}}
 \int d^3 p_1 d^3 p_2 \rho_{{\texttt N}_1}(p_1)\rho_{{\texttt N}_2}(p_2)
 v_{12}^{\tt Mol}\sigma_{{\texttt N}_1{\texttt N}_2\to l_x l_y}, 
\end{align}
where $\rho_{\texttt N}(p)$ is the  nucleon-${\texttt N}$ number density distribution in momentum space and $\rho_{\texttt N} \equiv \int \frac{\rho_{\texttt N}(p) d^3p}{(2\pi)^{3/2}}$.
The Moller velocity is defined as, 
\begin{align}
 v_{12}^{\tt Mol}
\equiv {\sqrt{(p_1\cdot p_2)^2 - m_1^2 m_2^2} \over E_1 E_2  }
= {\sqrt{ [\hat s - (m_1 - m_2)^2] [\hat s - (m_1 + m_2)^2]} \over 2 E_1 E_2}, 
\end{align}
where $E_{1}(E_{2})$ is the energy of the initial state nucleon ${\texttt N}_1 ({\texttt N}_2)$ and $\hat s\equiv (p_1 + p_2)^2$. 
The cross-section for the free nucleon annihilation is
\begin{align}
\sigma_{{\texttt N}_1{\texttt N}_2\to l_x l_y}
={1\over 1 + \delta_{xy}}{1\over 4 \sqrt{(p_1\cdot p_2)^2 - m_1^2 m_2^2} }\int d\Pi_2\overline{\left|{\cal M}_{{\texttt N}_1{\texttt N}_2\to l_x l_y}\right|}^{2}.
\label{eq:sigma}
\end{align}
Here the prefactor accounts for the double counting of phase space for identical final state leptons, and $d\Pi_2$ is the Lorentz invariant two-body phase space. The matrix elements are summarized in \cref{app:M2fordinucleon} for reference.

Working in the centre of mass frame of the two-nucleon system and neglecting the nucleons' velocity corrections and mass difference, 
the velocity-weighted total cross sections for all possible $\Delta B=2$ dinucleon decays are calculated to be 
\begin{subequations}
\begin{align}
v\sigma_{pp\to \ell_x^+\ell_y^+}^{\Delta L=2} 
&= {1\over 1 + \delta_{xy}}
\frac{ \lambda^{1/2}(1,z_x/4,z_y/4)}{64\pi m_{\texttt N}^2 [2( 1+ z_\varphi)-z_x-z_y]^2}
\nonumber\\
&\times \Big\{
\big[ 4(4-z_x-z_y) \big( |C_{p \ell_x}^\tL|^2\,|C_{p \ell_y}^\tL|^2
+ |C_{p \ell_x}^\tR|^2\,|C_{p \ell_y}^\tR|^2 \big)
\nonumber\\
& +  \big(4 (z_x+z_y) -(z_x -z_y)^2\big) 
\big(|C_{p \ell_x}^\tL|^2\,|C_{p \ell_y}^\tR|^2 
+ |C_{p \ell_x}^\tR|^2\,  |C_{p \ell_y}^\tL|^2  \big)
\nonumber\\
& + 4\sqrt{z_y} (4+z_x -z_y) \big(|C_{p \ell_x}^\tL|^2+|C_{p \ell_x}^\tR|^2\big)\Re( C_{p \ell_y}^\tL C_{p \ell_y}^{\tR*})
\nonumber\\
&+ 4\sqrt{z_x} (4-z_x + z_y) \big(|C_{p \ell_y}^\tL|^2+|C_{p \ell_y}^\tR|^2\big) \Re( C_{p \ell_x}^\tL C_{p \ell_x}^{\tR*}) 
\nonumber\\
&+ 32\sqrt{z_x z_y} \Re( C_{p \ell_x}^\tL C_{p \ell_x}^{\tR*}) \Re( C_{p \ell_y}^\tL C_{p \ell_y}^{\tR*})
\Big\},
\\
v\sigma_{pn\to \ell_x^+\bar\nu_y}^{\Delta L=2} 
& = \frac{(4- z_x)^2}{512\pi m_{\texttt N}^2 (2 + 2 z_\varphi - z_x)^2}
\nonumber\\
&\times\big[ (4 + z_x)(|C_{p \ell_x}^\tL|^2 + |C_{p \ell_x}^\tR|^2) + 8 \sqrt{z_x} \Re(C_{p \ell_x}^\tL C_{p \ell_x}^{\tR*})\big] |C_{n \nu_y}^\tL|^2,
\\
v\sigma_{nn\to \bar\nu_x\bar\nu_y}^{\Delta L=2}
& = {1\over 1 + \delta_{xy}}\frac{1}{16\pi m_{\texttt N}^2 (1 + z_\varphi)^2}|C_{n \nu_x}^\tL|^2|C_{n \nu_y}^\tL|^2,
\\
v\sigma_{pn\to \ell_x^+\nu_y}^{\Delta L=0} 
& = v\sigma_{pn\to \ell_x^+\bar\nu_y}^{\Delta L=2}\big|_{C_{n \nu_{y}}^\tL\to C_{n \nu_{y}}^\tR}, 
\\
v\sigma_{nn\to \nu_x\nu_y}^{\Delta L=-2}
& = v\sigma_{nn\to \bar\nu_x\bar\nu_y}^{\Delta L=2}\big|_{C_{n \nu_{x,y}}^\tL\to C_{n \nu_{x,y}}^\tR},
\end{align}
\end{subequations}
where $z_x=m_{\ell_x}^2/m_{\texttt N}^2$ and $z_\varphi=m_{\varphi}^2/m_{\texttt N}^2$.  
In terms of $\varphi$LEFT interactions, we have
\begin{align}
C_{p\ell_x}^{\tL(\tR)} = c_1 C_{\varphi \ell uud}^{\tL\tR(\tR\tL),x} +c_2  C_{\varphi \ell uud}^{\tL\tL(\tR\tR),x}, \quad 
C_{n\nu_x}^{\tL(\tR)} = c_1 C_{\varphi \nu(\bar\nu) dud}^{\tL\tR(\tR\tL),x} +c_2  C_{\varphi \nu(\bar\nu) dud}^{\tL\tL(\tR\tR),x}.
\label{eq:dinucleonWCs}
\end{align}

\begin{figure}[t]
\centering
\includegraphics[width=0.45\linewidth]{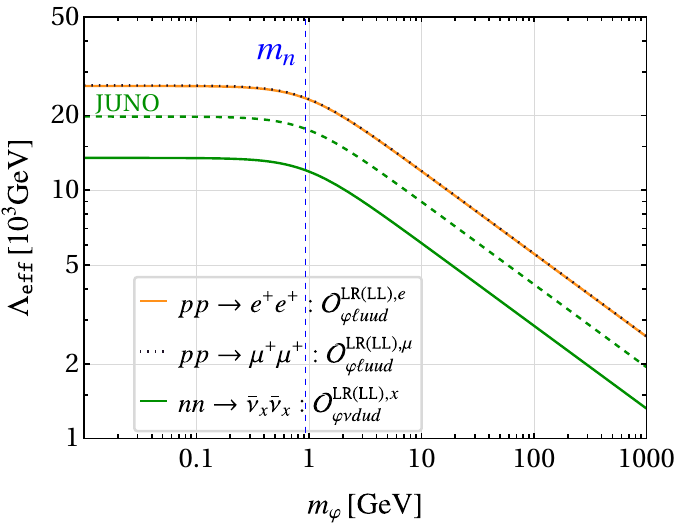}
\caption{Constraints on the effective scales associated with the dim-7 $\varphi$LEFT Wilson coefficients from dinucleon decays. The constraints apply equally to the chirality-flipped operators with $\tL\leftrightarrow\tR$. }
\label{fig:dinucleondecay}
\end{figure}

Experimentally, those dinucleon to dilepton decays have been searched in oxygen ${}^{16}$O and carbon
${}^{12}$C nuclei by Frejus~\cite{Frejus:1991ben}, KamLAND~\cite{KamLAND:2005pen}, and 
Super-Kamiokande~\cite{Super-Kamiokande:2015pys,Super-Kamiokande:2018apg}.
To a good approximation, we treat the nucleons in these nuclei to be quasi-free and neglect the small effects due to the nucleon Fermi motion and nuclear binding energy. $\rho_\texttt{N}$ approximately equals $0.25~{\rm fm}^{-3}$ for either proton or neutron. Then 
\begin{align}
\Gamma_{{\texttt N}_1{\texttt N}_2\to l_x l_y} = \rho_\texttt{N} (v\sigma)_{\texttt{N}_1\texttt{N}_2\to l_xl_y}. 
\end{align}
Similar to the single nucleon decay, we consider one effective operator at a time to set our constraints. Then the relevant dinucleon processes are $pp\to e^+ e^+$, $pp\to \mu^+ \mu^+$, and $nn\to \bar\nu_x\bar\nu_x (\nu_x\nu_x)$. The experimental bounds are 
$\Gamma_{pp\to e^+ e^+}^{-1,\tt exp}\geq 4.2 \times 10^{33}\,\rm yr$~\cite{Super-Kamiokande:2018apg}, 
$\Gamma_{pp\to \mu^+ \mu^+}^{-1,\tt exp}\geq 4.4 \times 10^{33}\,\rm yr$~\cite{Super-Kamiokande:2018apg}, 
and  $\Gamma_{nn\to inv.}^{-1,\tt exp}\geq 1.4 \times 10^{30}\,\rm yr$~\cite{KamLAND:2005pen}, respectively.
In \cref{fig:dinucleondecay}, we show the constraints on the effective scale associated with the relevant Wilson coefficients in \cref{eq:dinucleonWCs}. 
As shown, the effective scale is constrained to be above $\calO(10\,\rm TeV)$ by dinucleon decays when $m_\varphi$ is below the nucleon decay threshold. This constraint is significantly weaker than that from single nucleon decays. However, in the region where single nucleon decay constraints are absent, i.e., for $m_\varphi \gtrsim m_n$, dinucleon decays can impose a lower bound on the effective scale of $\calO(1\,\rm TeV)$. Thus, single nucleon decays and dinucleon decays provide complementary probes for these BNV interactions. 
For the dineutron mode, we also show the projected sensitivity of JUNO experiments with $\Gamma_{nn\to inv.}^{-1,\tt JUNO}\geq 1.4 \times 10^{32}\,\rm yr$~\cite{JUNO:2024pur}.  

%%%%%%%%%%%%%%%%%%%%%%%%%%%%%%
\section{UV-complete models}
\label{sec:UVmodels}
%%%%%%%%%%%%%%%%%%%%%%%%%%%%%%

Unlike the ALP case~\cite{Li:2024liy}, the effective BNV interactions involving a light scalar can be easily realized in UV-complete models. Below, we provide three representative models that can generate some of the $\varphi$SMEFT BNV operators at leading order without generating the corresponding SMEFT operator without the scalar $\varphi$.

%%%%%%%%%%%%%%%%%%%%%%%%%%%%%%
\subsection{A leptoquark model generating $\Delta(B-L)=0$ dim-7 interactions}
%%%%%%%%%%%%%%%%%%%%%%%%%%%%%%

We introduce the lepton-parity symmetry, $\mathbb{Z}_2^L$. Leptons are lepton-parity odd and all other SM particles even.  In addition, we introduce a light SM-singlet $\mathbb{Z}_2^L$-odd scalar field $\varphi\sim(1,1,0)_-$, one heavy scalar diquark $S_1\sim(3,1,-1/3)_+$, which is even under $\mathbb{Z}_2^L$, and one heavy scalar leptoquark $R_1\sim(3,1,-1/3)_-$, which is odd under $\mathbb{Z}_2^L$.
The three numbers in each bracket represent the associated field's representations under the SM gauge group $\rm SU(3)_c\otimes SU(2)_L\otimes U(1)_Y$, and the subscripts $\pm$ denote the $\mathbb{Z}_2^L$ parity. 
The relevant terms in the Lagrangian are
\begin{equation} 
\mathcal{L} \supset  R^\dagger_{1}( y_{Lpr}  \overline{Q_p^{i\C}} \epsilon_{ij} L^{j}_r  + y_{Rpr} \overline{u^\C_p} e_r)
+ S_{1}^\alpha (z_{Lpr}\overline{Q^{i \beta \C}_p} \epsilon_{ij} Q^{j\gamma}_r + z_{Rpr}\overline{u^{\beta \C}_p} d^{\gamma}_r)
\epsilon_{\alpha\beta\gamma}
-\kappa R_{1}^\dagger S_{1} \varphi
+ \mathrm{h.c.},
\end{equation}
where we largely follow the notation in~\cite{Dorsner:2016wpm}.
The coupling $z_{Lij}$ is symmetric under exchange of the flavor indices. Lepton parity is equivalent to the introduction of quark parity $\mathbb{Z}_2^Q$ with an $\mathbb{Z}_2^Q$-odd quark, since it results in the same Lagrangian. 

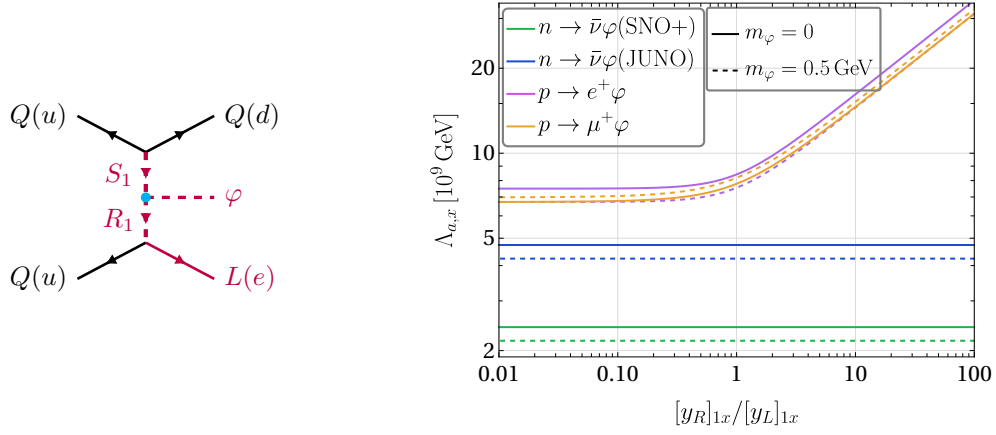
\begin{figure}[tb]
\centering
\begin{tikzpicture}[mystyle,scale=1]
\draw[f] (0, 0.7) -- (-1.5,1.5) node[left]{$Q(u)$};
\draw[f] (0, 0.7) -- (1.5,1.5) node[right]{$Q(d)$};
\draw[s,ultra thick] (0,0.7) -- (0,-0.3) node[midway,xshift = -10 pt]{$S_{1}$};
\draw[s,ultra thick, purple] (0,-0.3) -- (0,-1.3) node[midway,xshift = -10 pt]{$R_{1}$};
\draw[snar, purple] (0,-0.3) -- (1.5,-0.3) node[right]{$\varphi$};
\draw[f] (0,-1.3)--(-1.5,-2.1) node[left]{$Q(u)$};
\draw[f, purple](0,-1.3)--(1.5,-2.1)node[right]{$L(e)$};
\draw[draw=cyan,fill=cyan] (0,-0.3) circle (0.08cm);
\node[below right] at (6,4.3) {
\includegraphics[width=0.45\linewidth]{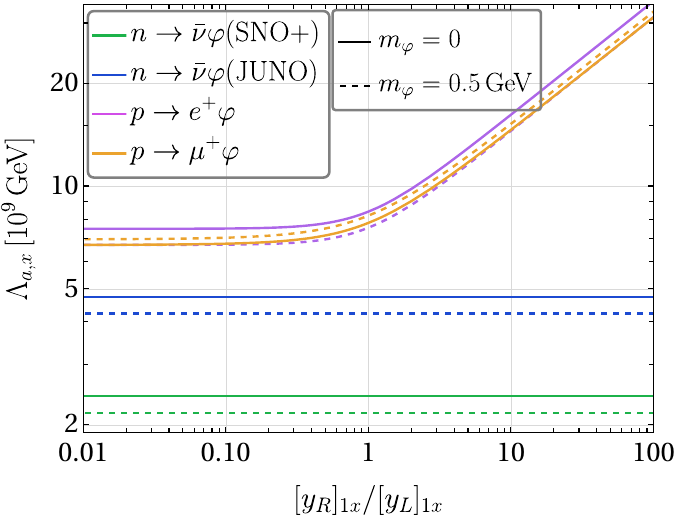} };
\end{tikzpicture}
\caption{
Left: Feynman diagram in the $\mathbb{Z}_2^L$ model that induces dim-7 BNV interactions. $\mathbb{Z}_2^L$-odd fields are highlighted in purple. Right: Constraints on the effective scale
$\Lambda_{a,x}$ as a function of the ratio $[y_R]_{1x}/[y_L]_{1x}$. 
The region below each curve is excluded  from current experimental bounds,
except for $n\to\bar\nu\varphi(\rm JUNO)$,
which indicates the sensitivity reach of the experiment.}
\label{fig:modelA}
\end{figure}

After integrating out the heavy leptoquarks in the left panel of \cref{fig:modelA}, we obtain four dim-7 $\varphi$SMEFT BNV operators 
\begin{subequations}
\begin{align}
C_{LQdu\varphi}^{prst} &= 
- \frac{\kappa^* [y_{L}]_{rp} [z_{R}]_{ts}  }{ m_S^2 m_R^2},
& C_{euQQ\varphi}^{prst} &= 
- \frac{\kappa^*  [y_{R}]_{rp} [z_{L}]_{st}  }{ m_S^2 m_R^2}, 
\\
C_{LQQQ\varphi}^{prst} &=
\frac{2 \kappa^* [y_{L}]_{rp} [z_L]_{st}  }{ m_S^2 m_R^2},
& C_{eudu\varphi}^{prst} & = 
\frac{\kappa^* [y_{R}]_{rp} [z_{R}]_{ts} }{ m_S^2 m_R^2}.
\end{align}
\end{subequations}
Restricting to first-generation quarks and approximating the CKM matrix as a unit matrix, we obtain the following nonzero $\varphi$LEFT Wilson coefficients that enter into the chiral matching at the chiral symmetry breaking scale,
\begin{subequations}
\begin{align}
C_{\varphi\nu dud}^{\tL\tR,x}  =    
1.78 \frac{\kappa^* [y_{L}]_{1x} [z_{R}]_{11} }{ m_S^2 m_R^2},\,
C_{\varphi\ell uud}^{\tL\tR,x}  =    
-C_{\varphi\nu dud}^{\tL\tR,x},\,
C_{\varphi\ell uud}^{\tR\tL,x}  =    
-1.78 \frac{2 \kappa^*[y_{R}]_{1x} [z_{L}]_{11} }{ m_S^2 m_R^2},
\\
C_{\varphi\nu dud}^{\tL\tL,x}  =    
1.78\frac{2\kappa^* [y_{L}]_{1x} [z_{L}]_{11} }{ m_S^2 m_R^2},\,
C_{\varphi\ell uud}^{\tL\tL,x}  =    
-C_{\varphi\nu dud}^{\tL\tL,x},\,
 C_{\varphi\ell uud}^{\tR\tR,x}  =    
-1.78 \frac{\kappa^*[y_{R}]_{1x} [z_{R}]_{11} }{ m_S^2 m_R^2},
\end{align}
\end{subequations}
where the prefactor 1.78 accounts for the RG running effects from a NP scale $\Lambda_{\tt NP} = 10^9\,\rm GeV$ to $\Lambda_\chi$. 

Numerically, we focus on the two-body nucleon decays $n\to \bar\nu\varphi$ and $p\to e^+(\mu^+)\varphi$, as these processes impose the most stringent constraints on the relevant parameter space as demonstrated by the EFT analysis above. The decay rates depend on the linear combinations 
$c_1 C^{\tL\tR,x}_{\varphi\ell uud} + c_2 C^{\tL\tL,x}_{\varphi\ell uud}$ and
$c_1 C^{\tR\tL,x}_{\varphi\ell uud} + c_2 C^{\tR\tR,x}_{\varphi\ell uud}$.
As $c_1\approx -c_2$, rather than assuming specific numerical values for the multidimensional parameter set, we analyze the constraints on two combined parameters: an effective scale defined as, $\Lambda_{a,x} \equiv (m_S^2 m_R^2/|\kappa^*[y_L]_{1x}([z_R]_{11}-2[z_L]_{11})|)^{1/3}$,
and the ratio $[y_R]_{1x}/[y_L]_{1x}$, which is taken to be real.
The right panel of \cref{fig:modelA} shows the constraints on the two-parameter plane derived from the three processes with two benchmark values of the scalar mass $m_\varphi$: 0 (solid curves) vs $0.5\,\rm GeV$ (dashed curves) for universal couplings.  
For the neutrino mode, both the current SNO+ limit (green) and the projected JUNO sensitivity (blue) are considered.

%%%%%%%%%%%%%%%%%%%%%%%%%%%%%%
\subsection{A vector-like-quark model generating $\Delta(B-L)=0$ dim-7 interactions} 
%%%%%%%%%%%%%%%%%%%%%%%%%%%%%%

Similarly to above, we introduce a lepton-parity symmetry, $\mathbb{Z}_2^L$. Leptons are lepton-parity odd and all other SM particles even. The SM particle content is extended with a light SM-singlet $\mathbb{Z}_2^L$-odd scalar field $\varphi\sim(1,1,0)_-$, one heavy scalar leptoquark $R_1\sim(3,1,-1/3)_-$, and one heavy vector-like quark $D(3,1,-1/3)_-$ [or $U(3,1,2/3)$], both are odd under $\mathbb{Z}_2^L$. The relevant terms in the Lagrangian are
\begin{equation} 
\mathcal{L}  \supset R^\dagger_{1}
( y_{Lpr}\overline{Q_p^{i\C}} \epsilon_{ij} L_r^{j}  
+ y_{Rpr} \overline{u_p^{\C}} e_r)
+ z_{Rp} R_{1}^\alpha 
(\overline{u_p^{\beta \C}} D_R^{\gamma} )
\epsilon_{\alpha\beta\gamma}
+ z_{Lp}
\varphi (\overline{d_p} D_L)
+ \mathrm{h.c.}.  
\end{equation}

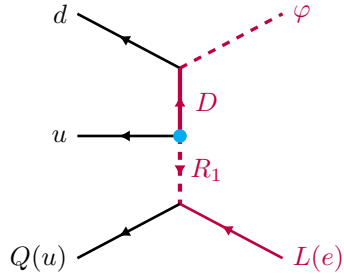
\begin{figure}[b]
\centering
\begin{tikzpicture}[mystyle,scale=1.5]
\begin{scope}[shift={(1,1)}]
\draw[f] (0, 0.7) -- (-1.5,1.5) node[left]{$d$};
\draw[snar, purple] (0, 0.7) -- (1.5,1.5) node[right]{$\varphi$};
\draw[f,ultra thick,purple]  (0,-0.3) --(0,0.7) node[midway,xshift = 10 pt]{$D$};
\draw[s,ultra thick, purple] (0,-0.3) -- (0,-1.3) node[midway,xshift = 10 pt]{$R_{1}$};
\draw[f] (0,-0.3) -- (-1.5,-0.3) node[left]{$u$};
\draw[f] (0,-1.3)--(-1.5,-2.1) node[left]{$Q(u)$};
\draw[f, purple](1.5,-2.1)node[right]{$L(e)$}--(0,-1.3);
\draw[draw=cyan,fill=cyan] (0,-0.3) circle (0.08cm);
\end{scope}
\end{tikzpicture}
\caption{Feynman diagram in the $\mathbb{Z}_2^L$ [or $\rm U(1)$] model that induces dim-7 BNV interactions. $\mathbb{Z}_2^L$-odd fields are highlighted in purple.}
\label{fig:BNVdim7M2}
\end{figure}

\cref{fig:BNVdim7M2} presents the Feynman diagram responsible for the dim-7 BNV interactions. After integrating out the heavy fields $D$ and $R_1$, we obtain the following non-vanishing Wilson coefficients for the BNV dim-7 $\varphi$SMEFT operators, 
\begin{align}
C_{LQdu\varphi}^{prst} = 
 \frac{[y_L]_{rp} z_{L s}^* z_{Rt}}{m_D m_R^2}, \quad 
 C_{eudu\varphi}^{prst}  = 
- \frac{[y_R]_{rp} z_{L s}^* z_{Rt}}{m_D m_R^2}. 
\end{align}

The $\mathbb{Z}_2^L$ lepton parity in this model can be extended to a global $U(1)$ symmetry, the generalized lepton number $U(1)_L$, under which $L,\,e,\,R_1,\,D_{L,R}$, and $\varphi$ carry charge 1, while all other particles are uncharged. 
In this case, the singlet $\varphi$ is a complex scalar. The above BNV matching results remain the same, except that $\varphi$ should be replaced by its conjugate $\varphi^*$ in the operators. 
Focusing on first-generation quarks, 
the above $\varphi$SMEFT Wilson coefficients result in the following $\varphi$LEFT Wilson coefficients,
\begin{align}
C_{\varphi\nu dud}^{\tL\tR,x}  =    
- 1.78 \frac{[y_{L}]_{1x} z_{L1}^* z_{R1} }{ m_D m_R^2},\,
C_{\varphi\ell uud}^{\tL\tR,x}  =    
-C_{\varphi\nu dud}^{\tL\tR,x},\,
 C_{\varphi\ell uud}^{\tR\tR,x}  =    
1.78 \frac{ [y_{R}]_{1x} z_{L1}^* z_{R1} }{ m_D m_R^2},
\end{align}
where the prefactor 1.78 corresponds to the RG running factor between $\Lambda_{\tt NP}=10^9\,\rm GeV$ and $\Lambda_\chi$.
Similarly, we define an effective scale, $\Lambda_{b,x} \equiv (m_D m_R^2/|[y_{L}]_{1x} z_{L1}^* z_{R1}|)^{1/3}$. 
The constraints in \cref{fig:modelA}
also apply to this model by replacing $\Lambda_{a,x}$ with $\Lambda_{b,x}$. 

%%%%%%%%%%%%%%%%%%%%%%%%%%%%%%
\subsection{A leptoquark model generating $\Delta(B+L)=0$ dim-8 interactions}
%%%%%%%%%%%%%%%%%%%%%%%%%%%%%%

The last model provides a UV completion for two of the $\Delta(B+L)=0$ dim-8 operators. As above we introduce a $\mathbb{Z}_2^L$ lepton parity and extend the SM by a light real scalar $\varphi\sim(1,1,0)_-$ odd under $\mathbb{Z}_2^L$, a heavy scalar diquark $S_1\sim(3,1,-1/3)_+$, even with respect to $\mathbb{Z}_2^L$, and a heavy scalar leptoquark $R_2\sim(3,2,1/6)_-$, odd with respect to $\mathbb{Z}_2^L$. The relevant terms in the Lagrangian are
\begin{equation} 
\mathcal{L}  \supset \left[ y_{2pr}  \overline{d_p} R^i_{2} \epsilon_{ij} L^{j}_r
+ S_{1}^\alpha (z_{Lpr} \overline{Q^{i\beta \C }_p} \epsilon_{ij} Q^{j\gamma}_r + z_{Rpr} \overline{u^{\beta \C}_p} d_r^\gamma )\epsilon_{\alpha\beta\gamma}
-\lambda_{\varphi SR} H^\dagger S_{1}^\dagger R_{2}  \varphi
+ \mathrm{h.c.}\right].
\end{equation}
The coupling $z_{Lij}$ is symmetric under the exchange of its flavor indices.

\begin{figure}[b]
\centering
\begin{tikzpicture}[mystyle,scale=1.5]
\begin{scope}[shift={(1,1)}]
\draw[f] (0, 0.7) -- (-1.5,1.5) node[left]{$Q(u)$};
\draw[f] (0, 0.7) -- (1.5,1.5) node[right]{$Q(d)$};
\draw[s,ultra thick] (0,0.7) -- (0,-0.3) node[midway,xshift = -10 pt]{$S_{1}$};
\draw[s,ultra thick, purple] (0,-0.3) -- (0,-1.3) node[midway,xshift = -10 pt]{$R_{2}$};
\draw[snar, purple] (0,-0.3) -- (1.5,-0.3) node[right]{$\varphi$};
\draw[s] (-1.5,-0.3)node[left]{$H$} -- (0,-0.3) ;
\draw[f] (0,-1.3)--(-1.5,-2.1) node[left]{$d$};
\draw[f, purple](1.5,-2.1)node[right]{$L$}--(0,-1.3);
\draw[draw=cyan,fill=cyan] (0,-0.3) circle (0.08cm);
\node[below right] at (4,2.8) {
\includegraphics[width=0.45\linewidth]{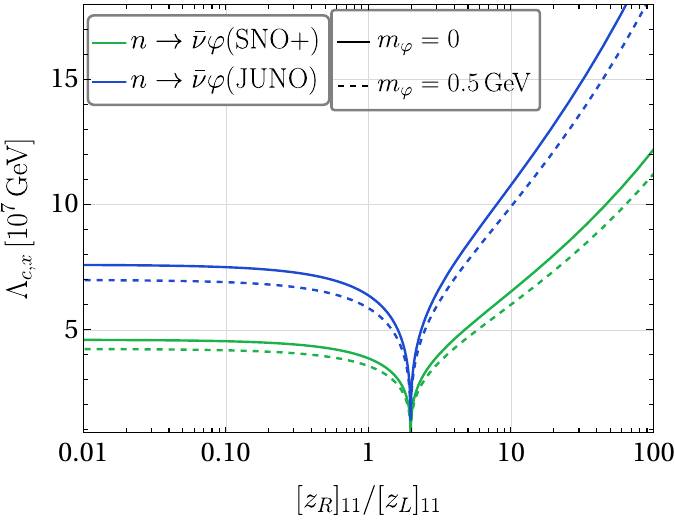}};
\end{scope}
\end{tikzpicture}
\caption{Left: Feynman diagram in the $\mathbb{Z}_2^L$ model that induces dim-8 BNV interactions. $\mathbb{Z}_2^L$-odd fields are highlighted in purple.
Right: Same as \cref{fig:modelA} but for the third model.}
\label{fig:modelC}
\end{figure}
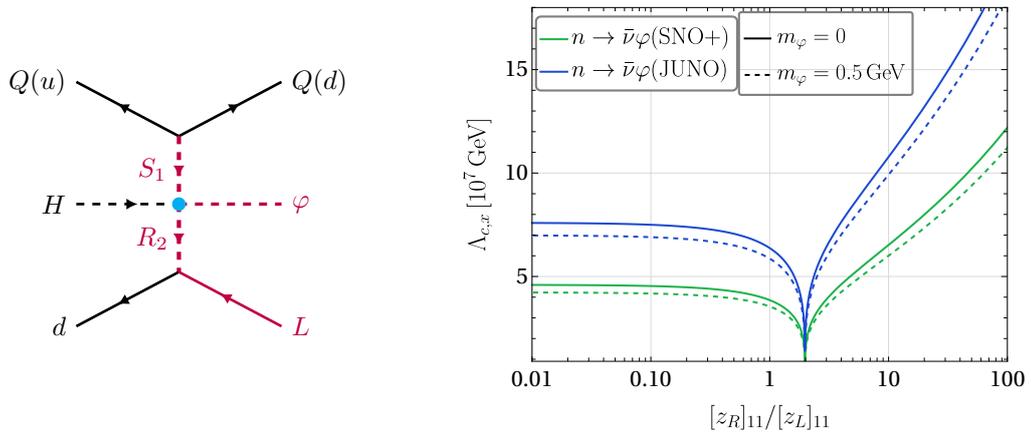

The left panel of \cref{fig:modelC} displays the tree-level Feynman diagram that generates the dim-8 BNV interactions. Integrating out $S_1$ and $R_2$ results into two nonvanishing Wilson coefficients at tree level
\begin{align}
C_{LdudH\varphi}^{prst} = 
\frac{[y_2]_{rp}^* [z_R]_{st} \lambda_{\varphi SR} }{m_S^2 m_R^2}, \quad
C_{LdQQH\varphi}^{prst} = 
\frac{2 [y_2]_{rp}^* [z_L]_{st} \lambda_{\varphi SR} }{m_S^2 m_R^2}.
\end{align}
Again, we restrict ourselves to first-generation quarks and approximate the CKM matrix by the unit matrix. 
The following two $\varphi$LEFT Wilson coefficients are generated by the matching condition in \cref{tab:dim7ope}, 
\begin{align}
C_{\varphi\bar\nu dud}^{\tR\tL,x} = 
1.61\frac{2 v [y_2]_{1x}^* [z_L]_{11} \lambda_{\varphi SR}}{\sqrt{2} m_S^2 m_R^2},\quad
C_{\varphi\bar\nu dud}^{\tR\tR,x} = 
1.61\frac{v [y_2]_{1x}^* [z_R]_{11} \lambda_{\varphi SR}}{\sqrt{2}m_S^2 m_R^2},
\end{align}
where the prefactor $1.61$ originates from the RG evolution from the NP scale $\Lambda_{\tt NP}=10^7\, \rm GeV$ to the scale $\Lambda_\chi$.
In contrast to the first two models, there are no two-body proton decays, but only the invisible two-body neutron decay. We define the effective scale $\Lambda_{c,x} \equiv (m_S^2 m_R^2/|[y_2]_{1x}^* [z_L]_{11} \lambda_{\varphi SR}|)^{1/4}$ and show the constraints in the plane $[z_R]_{11}/[z_L]_{11}$-$\Lambda_{c,x}$ in the right panel of \cref{fig:modelC}. Note that two-body neutron decays do not constrain $[z_R]_{11}\approx 2 [z_L]_{11}$ due to cancelation in the decay rate $n\to \bar\nu \varphi$. Three-body decays $n\to \bar \nu M \varphi$ will become the dominant constraint for $[z_R]_{11} \approx 2 [z_L]_{11}$.

%%%%%%%%%%%%%%%%%%%%%%%%%%%%%%
\section{Conclusion}
\label{sec:conclusion}
%%%%%%%%%%%%%%%%%%%%%%%%%%%%%%

We have systematically investigated BNV nucleon decays involving a light scalar particle $\varphi$ within the framework of EFTs. The construction of the leading-order BNV operators in the $\varphi$SMEFT framework shows that there are four dim-7 operators with $\Delta B =\Delta L =1$ and four dim-8 operators with $\Delta B = -\Delta L =1$, excluding operators with covariant derivatives. 
At the electroweak scale, these $\varphi$SMEFT operators are then matched to the dim-7 $\varphi$LEFT operators involving the scalar $\varphi$, SM leptons and light quarks, which describe the leading-order contributions to BNV nucleon decays. 
Employing the chiral perturbation theory framework, where the quark degrees of freedom are traded by hadronic octet baryons and pseudoscalar mesons,
we have formulated the general width expressions of two- and three-body nucleon decays involving the $\varphi$ in terms of the $\varphi$LEFT Wilson coefficients, and we have studied the momentum distribution of visible decay products such as charged leptons and pseudoscalar mesons.

For the two-body proton decay $p\to e^+(\mu^+)\varphi$ and the three-body neutron decay $n\to\bar\nu(\nu)\pi^0\varphi$, we reinterpret the Super-K experimental data and set stringent bounds on the occurrence of these modes over a wide range of the scalar mass. Together with the available experimental bounds on the invisible neutron decay, as well as those recasting bounds on three-body proton decays involving a $K^0$ or $K^+$, we have set stringent constraints on the Wilson coefficients of the relevant $\varphi$LEFT operators and their associated effective scale, which are further utilized to predict the lower bounds on some three-body nucleon decays that were not yet constrained.  
By including RG running effects and matching results, we further set constraints on the relevant $\varphi$SMEFT Wilson coefficients at a higher scale. The effective scales of BNV dim-7 and dim-8 $\varphi$SMEFT operators are constrained to be larger than $\calO(10^9\,\rm GeV)$ and $\calO(10^7\,\rm GeV)$, respectively.

In addition, we probe these exotic interactions through dinucleon to dilepton transitions, which provide complementary constraints. Although dinucleon decays only probe scales up to $\calO(10^4\,\rm GeV)$ and are thus less sensitive than nucleon decays which probe scales up to $\calO(10^9\,\rm GeV)$, they provide unique constraints, when the scalar mass exceeds the neutron threshold.  

Finally, we present several UV-complete models that can generate these BNV interactions and processes at leading order, while simultaneously forbidding the generation of the corresponding lower-dimensional SMEFT operators  without the scalar $\varphi$. Notably, the scalar particle in these models emerges as a compelling light dark matter candidate, owing to its kinematic stability against decay. 
A comprehensive study of the scalar dark matter phenomenology requires to go to a UV complete model and is beyond the scope of this study and will be presented in an upcoming publication~\cite{Ma:future}.

%%%%%%%%%%%%%%%%%%%%%%%%%%%%%%
\section*{Acknowledgements}
%%%%%%%%%%%%%%%%%%%%%%%%%%%%%%

X.-D.M. would like to thank the School of Physics of the University of New South Wales (UNSW) for its hospitality and financial support during the initial stage of this project. 
X.-D.M. is supported by Grant No.\,NSFC-12305110. W.Z. acknowledges support from an Australian Government Research Training Program Scholarship. This research includes computations using the computational cluster Katana~\cite{katana} supported by Research Technology Services at UNSW Sydney.

\appendix
%%%%%%%%%%%%%%%%%%%%%%%%%%%%%%
\section{Summary of decay widths in the massless limit}
\label{app:Gammaatm20}
%%%%%%%%%%%%%%%%%%%%%%%%%%%%%%

For the two- and three-body modes with a charged lepton, the decay widths in terms of $\varphi$LEFT Wilson coefficients take the forms:
\begin{subequations}
\begin{align}
\frac{\Gamma_{p\to e^+ \varphi}}{10^{-8}~\rm GeV^7} 
& =  
 147 |C_{\varphi \ell uud}^{\tL\tR,e}|^2
 +150 |C_{\varphi \ell uud}^{\tL\tL,e}|^2
 -298\Re(C_{\varphi \ell uud}^{\tL\tR,e} C_{\varphi \ell uud}^{\tL\tL,e*})
 - 0.161 C_{\varphi \ell uud}^{\tL\tR,e}C_{\varphi \ell uud}^{\tR\tL,e*}
\nonumber
\\
& + 0.162 (C_{\varphi \ell uud}^{\tL\tR,e}C_{\varphi \ell uud}^{\tR\tR,e* }
+ C_{\varphi \ell uud}^{\tL\tL,e}C_{\varphi \ell uud}^{\tR\tL,e* })
- 0.164 C_{\varphi \ell uud}^{\tL\tL,e} C_{\varphi \ell uud}^{\tR\tR,e*} +\tL\leftrightarrow \tR, 
 \\%
 \frac{\Gamma_{p\to \mu^+ \varphi}}{10^{-8}~\rm GeV^7} 
 & =  
 147 |C_{\varphi \ell uud}^{\tL\tR,\mu}|^2
 +150 |C_{\varphi \ell uud}^{\tL\tL,\mu}|^2
 -298\Re(C_{\varphi \ell uud}^{\tL\tR,\mu} C_{\varphi \ell uud}^{\tL\tL,\mu*})
 - 32.8 C_{\varphi \ell uud}^{\tL\tR,\mu}C_{\varphi \ell uud}^{\tR\tL,\mu*}
\nonumber
\\
& + 33.1 (C_{\varphi \ell uud}^{\tL\tR,\mu}C_{\varphi \ell uud}^{\tR\tR,\mu* }
+ C_{\varphi \ell uud}^{\tL\tL,\mu}C_{\varphi \ell uud}^{\tR\tL,\mu* })
- 33.4 C_{\varphi \ell uud}^{\tL\tL,\mu} C_{\varphi \ell uud}^{\tR\tR,\mu*} +\tL\leftrightarrow \tR, 
 \\%%
 \frac{\Gamma_{p\to e^+ \pi^0 \varphi}}{10^{-8}~\rm GeV^7} 
 & =   
 27.2 |C_{\varphi \ell uud}^{\tL\tR,e}|^2
 +27.7 |C_{\varphi \ell uud}^{\tL\tL,e}|^2
 - 55\Re(C_{\varphi \ell uud}^{\tL\tR,e} C_{\varphi \ell uud}^{\tL\tL,e*})
 - 0.0035 C_{\varphi \ell uud}^{\tL\tR,e}C_{\varphi \ell uud}^{\tR\tL,e*}
\nonumber
\\
& + 0.0035 (C_{\varphi \ell uud}^{\tL\tR,e}C_{\varphi \ell uud}^{\tR\tR,e* }
+ C_{\varphi \ell uud}^{\tL\tL,e}C_{\varphi \ell uud}^{\tR\tL,e* })
- 0.0036 C_{\varphi \ell uud}^{\tL\tL,e} C_{\varphi \ell uud}^{\tR\tR,e*} +\tL\leftrightarrow \tR,
\\%
 \frac{\Gamma_{p\to \mu^+ \pi^0 \varphi}}{10^{-8}~\rm GeV^7} 
 & =   
 25.6 |C_{\varphi \ell uud}^{\tL\tR,\mu}|^2
 +26.1 |C_{\varphi \ell uud}^{\tL\tL,\mu}|^2
 - 51.6\Re(C_{\varphi \ell uud}^{\tL\tR,\mu} C_{\varphi \ell uud}^{\tL\tL,\mu*})
 - 0.58 C_{\varphi \ell uud}^{\tL\tR,\mu}C_{\varphi \ell uud}^{\tR\tL,\mu*}
\nonumber
\\
& + 0.59 (C_{\varphi \ell uud}^{\tL\tR,\mu}C_{\varphi \ell uud}^{\tR\tR,\mu * }
+ C_{\varphi \ell uud}^{\tL\tL,\mu}C_{\varphi \ell uud}^{\tR\tL,\mu * })
- 0.60 C_{\varphi \ell uud}^{\tL\tL,\mu} C_{\varphi \ell uud}^{\tR\tR,\mu *} +\tL\leftrightarrow \tR,
\\%
 \frac{\Gamma_{p\to e^+ \eta \varphi}}{10^{-8}~\rm GeV^7} 
 & =   
 0.09 |C_{\varphi \ell uud}^{\tL\tR,e}|^2
 +3.02 |C_{\varphi \ell uud}^{\tL\tL,e}|^2
 +0.97\Re(C_{\varphi \ell uud}^{\tL\tR,e} C_{\varphi \ell uud}^{\tL\tL,e*})
 + 0.0004 C_{\varphi \ell uud}^{\tL\tR,e}C_{\varphi \ell uud}^{\tR\tL,e*}
\nonumber
\\
& + 0.002 (C_{\varphi \ell uud}^{\tL\tR,e}C_{\varphi \ell uud}^{\tR\tR,e* }
+ C_{\varphi \ell uud}^{\tL\tL,e}C_{\varphi \ell uud}^{\tR\tL,e* })
+ 0.0078 C_{\varphi \ell uud}^{\tL\tL,e} C_{\varphi \ell uud}^{\tR\tR,e*} +\tL\leftrightarrow \tR,
\\%
 \frac{\Gamma_{p\to \mu^+ \eta \varphi}}{10^{-8}~\rm GeV^7} 
 & =   
 0.08 |C_{\varphi \ell uud}^{\tL\tR,\mu}|^2
+2.42 |C_{\varphi \ell uud}^{\tL\tL,\mu}|^2
+0.82\Re(C_{\varphi \ell uud}^{\tL\tR,\mu} C_{\varphi \ell uud}^{\tL\tL,\mu*})
+0.055 C_{\varphi \ell uud}^{\tL\tR,\mu}C_{\varphi \ell uud}^{\tR\tL,\mu*}
\nonumber
\\
& + 0.27 (C_{\varphi \ell uud}^{\tL\tR,\mu}C_{\varphi \ell uud}^{\tR\tR,\mu * }
+ C_{\varphi \ell uud}^{\tL\tL,\mu}C_{\varphi \ell uud}^{\tR\tL,\mu * })
+1.08 C_{\varphi \ell uud}^{\tL\tL,\mu} C_{\varphi \ell uud}^{\tR\tR,\mu *} +\tL\leftrightarrow \tR,
\\%
 \frac{\Gamma_{p\to e^+ K^0\varphi}}{10^{-8}~\rm GeV^7} 
 & =   
 3.3 |C_{\varphi \ell usu}^{\tL\tR,e}|^2
 +1.55 |C_{\varphi \ell usu}^{\tL\tL,e}|^2
 +4.4\Re(C_{\varphi \ell usu}^{\tL\tR,e} C_{\varphi \ell usu}^{\tL\tL,e*})
+0.0074 C_{\varphi \ell usu}^{\tL\tR,e}C_{\varphi \ell usu}^{\tR\tL,e*}
\nonumber
\\
& + 0.0063 (C_{\varphi \ell usu}^{\tL\tR,e}C_{\varphi \ell usu}^{\tR\tR,e* }
+ C_{\varphi \ell usu}^{\tL\tL,e}C_{\varphi \ell usu}^{\tR\tL,e* })
+ 0.005 C_{\varphi \ell usu}^{\tL\tL,e} C_{\varphi \ell usu}^{\tR\tR,e*} +\tL\leftrightarrow \tR,
\\%
 \frac{\Gamma_{p\to \mu^+ K^0 \varphi}}{10^{-8}~\rm GeV^7} 
 & =   
2.77 |C_{\varphi \ell usu}^{\tL\tR,\mu}|^2
+1.36 |C_{\varphi \ell usu}^{\tL\tL,\mu}|^2
+3.8\Re(C_{\varphi \ell usu}^{\tL\tR,\mu} C_{\varphi \ell usu}^{\tL\tL,\mu*})
+1.1 C_{\varphi \ell usu}^{\tL\tR,\mu}C_{\varphi \ell usu}^{\tR\tL,\mu*}
\nonumber
\\
& + 0.94 (C_{\varphi \ell usu}^{\tL\tR,\mu}C_{\varphi \ell usu}^{\tR\tR,\mu * }
+ C_{\varphi \ell usu}^{\tL\tL,\mu}C_{\varphi \ell usu}^{\tR\tL,\mu * })
+0.75 C_{\varphi \ell usu}^{\tL\tL,\mu} C_{\varphi \ell usu}^{\tR\tR,\mu *} +\tL\leftrightarrow \tR,
\\%
 \frac{\Gamma_{n\to e^+ \pi^- \varphi}}{10^{-8}~\rm GeV^7} 
 & =   
 54.1 |C_{\varphi \ell uud}^{\tL\tR,e}|^2
 +55.1 |C_{\varphi \ell uud}^{\tL\tL,e}|^2
 - 109\Re(C_{\varphi \ell uud}^{\tL\tR,e} C_{\varphi \ell uud}^{\tL\tL,e*})
 - 0.0068 C_{\varphi \ell uud}^{\tL\tR,e}C_{\varphi \ell uud}^{\tR\tL,e*}
\nonumber
\\
& + 0.0069 (C_{\varphi \ell uud}^{\tL\tR,e}C_{\varphi \ell uud}^{\tR\tR,e* }
+ C_{\varphi \ell uud}^{\tL\tL,e}C_{\varphi \ell uud}^{\tR\tL,e* })
- 0.007 C_{\varphi \ell uud}^{\tL\tL,e} C_{\varphi \ell uud}^{\tR\tR,e*} +\tL\leftrightarrow \tR,
\\%
 \frac{\Gamma_{n\to \mu^+ \pi^- \varphi}}{10^{-8}~\rm GeV^7} 
 & =   
50.8 |C_{\varphi \ell uud}^{\tL\tR,\mu}|^2
 +51.8 |C_{\varphi \ell uud}^{\tL\tL,\mu}|^2
 - 103\Re(C_{\varphi \ell uud}^{\tL\tR,\mu} C_{\varphi \ell uud}^{\tL\tL,\mu*})
 - 1.14 C_{\varphi \ell uud}^{\tL\tR,\mu}C_{\varphi \ell uud}^{\tR\tL,\mu*}
\nonumber
\\
& + 1.15 (C_{\varphi \ell uud}^{\tL\tR,\mu}C_{\varphi \ell uud}^{\tR\tR,\mu * }
+ C_{\varphi \ell uud}^{\tL\tL,\mu}C_{\varphi \ell uud}^{\tR\tL,\mu * })
- 1.16 C_{\varphi \ell uud}^{\tL\tL,\mu} C_{\varphi \ell uud}^{\tR\tR,\mu *} +\tL\leftrightarrow \tR,
\\%
 \frac{\Gamma_{n\to e^- K^+ \varphi}}{10^{-8}~\rm GeV^7} 
 & =   
3.43 |C_{\varphi\bar\ell dds}^{\tL\tR,e}|^2
 +1.62 |C_{\varphi\bar\ell dds}^{\tL\tL,e}|^2
+4.6\Re(C_{\varphi\bar\ell dds}^{\tL\tR,e} C_{\varphi\bar\ell dds}^{\tL\tL,e*})
+0.0076 C_{\varphi\bar\ell dds}^{\tL\tR,e}C_{\varphi\bar\ell dds}^{\tR\tL,e*}
\nonumber
\\
& + 0.0065 (C_{\varphi\bar\ell dds}^{\tL\tR,e}C_{\varphi\bar\ell dds}^{\tR\tR,e* }
+ C_{\varphi\bar\ell dds}^{\tL\tL,e}C_{\varphi\bar\ell dds}^{\tR\tL,e* })
+0.0052 C_{\varphi\bar\ell dds}^{\tL\tL,e} C_{\varphi\bar\ell dds}^{\tR\tR,e*} +\tL\leftrightarrow \tR,
\\%
 \frac{\Gamma_{n\to \mu^- K^+ \varphi}}{10^{-8}~\rm GeV^7} 
 & =   
2.89 |C_{\varphi\bar\ell dds}^{\tL\tR,\mu}|^2
+1.42 |C_{\varphi\bar\ell dds}^{\tL\tL,\mu}|^2
+ 4\Re(C_{\varphi\bar\ell dds}^{\tL\tR,\mu} C_{\varphi\bar\ell dds}^{\tL\tL,\mu*})
+1.14 C_{\varphi\bar\ell dds}^{\tL\tR,\mu}C_{\varphi\bar\ell dds}^{\tR\tL,\mu*}
\nonumber
\\
& + 0.97 (C_{\varphi\bar\ell dds}^{\tL\tR,\mu}C_{\varphi\bar\ell dds}^{\tR\tR,\mu * }
+ C_{\varphi\bar\ell dds}^{\tL\tL,\mu}C_{\varphi\bar\ell dds}^{\tR\tL,\mu * })
+0.78 C_{\varphi\bar\ell dds}^{\tL\tL,\mu} C_{\varphi\bar\ell dds}^{\tR\tR,\mu *} +\tL\leftrightarrow \tR.
\end{align}
\end{subequations}

For the two- and three-body modes with a neutrino, the decay widths in terms of $\varphi$LEFT Wilson coefficients take the forms:
\begin{subequations}
\begin{align}
 \frac{\Gamma_{n\to \bar\nu_x \varphi}}{10^{-8}~\rm GeV^7} 
 & =   
148 |C_{\varphi\nu dud}^{\tL\tR,x}|^2
+151 |C_{\varphi\nu dud}^{\tL\tL,x}|^2
-298\Re(C_{\varphi\nu dud}^{\tL\tR,x} C_{\varphi\nu dud}^{\tL\tL,x*}),
\\%
\frac{\Gamma_{p\to \bar\nu_x \pi^+ \varphi}}{10^{-8}~\rm GeV^7} 
& =   
53.5 |C_{\varphi\nu dud}^{\tL\tR,x}|^2
+54.6 |C_{\varphi\nu dud}^{\tL\tL,x}|^2
-108 \Re(C_{\varphi\nu dud}^{\tL\tR,x} C_{\varphi\nu dud}^{\tL\tL,x*}),
\\%
\frac{\Gamma_{p\to \bar\nu_x K^+ \varphi}}{10^{-8}~\rm GeV^7} 
& =   
1.84 |C_{\varphi\nu uds}^{\tL\tR,x}|^2
+ 0.3 |C_{\varphi\nu dsu}^{\tL\tR,x}|^2
+ 5.04 |C_{\varphi\nu sud }^{\tL\tR,x}|^2
+ 0.3 |C_{\varphi\nu dsu}^{\tL\tL,x}|^2
+ 5.13 |C_{\varphi\nu sud }^{\tL\tL,x}|^2
\nonumber
\\
& - 1.26 \Re(C_{\varphi\nu uds}^{\tL\tR,x} C_{\varphi\nu dsu}^{\tL\tR,x*})
+5.93 \Re(C_{\varphi\nu uds}^{\tL\tR,x} C_{\varphi\nu sud}^{\tL\tR,x*})
-2.33 \Re(C_{\varphi\nu dsu}^{\tL\tR,x} C_{\varphi\nu sud}^{\tL\tR,x*})
\nonumber
\\
&+1.27 \Re(C_{\varphi\nu uds}^{\tL\tR,x} C_{\varphi\nu dsu}^{\tL\tL,x*})
-5.98 \Re(C_{\varphi\nu uds}^{\tL\tR,x} C_{\varphi\nu sud}^{\tL\tL,x*})
-0.6 \Re(C_{\varphi\nu dsu}^{\tL\tR,x} C_{\varphi\nu dsu}^{\tL\tL,x*})
\nonumber
\\
& +2.35 \Re(C_{\varphi\nu dsu}^{\tL\tR,x} C_{\varphi\nu sud}^{\tL\tL,x*} + C_{\varphi\nu sud}^{\tL\tR,x} C_{\varphi\nu dsu}^{\tL\tL,x*})
-10.2 \Re(C_{\varphi\nu sud}^{\tL\tR,x} C_{\varphi\nu sud}^{\tL\tL,x*})
\nonumber
\\
&
-2.37 \Re(C_{\varphi\nu dsu }^{\tL\tL,x} C_{\varphi\nu sud}^{\tL\tL,x*}),
\\%
 \frac{\Gamma_{n\to \bar\nu_x \pi^0 \varphi}}{10^{-8}~\rm GeV^7} 
 & =   
27.3 |C_{\varphi\nu dud}^{\tL\tR,x}|^2
+27.9 |C_{\varphi\nu dud}^{\tL\tL,x}|^2
-55.2 \Re(C_{\varphi\nu dud}^{\tL\tR,x} C_{\varphi\nu dud}^{\tL\tL,x*}),
\\%
 \frac{\Gamma_{n\to \bar\nu_x \eta \varphi}}{10^{-8}~\rm GeV^7} 
 & =   
0.094 |C_{\varphi\nu dud}^{\tL\tR,x}|^2
+3.05 |C_{\varphi\nu dud}^{\tL\tL,x}|^2
+0.98 \Re(C_{\varphi\nu dud}^{\tL\tR,x} C_{\varphi\nu dud}^{\tL\tL,x*}),
\\%
\frac{\Gamma_{n\to \bar\nu_x K^0 \varphi}}{10^{-8}~\rm GeV^7} 
 & =   
0.29 |C_{\varphi\nu uds}^{\tL\tR,x}|^2
+ 1.8 |C_{\varphi\nu dsu}^{\tL\tR,x}|^2
+ 4.94 |C_{\varphi\nu sud }^{\tL\tR,x}|^2
+ 3 |C_{\varphi\nu dsu}^{\tL\tL,x}|^2
+ 5.04 |C_{\varphi\nu sud }^{\tL\tL,x}|^2
\nonumber
\\
& - 1.24 \Re(C_{\varphi\nu uds}^{\tL\tR,x} C_{\varphi\nu dsu}^{\tL\tR,x*})
-2.29 \Re(C_{\varphi\nu uds}^{\tL\tR,x} C_{\varphi\nu sud}^{\tL\tR,x*})
+5.81 \Re(C_{\varphi\nu dsu}^{\tL\tR,x} C_{\varphi\nu sud}^{\tL\tR,x*})
\nonumber
\\
&-1.72 \Re(C_{\varphi\nu uds}^{\tL\tR,x} C_{\varphi\nu dsu}^{\tL\tL,x*})
+2.31 \Re(C_{\varphi\nu uds}^{\tL\tR,x} C_{\varphi\nu sud}^{\tL\tL,x*})
+4.61 \Re(C_{\varphi\nu dsu}^{\tL\tR,x} C_{\varphi\nu dsu}^{\tL\tL,x*})
\nonumber
\\
& - 5.86 \Re(C_{\varphi\nu dsu}^{\tL\tR,x} C_{\varphi\nu sud}^{\tL\tL,x*})
+ 7.67 \Re(C_{\varphi\nu sud}^{\tL\tR,x} C_{\varphi\nu dsu}^{\tL\tL,x*})
-9.98 \Re(C_{\varphi\nu sud}^{\tL\tR,x} C_{\varphi\nu sud}^{\tL\tL,x*})
\nonumber
\\
&
-7.74 \Re(C_{\varphi\nu dsu }^{\tL\tL,x} C_{\varphi\nu sud}^{\tL\tL,x*}),
\end{align}
\end{subequations}
and these modes with a neutrino in the final state can be directly obtained from the above expressions by interchanging $\bar\nu \leftrightarrow \nu$ and $\tL\leftrightarrow\tR$ simultaneously. 

%%%%%%%%%%%%%%%%%%%%%%%%%%%%%%
\section{Matrix elements for dinucleon decays}
\label{app:M2fordinucleon}
%%%%%%%%%%%%%%%%%%%%%%%%%%%%%%

The amplitudes for the three types of dinucleon processes depicted in \cref{fig:NN2ll} are  
\begin{subequations}
\begin{align}
 {\cal M}(p p \to \ell_x^+ \ell_y^+) 
 & = 
 \frac{1}{m_\varphi^2 - \hat t} 
 \overline{v^\C_{x}}(C_{p \ell_x}^\tL P_\tL + C_{p \ell_x}^\tR P_\tR)u(p_1) \,
\overline{v^\C_{y}}(C_{p \ell_y}^\tL P_\tL + C_{p \ell_y}^\tR P_\tR)u(p_2) 
\nonumber
\\
& - \frac{1}{m_\varphi^2 - \hat u} 
 \overline{v^\C_{x}}(C_{p \ell_x}^\tL P_\tL + C_{p \ell_x}^\tR P_\tR)u(p_2) \,
\overline{v^\C_{y}}(C_{p \ell_y}^\tL P_\tL + C_{p \ell_y}^\tR P_\tR)u(p_1) , 
\\%
 {\cal M}(p n \to \ell_x^+ \hat \nu_y) 
 & = 
\frac{1}{m_\varphi^2 - \hat t} 
\overline{v^\C_{x}}(C_{p \ell_x}^\tL P_\tL + C_{p \ell_x}^\tR P_\tR)u(p_1) \,
\overline{v^\C_{y}}(C_{n \nu_y}^\tL P_\tL + C_{n \nu_y}^\tR P_\tR)u(p_2),
\\
{\cal M}(nn \to \hat\nu_x \hat\nu_y) 
 & = 
 \frac{1}{m_\varphi^2 - \hat t} 
 \overline{v^\C_{x}}(C_{n \nu_x}^\tL P_\tL + C_{n \nu_x}^\tR P_\tR)u(p_1) \,
\overline{v^\C_{y}}(C_{n \nu_y}^\tL P_\tL + C_{n \nu_y}^\tR P_\tR)u(p_2) 
\nonumber
\\
& - \frac{1}{m_\varphi^2 - \hat u} 
 \overline{v^\C_{x}}(C_{n \nu_x}^\tL P_\tL + C_{n \nu_x}^\tR P_\tR)u(p_2) \,
\overline{v^\C_{y}}(C_{n \nu_y}^\tL P_\tL + C_{n \nu_y}^\tR P_\tR)u(p_1) , 
\end{align}
\end{subequations}
where $\hat t =(p_1-k_1)^2$ and $\hat u = (p_1-k_2)^2$.
$\hat\nu=\nu,\bar\nu$ represents both neutrino and antineutrino states.

\bibliography{refs}
\bibliographystyle{utphys28mod}

\end{document}